\documentclass[a4paper,12pt]{article}
\usepackage{graphicx}
\usepackage{tikz}
\usetikzlibrary{plotmarks}
\usepackage{tikzscale}
\usepackage[bf,small]{caption}
\usepackage{subcaption}
\usepackage{epstopdf}
\usepackage{chngcntr}

\usepackage{multirow}
\usepackage{bbm}
\usepackage{float}
\usepackage{footnote}
\usepackage{amssymb}
\usepackage{pifont}
\usepackage{ifpdf}
\ifpdf
\else
\usepackage{pstcol,pst-fill,pstricks}
\fi
\usepackage{natbib}
\usepackage{mathpazo}
\usepackage{import}
\usepackage{dsfont}
\usepackage{enumerate}
\usepackage{a4}
\usepackage{lscape}
\usepackage{epsfig}
\usepackage[latin1]{inputenc}
\usepackage[OT1]{fontenc}
\usepackage[T1]{fontenc}
\usepackage{color}
\usepackage{geometry}
\usepackage{mdframed}
\usepackage{tikz}
\usepackage{tkz-graph}
\usepackage[bottom,flushmargin]{footmisc}

\makeatletter
\newcommand*\bigcdot{\mathpalette\bigcdot@{.5}}
\newcommand*\bigcdot@[2]{\mathbin{\vcenter{\hbox{\scalebox{#2}{$\m@th#1\bullet$}}}}}
\makeatother
\usepackage{authblk}
\usepackage{setspace}
 \newcommand{\be}{\begin{equation}}
\newcommand{\ee}{\end{equation}}
\newcommand{\E}{\mathbb{E}}
\usepackage{mathtools}
\usepackage{amsmath}
\usepackage{amsthm}
\usepackage{thmtools}

\DeclareMathAlphabet{\pazocal}{OMS}{zplm}{m}{n}
\allowdisplaybreaks

\usepackage[hyperfootnotes=false]{hyperref}
 \hypersetup{colorlinks=true,
 linkcolor=blue,
 citecolor=blue,
 filecolor=black,
 urlcolor=black,
 pdfstartview={Fit},
pdfpagemode=UseNone
 }
\usepackage{cleveref}

\declaretheoremstyle[
  headfont=\normalfont\scshape,
  numbered=unless unique,
  bodyfont=\normalfont,
  spaceabove=1em,
  prefoothook=\newline\rule{\linewidth}{1pt},
  spacebelow=1em,
]{exmpstyle}

\declaretheorem[
  style=exmpstyle,
  title=\textbf{Definition},
  refname={example,examples},
  Refname={Example,Examples}
]{ddef}

\newcommand{\defeq}{\vcentcolon=}

\def\ee{\mathsf{e}}

\DeclareMathOperator*{\argmin}{arg\,min}

\newtheorem{Ass}{Assumption}[section]
\newtheorem{prop}{Proposition}
\newtheorem{lemma}{Lemma}[section]
\theoremstyle{definition}
\newtheorem{definition}{Definition}[section]
\newtheorem{theorem}{Theorem}
\newtheorem{corollary}{Corollary}

\title{Endogenous Persistence at the Effective Lower Bound\thanks{Corresponding author: Jordan Roulleau-Pasdeloup: \href{mailto:ecsjr@nus.edu.sg}{ecsjr@nus.edu.sg}. We thank Guido Ascari, Christian Bayer, Gauti Eggertsson, Martin Eichenbaum, Tom Holden, Mariano Kulish, Nigel McClung, Karel Mertens, Taisuke Nakata, Tom Sargent, Sebastian Schmidt, Sarah Zubairy as well as seminar participants from the National University of Singapore, Nazarbayev University, Wuhan University, Academia Sinica, the Bank of England, Oxford University, the Central Bank of Ireland, De Nederlandsche Bank, the European Central Bank as well as participants from the 2024 ICMAIF, CEF, IAAE, SRS and the 2025 EEA conferences for interesting comments. We also thank Shinji Takenaka and the Japan Center for Economic Research for providing individual-level ESP forecast data, which were used in parts of the empirical analyses reported in Section 4 of the paper. All remaining errors are ours.}}
\author[1]{Chunbing Cai}
\author[2]{Jordan Roulleau-Pasdeloup}
\author[3]{Zhongxi Zheng}
\affil[1,2,3]{Department of Economics, National University of Singapore}
\date{\today}

\begin{document}

\begin{titlepage}
\maketitle
\thispagestyle{empty}

\abstract{We develop a perfect foresight method to solve models with an interest rate lower bound constraint that nests \cite{Eggertsson2011}'s as well as \cite{Mertens2014}'s pen and paper solution methods as special cases. Our method generalizes these solution methods by allowing for endogenous persistence while maintaining tractability and interpretability. We prove that our method necessarily gives stable multipliers. We use it to solve a New Keynesian model with habit formation and government spending, which we match to expectations data from the Great Recession. We find an output multiplier of government spending close to 1 for both the US and Japan.}
\vspace{.5cm}
\end{titlepage}
\setstretch{1.5}
\setlength{\parskip}{1em}

\setlength{\abovedisplayskip}{1.5ex}
\setlength{\belowdisplayskip}{1.5ex}

\section{Introduction}

\begin{quote}
``Two views compete in macro when it comes to the use of models. One view is that models should be simple so as to yield insight. Another view is that the goal of modelling is to be able to do policy experiments. Trouble is that these two views are strongly conflicting.''   \hfill
\textemdash\ J\'{o}n Steinsson\footnote{See \url{https://x.com/jonsteinsson/status/1508671116801282053?s=46&t=hy0jETnoyf4aKU2ip8dm7g} (Accessed on September 1st, 2025.)}
\end{quote}

As evidenced by this quote, there is a tension between simple models used for insight and medium- to large-scale ones used for policy experiments. There are two ways to resolve this tension: making simple models more empirically relevant or making medium- to large-scale models more interpretable. The last option is fraught with difficulty due to the sheer complexity of this class of models. With an emphasis on interpretability, the main objective in this paper is to bridge this gap by extending the class of models used for insight and make them more amenable to policy experiments.  

This consideration is especially relevant when one wants to make policy recommendations in the context of large recessions with a magnitude comparable to the recent Great Recession or the Covid-19 crisis. Indeed, this tension is even more true when the model in question is used to study a large recession with an occasionally binding constraint: this usually limits models used for insights to ones without endogenous propagation mechanisms. Take the effective lower bound \textemdash\ henceforth ELB. In that context, there is a large and growing literature kickstarted by \cite{Eggertsson2011}, \cite{Christiano2011}, \cite{Woodford2011} and \cite{Mertens2014} that has sought to gain insights about the effects of policy at the ELB. The main insight about the effects of policy at the ELB coming from these models is that expectations conditional on being in a recession matter a lot. If recessionary dynamics are expected to be short-lived, we are in a world where fiscal policy has more stimulative power compared to normal times \textemdash\ see \cite{Eggertsson2011}. If recessionary dynamics are expected to be long-lived instead, we are in a world where fiscal policy has less stimulative power compared to normal times\textemdash see \cite{Mertens2014}. As a testament to the insightful nature of these models, one can produce simple aggregate supply/demand graphs at the ELB and use these to tell those two situations apart\textemdash see \cite{Bilbiie2022neo}. 

In order to take these policy prescriptions seriously, the underlying model should be able to replicate the salient features of expectations in a large recession. Using professional forecasters' expectations data for the U.S. and Japan, we document that these usually display a hump-shape at the onset of a large recession: forecasters expect things to get worse before they get better. We show that while expectations are crucial in the literature cited above, the models used cannot match this hump-shape by construction: these models need to be purely forward-looking in order for the clever tricks used to get a pen and paper solution to work. 

One solution would be to augment these models with a mechanism that injects endogenous persistence. Unfortunately, there does not exist a tractable/interpretable analytical solution method that allows for occasionally binding constraints and generalizes the one used in this literature yet. Currently available alternatives include piecewise linear deterministic algorithms (OccBin (\cite{Guerrieri2014}), DynareOBC (\cite{Holden2016computation,Holden2023existence})) the piecewise linear stochastic algorithms developed in \cite{Eggertsson2003}, \cite{Eggertsson2021toolkit} for Markov chain shocks as well as  \cite{Adam2006,Adam2007discretionary} for $AR(1)$ shocks, or a fully global stochastic solution method (\cite{Fernandez2015nonlinear}, \cite{Cao2023global}). As it currently stands, these algorithms are used to find a numerical approximation of the solution. 

Accordingly, our main goal in this paper is to develop an easily interpretable analytical solution method that generalizes the one used in the existing literature and thus can handle models that feature endogenous persistence in order to match conditional expectations in the data. To do so, we will build on \cite{Roulleau2023analyzing} who shows that one can recast a \textit{linear} DSGE model with endogenous persistence as a suitably defined finite-state Markov Chain. This result holds for linear models and thus precludes the analysis with an occasionally binding ELB constraint. On the other hand, the literature on the standard New Keynesian (henceforth NK) model without endogenous persistence at the ELB that followed \cite{Eggertsson2011} has made a heavy use of Markov chains. We show that the simple NK model developed in \cite{Eggertsson2011} is isomorphic (in expectations) to a perfect foresight model with an endogenous peg for the nominal interest rate. We then extend that insight to construct an endogenous peg for a model \textit{with} endogenous persistence which explicitly nests \cite{Eggertsson2011} (and the subsequent literature) as a special case. This peg will give us a terminal condition upon exit from the ELB and we follow \cite{Guerrieri2014} in assuming perfect foresight in previous periods. Just as in \cite{Eggertsson2011}, our approach will lend itself to an insightful graphical representation in terms of aggregate demand (AD) and supply (AS) curves. As a result our method will be different from the one developed in \cite{Eggertsson2021toolkit} in that it will lend itself to an amenable closed form solution and will nest the dynamics featured in \cite{Mertens2014} as a special case.\footnote{In \cite{Eggertsson2021toolkit}, if the persistence is above threshold then the equilibrium effect will be undefined. See \cite{RoulleauZheng2025} for details.} Note that, because we consider a perfect foresight equilibrium these dynamics will not be the result of a sunspot.

We show that our endogenous peg method is able to exactly replicate the results obtained with existing methods and we use it to evaluate policy prescriptions at the ELB. Given the extensive literature on the topic, we choose to focus on the government spending multiplier. Beyond being able to replicate the salient features of a large recession, we take it as a requirement that the model should not produce policy multipliers that can be arbitrarily large\textemdash see \cite{Cao2023nonlinear}. This feature is usually referred to as a "puzzle" and there is a large literature on the topic\textemdash see \cite{Michaillat2019resolving} and \cite{Gibbs2023does} as well as references therein. It has been shown in the literature that existing standard NK models can produce policy multipliers that flip qualitatively. More precisely, \cite{Mertens2014} and \cite{Bilbiie2022neo} have shown that this happens if the persistence $p\in (0,1)$ of the structural/sunspot shock that brings the economy at the ELB is more than a threshold $\overline{p}\in (0,1)$. In that case, the policy multiplier can be arbitrarily large if $p$ is in a neighborhood of $\overline{p}$. Using our method, we show that if one were to solve the \textit{same model} with either OccBin or DynareOBC, the policy prescription would \textit{also} switch if $p$ crosses $\overline{p}$. In contrast however, the policy multiplier can now be arbitrarily large \textit{for all} $p>\overline{p}$: a much bigger region of the parameter space.

Here is the intuition for why policy multipliers can become arbitrarily large. When solving the model using OccBin or DynareOBC, a persistent policy enacted at the ELB will modify the allocation upon exit. As a result, the Central Bank will adjust its interest rate accordingly upon exit. For example, assume that the policy causes the Central Bank to increase its interest rate \textit{ceteris paribus}. If the persistence $p$ is above threshold, then that decrease will decrease consumption upon exit and consumption in the preceding period will decrease even more: the further the exit, the stronger this effect. In our solution method, the endogenous peg rules out such a feedback loop. 

As an application, we use a New Keynesian model with consumption habits to study the effects of government spending at the ELB. In order to discipline the model, we develop a penalized minimum-distance estimation procedure to replicate the measured expectations from professional forecasters at the onset of the Great Recession in both the U.S and Japan. Using our method, we find that the effects of government spending at the ELB in the U.S is best represented by an AS line that slides along a less steep AD line: consumption is crowded in as in \cite{Eggertsson2011}. In Japan, we find that the economy is best represented by an AS line that slides along a \textit{steeper} AD line: consumption is crowded out as in \cite{Mertens2014}. In both cases however, the implied output multiplier is quite close to 1. Given our estimated parameter values, we compute the multiplier using the algorithm in OccBin/DynareOBC and find that the government spending multiplier grows without bounds with the expected ELB duration in the U.S case, but converges to a finite value in the Japanese case. 

\noindent \textbf{Related Literature}\textemdash Given our focus on computing an equilibrium at the ELB using a piece-wise linear model, our paper is related to \cite{Eggertsson2003}, \cite{Adam2006,Adam2007discretionary}, \cite{Eggertsson2012new},  \cite{Cagliarini2013solving}, \cite{Guerrieri2014}, \cite{Boneva2016},  \cite{Kulish2017estimating}, \cite{Aruoba2018}, \cite{Eggertsson2019log}, \cite{Holden2016computation,Holden2023existence}, \cite{Eggertsson2021toolkit}, \cite{Cao2023nonlinear}, \cite{Gibbs2023does} and \cite{Cuba2024understanding}. 

We use our piece-wise linear model to study the effects of government expenses at the ELB. As a result, we are related to a large stream of papers that includes \cite{Eggertsson2011}, \cite{Christiano2011}, \cite{Woodford2011}, \cite{Mertens2014}, \cite{Schmidt2017fiscal}, \cite{Leeper2017clearing}, \cite{wieland2018state}, \cite{Hills2018fiscal}, \cite{Miyamoto2018}, \cite{wieland2019zero}, \cite{Nakata2022expectations} and \cite{Bilbiie2022neo}.  

In order to derive stability conditions for policy multipliers at the ELB we use results from the theory of quadratic matrix equations. In particular, we rely on \cite{Higham2000numerical} and \cite{Gohberg2009matrix}. We share this mathematical reference with \cite{Rendahl2017linear},  \cite{Meyer2022solving} and \cite{Meyer2024solving} who use it to solve linear models that abstract from any occasionally binding constraints.\footnote{In that regard, \cite{Rendahl2017linear} does apply his Linear Time Iteration method to a model that features an ELB constraint, but the model doesn't feature endogenous persistence.}

Finally, in using data from the Survey of Professional Forecasters to evaluate expectations, our approach relates to \cite{Coibion2015}, \cite{coibion2015information}, \cite{Bordalo2018diagnostic}, \cite{Angeletos2021imperfect} and \cite{Gorodnichenko2021zero}. See \cite{Coibion2018formation} for a recent survey of this literature. 

Our paper is structured as follows. In Section \ref{sec:general_framework}, we develop a general framework to solve for the impulse response in a class of piece-wise linear DSGE models. In Section \ref{sec:application}, we apply our framework to a New Keynesian model with habit formation and an occasionally binding ELB constraint. We match it with expectations data from the U.S Great Recession and then study the government spending multiplier at the ELB. In Section \ref{sec:Japan}, we conduct a similar analysis for the case of Japan. Section \ref{sec:ccl} concludes.

\section{Multipliers at the ELB: an equivalence result and a recursive representation}\label{sec:general_framework}
\vspace*{-0.8em}

In this section, we focus on a class of piece-wise linear DSGE models. Historically, these models have been studied with two popular methods: (\textit{i}) the Markov chain approach pioneered in \cite{Eggertsson2011}, \cite{Christiano2011}, \cite{Woodford2011}, \cite{Mertens2014} and \cite{Bilbiie2022neo} as well as (\textit{ii}) the perfect foresight  numerical approaches with $AR(1)$ shocks developed in \cite{Cagliarini2013solving}, \cite{Guerrieri2014} (OccBin) and \cite{Holden2016computation,Holden2023existence} (DynareOBC). For future reference, we let MC-CF (for Markov Chain - Closed Form) refer to the literature cited in (\textit{i}) and AR-NA (for Auto Regressive - Numerical Approximation) refer to the literature cited in (\textit{ii}). The MC-CF method has been applied to solve models without endogenous persistence while the AR-NA method applies to a broader class of models. 

We show that the impact multipliers obtained with the MC-CF methods can be exactly replicated with the AR-NA method if one assumes a specific peg for the nominal interest rate. We leverage that insight and then show how to generalize this construction to a model with endogenous persistence using a variation of the method developed in \cite{Roulleau2023analyzing}. In turn, that construction allows us to derive a recursive representation for impact multipliers. We then use that representation to prove that, while AR-NA methods are prone to instability as the duration of the ELB increases, a suitable peg is guaranteed to yield stable impact multipliers regardless of the duration of the ELB episode.

\subsection{A class of piece-wise linear DSGE models}
\vspace*{-0.5em}

We assume that the vector of forward-looking variables is given by a vector $Y_t$ of size $N\times 1$, all in log-deviations from the non-stochastic steady state. There is a single endogenous backward-looking variable $x_t$. We collect all the structural parameters of our model in a vector $\theta$. We consider experiments where an exogenous, auto-regressive \textit{baseline} shock $w_{b,t}$ with persistence $p_{b}\in (0,1)$ makes the constraint bind for the first $\ell\geq 1$ periods. When that happens, we assume a \textit{scenario} where another shock $w_{s,t}$ with persistence $p_{s}\in (0,1)$ is implemented. This shock could be a policy like in the literature on the government spending multiplier or a technology shock as in \cite{Garin2016} and \cite{Wieland2018}. In line with AR-NA but in sharp contrast with the majority of the MC-CF literature, we allow for the possibility that $p_{b}\neq p_{s}$.\footnote{A few notable exceptions of the MC-CF literature with different persistence parameters are \cite{Eggertsson2012new}, \cite{wieland2018state} and \cite{wieland2019zero}.} Under these assumptions, the forward-looking block of the model is given by:
\begin{equation}
\label{eq:fwd_1}
Y_{t+n} = \mathbf{A}^\ast \E_t Y_{t+n+1} + B^\ast x_{t+n} + C_{b}^\ast w_{b,t+n}+ C_{s}^\ast w_{s,t+n} + E_{t+n}^\ast,
\end{equation}
for $n=0,\dots,\ell-1$, where all the matrices and vectors of parameters are conformable.\footnote{In principle, the first order conditions are written as $\mathbf{A}_0^\ast Y_{t+n} = \mathbf{A}_1^\ast \E_t Y_{t+n+1} + B_0^\ast x_{t+n} + C_{0,b}^\ast w_{b,t+n}+ C_{0,s}^\ast w_{s,t+n} + E_{0,t+n}^\ast$. We are effectively assuming here that $\mathbf{A}_0^\ast$ is non-singular and thus invertible. We assume the same for $\mathbf{A}_0$ outside the ELB. We effectively rule out cases where the OBC binds with a lag after the shock hits for analytical tractability and for a better comparison with the existing literature. Indeed, papers in the tractable DSGE literature at the ELB focus on variants of the perfectly forward looking standard New Keynesian model in which the ELB necessarily binds on impact.} Following the AR-NA literature, we assume perfect foresight. The time-varying term $E_{t+n}^\ast$ arises when monetary policy is passive. When the ELB is binding, this term will be given by a constant $E_{t+n}^\ast=E^\ast$. When one assumes a peg, this term will be potentially time-varying outside the ELB. The main contribution of this paper will be to show how to construct this peg so that it exactly nests the existing MC-CF literature as a special case. If the same model is solved with OccBin/DynareOBC, then the Taylor rule kicks back in immediately upon exit and $E_{t+n}^\ast=0_{N\times 1}$ outside the ELB. When the ELB isn't binding anymore, we then have:
\begin{equation}
\label{eq:fwd_2}
Y_{t+n} = \mathbf{A} \E_t Y_{t+n+1} + B x_{t+n} + C_{b} w_{b,t+n}+ C_{s} w_{s,t+n},
\end{equation}
from $n=\ell$ onward. We consider experiments where the path of the nominal interest rate can be written as $r_{t+n} = \underline{r}$ for $n=0,\dots,\ell-1$ and $r_{t+n}=f(n;\theta)$ for $n\geq \ell$,
where $\underline{r}<0$ is the effective lower bound expressed in deviations from the intended steady state. That formulation nests the usual Taylor rule if one sets $f(n;\theta) \defeq \phi Y_{t+n}$, where $\phi$ is such that the \cite{Blanchard1980solution} condition holds. In our method, we set $f(n;\theta)$ in such a way that $(i)$ it nests the Taylor rule if the ELB is not binding and $(ii)$ it also nests the MC-CF literature if we get rid of endogenous persistence. With some slight abuse of language, our formulation amounts to an endogenous peg. We will describe in detail later how we parameterize it. The backward equation is independent of the constraint and is governed by:
\begin{equation}
\label{eq:bwd}
x_{t+n} = \varrho x_{t+n-1} + DY_{t+n},
\end{equation}
where we have assumed that the presence of the OBC does not change the backward equation for simplicity.\footnote{There are cases where this assumption does not hold: if the endogenous state variable is public debt, then the backward equation will include the nominal interest rate and thus change at the lower bound.} We keep the dependence on the vectors/matrices of parameters $\theta$ implicit for expositional clarity.

With these in mind, our main objective is to derive an expression for the impact effect of the shock $w_{s,t}$ when the constraint is binding for $\ell\geq 1$ periods. In the class of models that we consider, defining the impact effect is far from straightforward. In principle, we want to simulate our model twice: once for a given value of the baseline shock $w_{b,t}$, and a second time with the same shock, but with $w_{s,t}$ in addition. The difference (scaled by $w_{s,t}$) between the two will be our impact multiplier. Throughout the paper, we maintain the assumption that the second shock $w_{s,t}$ is small enough so as to not influence the duration of the ELB period.

\subsection{Computing exit dynamics with Markov chains}
\vspace*{-0.5em}

Building on \cite{Roulleau2023analyzing}, we will exploit the fact that dynamics upon exit from the ELB and back to the steady state can be written in terms of suitably specified Markov chains. That will allow us to make a connection with the MC-CF literature, which has developed tools to compute the impulse response of simple NK models without endogenous persistence and an occasionally binding ELB constraint in closed form using Markov chains. These exit dynamics are described as follows:

\begin{ddef}[Markov chain representation]
\label{ddef:MChains_inertial}
Let us define a Markov chain $\mathbf{Z}_t$ for variable $z_{t+n}\in \{Y_{t+n},x_{t+n},w_{b,t+n},w_{s,t+n}\}$ for $n\geq \ell-1$. All Markov chains are characterized by an initial distribution $u$, transition matrix $\mathcal{P}_\ell$, and a vector of states $S_z$:
\begin{align*}
u^\top = \begin{bmatrix}
1 \\
0 \\
0 \\
0
\end{bmatrix}
\mathcal{P} =
\begin{bmatrix}
p_s & 1-p_s & 0      & 0\\
0   & p_b   & 1-p_b  & 0\\
0   &  0    & q      & 1-q\\
0   & 0     & 0      & 1
\end{bmatrix}
S_z = 
\begin{bmatrix}
s_{z,1}\\
s_{z,2}\\
s_{z,3}\\
0
\end{bmatrix}
\end{align*}
where $\top$ is the transpose operator. The initial distribution ensures that they start in the first state. The matrix $S_z$ collects all the vectors of Markov states. Both $q$ and $S_z$ have to be solved for. Importantly, $s_{z,1}$ is constructed assuming the ELB constraint binds. 
\end{ddef}

In order to solve for $q$, as well as the Markov states $s_{z,2}$ and $s_{z,3}$, we use the method in \cite{Roulleau2023analyzing} for a linear model with the ELB constraint.\footnote{In a model with multiple endogenous states such as a typical medium-scale model, we would have several values of $q$ to compute. Given that these values correspond to the policy function parameters in a typical state space solution, it is highly likely that we end up with a pair of complex conjugates. In that case, we cannot case the exit dynamics in terms of a Markov chain. For this reason, we focus on the case where there is only one endogenous state so that we can guarantee that $q$ takes on a real value.} In contrast, $s_{z,1}$ is solved for by guessing, and then verifying, that the ELB binds: this will be the last period of the ELB episode.\footnote{In the MC-CF literature, the ELB only binds for $\ell=1$ period \textit{in expectations}, so in that case $s_{z,1}$ is such that the ELB is binding and $s_{z,2}$ is such that it is not.} We compute the dynamics for the $\ell-1$ initial periods at the ELB under perfect foresight by enlarging the size of our Markov chain and consider transition probabilities of 1 to replicate deterministic dynamics. This yields a total of $\ell+3$ Markov states: $\ell-1$ for the initial periods, 1 for the last period of binding ELB and the last 3 for the dynamics back to steady state under the peg. These Markov states are solved by deriving a set of Markov restrictions that we describe in Appendix \ref{apsec:proof_prop_1}.

Let us denote by $\mathcal{P}_\ell$ the transition matrix for such an extended Markov chain and $S_r$ the vector of Markov states for the nominal interest rate with the first $\ell$ states given by $\underline{r}<0$. In that case, the nominal interest rate is given by $f(n;\theta) \defeq u\cdot\mathcal{P}_\ell^n\cdot S_r$. Given this peg, we can compute a minimum state variable (MSV) solution. In turn, this MSV solution gives us a terminal condition while at the ELB. Following \cite{Cagliarini2013solving}, given a terminal condition we can compute a unique path for endogenous variables while at the ELB. Once one has chosen a peg and thus a terminal condition, the initial impact of the shocks on inflation and other endogenous variables is then uniquely determined. This is in contrast with \cite{Cochrane2017} where, for a given interest rate path, one can select an equilibrium level of inflation on impact.

Such a choice for the monetary policy rule may seem arbitrary at first glance. The main reason for this choice is that the equilibrium we compute will exhibit desirable properties. Indeed, we can guarantee that the equilibrium we compute under our monetary policy rule is such that: (1) it nests the MC-CF literature as a special case and, perhaps more importantly, (2) it will give impact policy multipliers that are guaranteed to be finite. Neither (1) nor (2) holds in AR-NA and the models used in MC-CF cannot accommodate for endogenous persistence.\footnote{One notable exception is \cite{Eggertsson2012new}, who considers a model with external habit formation in consumption and leisure. In this special case where both habits have the exact same degree of inertia, the model boils down to one where a quasi-growth rate of consumption replaces actual consumption in the Euler equation. Rewritten in this way, the model is perfectly forward-looking and can be solved with the tools from the MC-CF literature. We will discuss this in more detail in Section \ref{sec:application}.}

Given the fact that Markov chains are step functions, it is not a guarantee that they do match the equilibrium conditions of the model with endogenous persistence. The key intuition here is that even though any single run of a Markov chain is a step function, the expectation across all possible runs is a deterministic, auto-regressive process. In that context, irrespective of the nature of the endogenous peg, the conditional expectations from the Markov chain approach are consistent with the model equilibrium conditions by construction: $\E\left(\mathbf{Z}_{t+n}|w_{b,t},w_{s,t}\right)=z_{t+n}(w_{b,t},w_{s,t};\theta)$.

Let us now turn to the case where the nominal interest rate is set according to a standard Taylor-type rule. This is the case considered in the AR-NA literature. It turns out that we can also exactly replicate the methods employed in that literature with an extended Markov chain. We describe in the online appendix how to derive a set of necessary Markov restrictions to achieve this. The main advantage of re-casting this well-known method is that it will make it possible to derive a recursive representation for the impact multipliers. We define these as follows:

\begin{ddef}\label{def:impact_multiplier}
The impact multiplier effect for variable $z$ is defined as:
\begin{align*}
\pazocal{M}_z(\ell;\theta) \equiv \lim_{w_{s,t}\to 0}\frac{\E\left(\mathbf{Z}_{t}|w_{b,t},w_{s,t}\right)-\E\left(\mathbf{Z}_{t}|w_{b,t},0\right)}{w_{s,t}},  
\end{align*}    
which can also be interpreted as $\partial \E\left(\mathbf{Z}_{t}|w_{b,t},w_{s,t}\right)/\partial w_{s,t}$. The vector of stacked impact multipliers is defined as $\pazocal{M}(\ell;\theta)=\left[\pazocal{M}_{y_1}(\ell;\theta),\ \pazocal{M}_{y_2}(\ell;\theta),\dots,\ \pazocal{M}_{y_N}(\ell;\theta)\right]^\top$.
\end{ddef}

We are now ready to derive one of the main results of the paper: the impact multiplier effect for a duration of $\ell$ periods can be expressed recursively for AR-NA, MC-CF, and our method with a peg. 

\subsection{A recursive representation for policy multipliers}
\vspace*{-0.5em}

The spirit behind that recursive representation is that if one can compute impact multipliers under both methods for a ELB duration of $\ell=1$, then our result enables a straightforward computation of multipliers for a duration of $\ell\geq 2$. This is useful for someone using AR-NA as our method bypasses the need to simulate the model for different values of the baseline shock $w_{b,t}$. Perhaps more importantly, our result will allow us to derive clear stability conditions for how impact multipliers vary with $\ell$. 

\begin{prop}[Impact multiplier]
\label{prop:impact_multiplier}
Suppose $p_b$ and $w_{b,t}$ are defined such that the constraint binds for $\ell$ periods. Then the sequence of impact multipliers for $\ell\geq 2$ obeys
\begin{align}
\label{eq:M_ell}
\mathcal{M}(\ell;\theta) &= \left(\mathbf{A}^\ast\right)^{-1}\mathcal{X}_{\ell-1} \left[C_s^\ast + p_s\mathbf{A}^\ast\mathcal{M}(\ell-1;\theta)\right]\\
\label{eq:x_ell}
\mathcal{X}_{\ell} \defeq F(\mathcal{X}_{\ell-1};\theta) &= \mathbf{A}^\ast\big(\mathbf{I}_{N}-B^\ast D + \varrho\mathbf{A}^\ast -\varrho\mathcal{X}_{\ell-1}\big)^{-1},
\end{align}
given initial conditions $\pazocal{M}(1;\theta)$ and $\mathcal{X}_1$, where $\mathbf{I}_{N}$ is the identity matrix of size $N$.
\end{prop}
\vspace*{-2em}
\begin{proof}
See Appendix \ref{apsec:proof_prop_1}.
\end{proof}
Taking stock, one can see from equation \eqref{eq:M_ell} that the sequence $\left\{\pazocal{M}(\ell;\theta)\right\}_{\ell\geq 1}$ follows a linear, discrete, time-varying parameter dynamical system. From that equation, one also notices that only the persistence of the second shock $w_{s,t}$ in the scenario appears explicitly.\footnote{This echoes the findings of \cite{wieland2018state}, where he shows that the persistence of government spending and not the demand shock that matters for the government spending multiplier.} The time-varying part comes from the fact that we have a time-varying matrix $\mathcal{X}_{\ell-1}$ in front of both the "drift" vector $C^\ast_s$ and the past multiplier. From equation \eqref{eq:x_ell}, we see that the sequence $\left\{\mathcal{X}_{\ell}\right\}_{\ell\geq 1}$ also obeys a discrete dynamical system, but a non-linear one. While there are many general results for linear, discrete \textit{constant parameters} dynamical systems, there are much less for time-varying parameters or non-linear systems. As a result, there are no results that we can import from the mathematics literature on dynamical systems to solve \eqref{eq:M_ell} \textit{analytically}. 

However, Proposition \ref{prop:impact_multiplier} provides some clues about how to go about solving for the sequence of impact multipliers. Indeed, notice that the dynamics of $\mathcal{X}_{\ell}$ are completely autonomous. So in principle, we can solve for these dynamics and then use them to solve for the dynamics of $\pazocal{M}(\ell;\theta)$ as a second step. Ideally, we want to know whether the sequence $\left\{\pazocal{M}(\ell;\theta)\right\}_{\ell\geq 1}$ has a well defined limit $\pazocal{M} <\infty$. If it does, then we would like to know under which conditions the sequence actually converges to that limit. 

Before going further, we note that one can find a related recursive expression in \cite{Guerrieri2014}. It is however different from us in a few aspects. To understand why, note that the state-space solution that they consider for a model without an ELB is written as $X_t = P X_{t-1} + Q\epsilon_t$, where $\epsilon_t$ regroups the innovation terms and $X_t$ includes both the scenario and baseline processes. The solution at the ELB is a generalization where $P_t$ will be time-varying and depends on the duration of the ELB. In that context, \cite{Guerrieri2014} derive a recursion for $P_t$ which doesn't lend itself to a recursion for the impact multiplier, which is the object of interest for us. 

\subsection{A stability condition for the sequence of multipliers}
\vspace*{-0.5em}

It turns out that a necessary condition for the sequence of multipliers to have a well-defined limit as $\ell\to\infty$ is that the sequence $\left\{\mathcal{X}_\ell\right\}_{\ell\geq 1}$ converges to a real-valued matrix. We prove in Appendix \ref{apsec:dynamics_X} that this sequence is guaranteed to converge to its minimal solution. We further assume that the structural parameters of the model are such that this minimal solution is real-valued.\footnote{If that minimal solution is complex valued instead, we end up with a "reversal puzzle" as in \cite{Carlstrom2015inflation}. We leave this avenue for future research.} Given this, it is quite straightforward to construct a fixed point of equation \eqref{eq:M_ell}. Our next objective is to study whether the sequence $\left\{\pazocal{M}(\ell;\theta)\right\}_{\ell\geq 1}$ does converge to such a fixed point. This is a difficult question because $\pazocal{M}(\ell;\theta)$ depends on the product $\Pi_{i=1}^{\ell}\mathcal{X}_i$ \textemdash\ this can be seen by repeated substitution of equation \eqref{eq:M_ell}. We show in the following theorem that this question can be given a definitive answer:
\begin{theorem}
\label{thm:saddle}
Let $\left\{\pazocal{M}(\ell;\theta)\right\}_{\ell\geq 1}$ be the sequence defined recursively in Proposition \ref{prop:impact_multiplier}. Assume that the minimal solution $\underline{\mathcal{X}}$ is real-valued. If it is such that the eigenvalues of $p_s\underline{\mathcal{X}}$ are all in the unit circle, then:
\begin{align*}
\lim_{\ell\to\infty}\pazocal{M}(\ell;\theta) = \pazocal{M} <\infty,
\end{align*}
regardless of the initial condition. If the limit $\underline{\mathcal{X}}$ of sequence $\left\{\mathcal{X}_\ell\right\}_{\ell\geq 1}$ is such that at least one of the eigenvalues of $p_s\underline{\mathcal{X}}$ is larger than 1, then:
\begin{align*}
\lim_{\ell\to\infty}\pazocal{M}(\ell;\theta) = \pazocal{M} <\infty
\end{align*}
if $\pazocal{M}(1,\theta)=\pazocal{M}^f(1;\theta)$ as well as $\mathcal{X}_1=\mathcal{X}^f_1$, where the superscript $f$ denotes our solution in which the interest rate follows the endogenous peg $f(n;\theta) \defeq u\cdot\mathcal{P}_\ell^n\cdot S_r$ and its $\ell-$th Markov state is such that $S_{r,\ell}=\underline{r}$. Otherwise, the sequence of impact multipliers diverges. Furthermore, provided it exists, the limit is given by:
\begin{align*}
\pazocal{M}  =  \left(\mathbf{I}_N-p_s\tilde{\underline{\mathcal{X}}} \mathbf{A}^\ast \right)^{-1}\tilde{\underline{\mathcal{X}}}C^\ast_s,  \quad \text{where}\quad \tilde{\underline{\mathcal{X}}} = \left(\mathbf{A}^\ast\right)^{-1}\underline{\mathcal{X}}.   
\end{align*}
In the absence of endogenous persistence, the expressions for $\underline{\mathcal{X}}$ and $\pazocal{M}$ boil down to the one obtained in the MC-CF literature.
\end{theorem}
\vspace*{-2em}
\begin{proof}
See Appendix \ref{apsec:proof_thm_1}.
\end{proof}
The main intuition behind Theorem \ref{thm:saddle} is that, if the sequence $\left\{\mathcal{X}_\ell\right\}_{\ell\geq 1}$ is guaranteed to converge to a real-valued fixed point, one can always construct the fixed point for the sequence of impact multipliers. There is however no guarantee that the sequence of impact multipliers will converge to this fixed point. If the maximum absolute eigenvalue of $p_s\underline{\mathcal{X}}$ is below one, then the dynamical system behaves like a \textit{sink}: regardless of the starting value, it has a limit and will converge to this fixed point. In that case, impact multipliers under both AR-NA or our method will be equivalent for a long enough time at the constraint. They might disagree over a short duration however.   

If the maximum absolute eigenvalue of $p_s\underline{\mathcal{X}}$ is above one  instead, then the system behaves like a \textit{saddle}. In that case, the starting value becomes crucial. Just like in the standard Ramsey-Cass-Koopmans model, there is a starting value for which the recursion will converge to a well defined steady state. We show that assuming a Taylor rule with $f(n;\theta)=\phi Y_{t+n}$ upon exit amounts to choosing a starting value that is off the saddle path: the sequence of impact multipliers necessarily diverges. 

The main take-away from Theorem \ref{thm:saddle} is that assuming our endogenous peg amounts to choosing a starting value that puts the system on its saddle path. Therefore, our method produces a stable multiplier regardless of the maximum eigenvalue of $p_s\underline{\mathcal{X}}$. If that maximum eigenvalue is larger than 1 in magnitude, the last part of Theorem \ref{thm:saddle} guarantees that our multiplier effectively generalizes the one developed in \cite{Mertens2014} to a model with endogenous persistence, while existing piecewise linear methods give a qualitatively different answer. Note that we have followed \cite{Bilbiie2022neo} and assumed that if the eigenvalue condition isn't met, then we switch to a MSV sunspot equilibrium. If one uses the method developed in \cite{Eggertsson2003} instead, the multiplier would diverge just like the AR-NA method.

Given the results in \cite{Eggertsson2019log}, one might expect the non-linear version of the model under that configuration to display no equilibrium. We argue that this point does not affect our results for two reasons. First, we compute the model under a peg, which is different from the two-state Markov structure considered in \cite{Eggertsson2019log}. Second, we check in our empirical application that all the equilibria that we compute feature low enough non-linear Euler equation errors.

The stability of the obtained sequence of deterministic multipliers hinges crucially on the eigenvalues of $p_s\underline{\mathcal{X}}$. Ideally, one would like to know whether the underlying system is a saddle or a sink. In light of our results, this condition is straightforward to check: Given that $\underline{\mathcal{X}}$ is independent of $p_s$, we immediately have:
\begin{corollary}
\label{cor:p^D}
Let $\rho(\underline{\mathcal{X}})$ denote the spectral radius of $\underline{\mathcal{X}}$. There exists a threshold
\begin{align*}
p^D \defeq \frac{1}{\rho(\underline{\mathcal{X}})}    
\end{align*}
such that the sequence of multipliers under a Taylor rule diverges if $p_s>p^D$.
\end{corollary}
This condition can be readily checked numerically. Ideally however, we would want to have some economic intuition to understand when AR-NA methods are producing a diverging sequence and when they are not. Following \cite{Eggertsson2011}, we would like to have an exact graphical representation to guide this process. One of the main advantages of our approach is that, by construction, it lends itself to such an exact representation: it will then be sufficient to look at the slopes of aggregate demand and supply equations at the ELB. In the next section, we use data from the U.S Survey of Professional Forecasters to recover the structural parameters underlying these slopes. To apply our stability criterion, we also consider a scenario that has received considerable attention fairly recently: the fiscal multiplier at the ELB. 

\section{Application: the Fiscal Multiplier at the ELB}\label{sec:application}
\vspace*{-0.8em}

Throughout this section we work with a standard New Keynesian model that we extend to include external habit formation in consumption. We study the properties of this model in depth and then compare it with the NK models considered in \cite{Eggertsson2011,Eggertsson2012new} as well as with forecast data from the Great Recession. 

\subsection{A model with consumption habits}
\label{subsec:application}
\vspace*{-0.5em}

Given our general formulation in Section \ref{sec:general_framework}, several kinds of endogenous propagation mechanisms can be considered and we have to make a choice. As alluded to before, we will make an effort to bring the model to the data, which may display a hump-shaped behavior for some variables. Because of this, we will consider one type of endogenous propagation mechanism: external habit formation in consumption. More precisely, we consider a New Keynesian model where households work and consume ($c_{t+n}$), while firms set prices in a monopolistically competitive environment, giving rise to inflation ($\pi_{t+n}$). The Central Bank sets the nominal rate ($r_{t+n}$) according to the endogenous peg developed earlier. We assume that, at time $t$, the economy is hit by a "risk premium" shock (see \citeauthor{Amano2012risk}, \citeyear{Amano2012risk}) and a government spending shock\textemdash denoted by $\xi_{t+n}$ and $g_{t+n}$, respectively. We relegate the full derivation to the online appendix and focus here on the linearized version of the first order conditions:
\begin{align}
\label{eq:backward}
c_{t+n} &= hc_{t+n-1} + \frac{1-h}{\sigma}\lambda_{t+n},\\
\label{eq:Euler_h}
\lambda_{t+n} &= \E_t\lambda_{t+n+1} - (r_{t+n}-\E_t\pi_{t+n+1}-\xi_{t+n}),\\
\label{eq:NKPC_h}
\pi_{t+n} &= \beta\E_t\pi_{t+n+1}+\kappa\eta(s_cc_{t+n}+s_gg_{t+n})+\kappa\lambda_{t+n}.
\end{align}
Here, $\lambda_{t+n}$ is the inverse of the marginal utility of consumption, and $h\in(0,1)$ governs the degree of habit formation.\footnote{This ensures that $\lambda_{t+n}=c_{t+n}/\sigma$ in the absence of endogenous persistence.
} $\sigma$ governs the curvature of the utility with respect to consumption, $\beta$ denotes the discount factor, and $\kappa$ represents the elasticity of inflation with respect to marginal costs. $s_c$ and $s_g$ denote the steady-state shares of consumption and government spending in output, respectively. The ELB will become a binding constraint as a result of a decrease in $\xi_{t+n}$ on impact (i..e., at $n=0$). At the same time, the government is assumed to step in and increase $g_t$ in an effort to stabilize the economy. The main goal of this section is to understand how the presence of habits shapes the government spending multiplier and how it crucially depends on the number of periods $\ell$ this economy is expected to spend at the ELB. 

At this point, we should note that our model is close in spirit to the one developed in \cite{Eggertsson2012new}. In that model, households are assumed to exhibit habit formation in both consumption and leisure. Let us denote these two degrees of habit formation as $h_c$ and $h_n$. In \cite{Eggertsson2012new}, the fact that $h_c=h_n$ allows the author to rewrite equations \eqref{eq:Euler_h}--\eqref{eq:NKPC_h} in terms of a quasi-growth rate $\tilde{c}_{t+n} \defeq c_{t+n} - h_c \cdot c_{t+n-1}$: that clever trick allows one to have a model that is forward looking in $\tilde{c}_{t+n}$. Our approach is more general, as we allow $h_c\neq h_n$. In fact, we follow most of the literature and assume that there is no habit formation in leisure.\footnote{Relatedly, \cite{Uhlig2007explaining} calibrates a model with both types of habits in order to match some empirical asset pricing moments and finds clear differences in degrees of habits between the two.
} Going further, our method enables us to consider models that cannot be rewritten in quasi-growth rates, for example, the model with private capital. Our model is also related to \cite{Nakata2017uncertainty}, who studies a numerical solution of a model with consumption habits at the ELB using projection methods but focuses on the role of uncertainty rather than the stability properties of the model. 

In order to develop the intuition behind our method, we consider a case where the risk premium shock is consistent with an expected ELB duration of one period. To this end, we use the Markov chain framework described in Definition \ref{ddef:MChains_inertial}. We claim that this framework is a \textit{bona fide} generalization of the two-state Markov chain approach found in \cite{Eggertsson2011} and the literature that followed. The two extra states in our setup reflect $(i)$ the different persistence of risk premium and government spending shocks, and $(ii)$ the presence of endogenous persistence.\footnote{\cite{Eggertsson2012new} also considers a setup where the policy shock can last longer than the demand shock. In that case, once the demand shock reverts back the policy decays back to its steady state value after having been constant throughout the recession. Our policy shock follows an AR(1) in expectations and thus decays immediately after impact.
} The associated Markov restrictions in this context will have to be written such that the ELB is binding for state $s_{z,1}$ but not for the rest. Moreover, the nominal rate will be given by $r_{t+n} = u\cdot\mathcal{P}^{n}\cdot S_r$ for $n\geq0$, where $r_{t+n} = \underline{r}$ for $n=0$ and $r_{t+n}>\underline{r}$ for $n>0$\textemdash see details in the online appendix.

The endogenous peg injects a backward-looking element in the interest rate. As a result, if the government increases spending at the ELB, its effect on monetary policy outside the ELB will be dampened. In this context, the results established in Section \ref{sec:general_framework} guarantee that, even if the underlying shocks are very persistent, our peg is such that these anticipated effects will \textit{not} lead to an arbitrarily large multiplier. This is however very much a possibility if the model is solved using existing AR-NA methods.

Beyond the $\ell=1$ case just covered, our framework can also accommodate an ELB of an arbitrarily long duration $\ell\to\infty$. In that case, the associated Markov restrictions will have to be written such that the ELB is binding for all states $s_{z,1}$ to $s_{z,3}$. In addition, the transition probability $q$ for the third state will have to reflect that as well: the degree of endogenous persistence will be different in an economy where the ELB essentially binds forever \textemdash see the online appendix. This case will turn out to be very informative: it will first inform us on the mechanisms behind the impact effect of a government spending shock in the short run. As in MC-CF, these mechanisms will be tied to a set of supply and demand curves. In addition, whether or not these curves can cross a second time at the ELB as in \cite{Bilbiie2022neo} will inform us on whether AR-NA would produce a diverging sequence of multipliers for the same first order conditions \eqref{eq:backward}--\eqref{eq:NKPC_h}. 

Under both $\ell=\left\{1,\infty\right\}$, the Markov states can be solved according to a very simple cookbook-like recipe. Let us work with the assumption that we have solved for $q$ already.\footnote{In the case where $\ell=1$, $q$ is the exact same as the one that would arise in a linear version of the model. As a result, it can be solved using standard methods such as \cite{Klein2000using}. In the $\ell\to\infty$ case, one has to use the Markov chain restrictions. We detail how to do this in the online appendix.} The model can then be solved backward from states $s_{z,3}$. In this process, computing the expectations of the underlying Markov chains will be especially simple. Let us assume that we are focusing on the Euler equation. In that case, we will be able to write that $\E_{t+n,3} \mathbf{\Lambda}_{t+n+1} = q s_{\lambda,3} + (1-q)\cdot 0$, where $s_{\lambda,3}$ is the third state for the marginal utility variable and $\E_{t+n,3}$ denotes expectations conditional on being in state $3$ at time $t+n$. The same procedure can be applied to expected inflation. For a given $q$, this will yield a system of \textit{linear} equations involving the third states of all the variables.

After that, one just has to move one step back. In that case, the same conditional expectation will be computed as $\E_{t+n,2} \mathbf{\Lambda}_{t+n+1} = p_b s_{\lambda,2} + (1-p_b) s_{\lambda,3}$. From the previous step, we do have an expression for $s_{\lambda,3}$. Finally, one can compute the conditional expectation on impact as $\E_{t} \mathbf{\Lambda}_{t+1} = p_s s_{\lambda,1} + (1-p_s) s_{\lambda,2}$. In both cases, the same applies to the conditional expectation for inflation. Using this method, we can recast both the Phillips curve and the Euler equations on impact as:
\begin{align*}
s_{\lambda,1} & = p_ss_{\lambda,1}+(1-p_s)s_{\lambda,2} - \underline{r} + p_ss_{\pi,1}+(1-p_s)s_{\pi,2}  + s_{\xi,1}\\
s_{\pi,1} & = \beta p_ss_{\pi,1} + \beta(1-p_s)s_{\pi,2} + \kappa s_{\lambda,1} + \kappa\eta s_cs_{c,1} + \kappa\eta s_gs_{g,1},
\end{align*}
which clearly nests the MC-CF literature whenever $s_{\lambda,2}=s_{\pi,2}=0$ and $s_{\lambda,1}=\sigma s_{c,1}$. We will expand on this case in more detail in the next subsection. In our case, these second states will be tightly linked to $s_{\lambda,1}, s_{c,1}$ and $s_{\pi,1}$ through the remaining Markov restrictions. These are described in Appendix \ref{apsec:app_markovres_shortspell}.  

\subsection{The existing MC-CF literature as a special case: intuition}
\vspace*{-0.5em}

Readers familiar with the procedure developed in \cite{Eggertsson2011} and used in the MC-CF literature may see how our method relates to and generalizes it. In the standard New Keynesian model used in MC-CF, the economy returns to its intended steady state as soon as the shock is over. Thus, for all intents and purposes, $s_{\lambda,2}=s_{c,2}=0$ in MC-CF. In the absence of habits and $p_s=p_b$, this implies that one can write expected consumption as $\E_t \mathbf{C}_{t+n} = p_s^ns_{c,1} = p_s^nc_t$: consumers cannot expect anything other than a recovery back to steady state. We will show later that this is clearly at odds with the expectations measured in the data. In our more general case, $s_{\lambda,2}$ will be different from zero both because of the different persistence of exogenous shocks and the presence of habits: this will allow us to replicate the hump-shape features of the data. In turn, expectations consistent with a hump-shape path for consumption may qualitatively change the effects of government spending on consumption.\footnote{Indeed, the existing literature has shown that if $0>\E_t \mathbf{C}_{t+1}=p_s\cdot c_t>c_t$ is persistent enough, then that opens up the door to sunspot ELBs \textemdash\ see \cite{Bilbiie2022neo}. In our case, to replicate a hump-shape we will need to have $\E_t \mathbf{C}_{t+1}=(p_s+\psi)\cdot c_t<c_t<0$ with $p_s+\psi>1$ for some $\psi>0$.} Indeed, in the online appendix we run an experiment in which we don't target the expected consumption path and we show that this results in a qualitatively different effect of government spending \textemdash the response of consumption switches from crowding-in to crowding-out.

In addition, in \cite{Eggertsson2011} the expected path for the interest takes the following simple form: $\E_t \mathbf{R}_{t+n} = p_s^n\underline{r}$. The nominal interest rate is then expected to equal its ELB on impact, but not after. In that very specific case where the ELB duration in our setup is $\ell=1$, one can just compute the equilibrium just as in \cite{Eggertsson2011} and average across multiple runs of the Markov chain. That method doesn't work anymore as soon as the ELB lasts for $\ell\geq 2$ periods: as a result, our method is more than just averaging across runs of \cite{Eggertsson2011}'s method. We provide numerical experiments supporting this claim in the online appendix. This is the insight that we leverage in this paper: in our model with habits, the endogenous peg given by $r_{t+n} = u\cdot\mathcal{P}^{n}\cdot S_r$ is a generalization that nests the one used in the MC-CF literature.

In our model with habits, the government spending multiplier at the ELB potentially depends on many parameters. Instead of providing a detailed theoretical discussion of how the multiplier depends on our set of parameters, we follow a different approach here that is more empirical. We take as a starting point that the exercise that is usually being considered in theory is one where a large enough demand shock hits the economy and sends it to the ELB. Besides the contemporaneous government spending shock, no other shock is assumed to occur beyond the first time period $t$. As a result, we argue that this kind of experiment cannot be expected to replicate the path of \textit{realized} data after the shocks have occurred. However, we can entertain the fact that the \textit{expectations} from the model potentially match the ones from the data.

Given that the class of models we are interested in are typically used to study the effects of policy in a deep recession, we will match the model with expectations measured during the early stages of the Great Recession of 2009.\footnote{Given that we rely on a piece-wise linear model, the Covid-19 recession entails a deviation from steady state that is certainly too big to be handled. That would require a full global solution of the underlying model. This is an interesting question that we leave for future research.} In order to map the model to the expectations data, we need the conditional expectation of both consumption and inflation next quarter. For this reason, we will focus on the U.S Survey of Professional Forecasters. Later, we also consider forecast data for Japan. This exercise will allow us to kill two birds with one stone. First, we will be able to contrast expectations from the data with expectations from the standard New Keynesian model typically used in the MC-CF literature. Second, we will use these to discipline the parameters of the model with habits by ensuring that the model delivers expectations in the early stages of the recession that matches those from the data. In order to ensure that they do match, we will use a minimum distance estimation procedure. We note that one could also match this expectation data using the model developed in \cite{Eggertsson2012new} with equally persistent habits in consumption and leisure. Our goal is to showcase that our method is able to handle a standard model with habits in consumption only that cannot be rewritten in a manner that is purely forward-looking.

\subsection{Not so Great Expectations during the Great Recession}
\vspace*{-0.5em}

The title of this subsection is a hat-tip to the celebrated paper  by \cite{Eggertsson2008great}, where the recovery from the Great Depression was shown to work through optimistic expectations about the future. The main result of this subsection is that data from professional forecasters at the onset of the Great Recession tells a very different story: forecasters expect the recession to worsen for several quarters before things start to look brighter. We will show that, while standard New Keynesian models used in the MC-CF literature cannot match this feature, our extension with habits can. In addition, our new solution method ensures that this improved empirical fit will not come at the expense of analytical tractability or interpretability. It should be noted here that both \cite{Eggertsson2008great,Eggertsson2012new} feature equal habit formation in consumption and leisure. Our model is one example with different habit formation in consumption and leisure which cannot be rewritten as a purely forward looking model. 

In order to map the model to the data, we have to take into account that our model is written in deviations from a potential path that is growing over time. To deal with this, we use long-run projections from the Survey of Professional Forecasters to compute a potential trend. We then compute the expected deviations from potential as the reported expectations minus the expected potential. We explain in detail in the online appendix how we compute the potential for each variable. We work under the assumption that the sizable decline in GDP/consumption observed in 2009:Q1 is due to a large negative realization of the risk premium shock $\xi_t$ that forced the Federal Reserve to set its main interest rate to zero.\footnote{An implicit assumption here is that the path of expectations starting in that date can be seen as an impulse response given that this large negative demand shock trumps all other possible shocks. That being said, we provide a more rigorous approach in the online appendix where we study how the U.S economy reacts after being hit with the "main business cycle shock" estimated in \cite{Angeletos2020business}.} Loosely speaking, we want to see whether our model can reproduce expectations during the early stages of the Great Recession. 

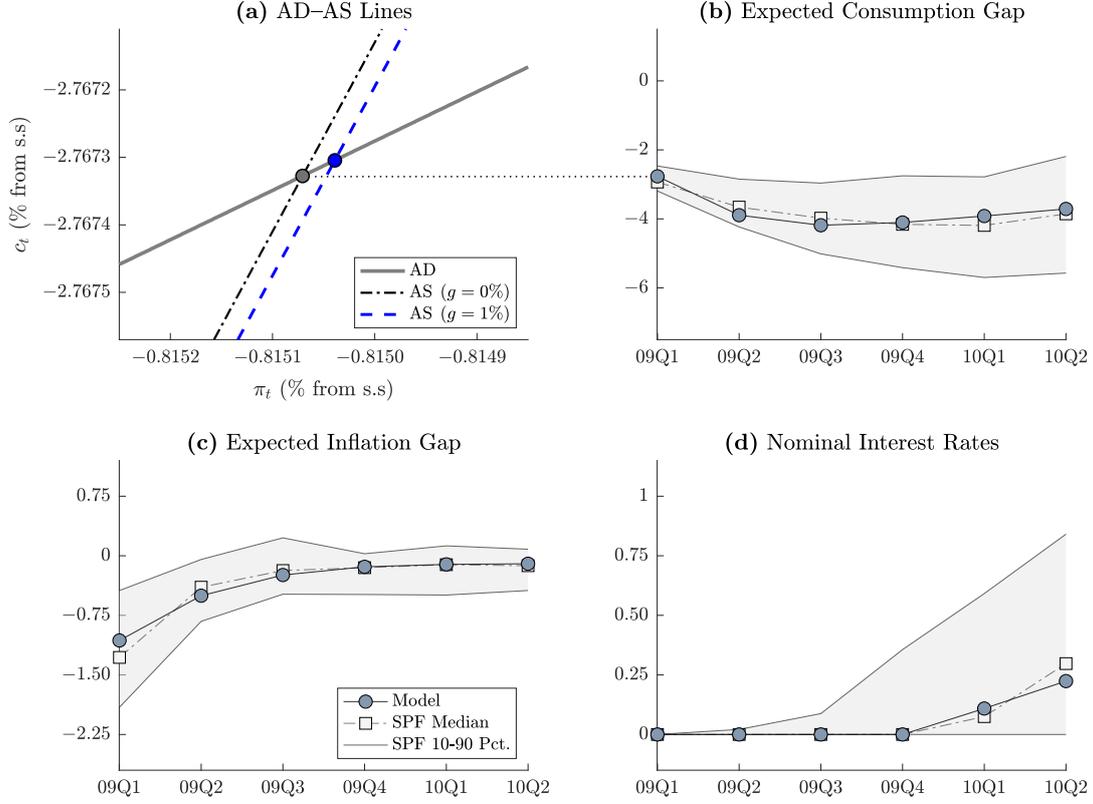
\begin{figure}[ht]
	\centering
	\caption{Conditional expectations with endogenous persistence}\label{fig:AS_AD_US}
	\begin{tikzpicture}
	\node (boxes)[inner sep=3.5pt] {
	\begin{minipage}{1\textwidth}
	\centering
	\hspace*{-.4cm}\subcaptionbox{AD and AS lines\hspace*{-5em}}{
            \centering
            \begin{tikzpicture}[scale=3.15]
            
                \draw[black,thick] (0,0)--(0,1.2);
                \draw[black,thick] (0,0)--(2,0);
                
               \draw ({0.2},0.1pt) -- ({0.2},-1.2pt) node[anchor=north,font=\scriptsize] {$-0.7076$};
               \draw ({0.75},0.1pt) -- ({0.75},-1.2pt) node[anchor=north,font=\scriptsize] {$-0.7072$};
               \draw ({1.3},0.1pt) -- ({1.3},-1.2pt) node[anchor=north,font=\scriptsize] {$-0.7068$};
               \draw ({1.85},0.1pt) -- ({1.85},-1.2pt) node[anchor=north,font=\scriptsize] {$-0.7064$};
            
               \draw (0.1pt,{0.1}) -- (-0.9pt,{0.1}) node[left,font=\scriptsize] at (-0.01,{0.1}) {$-2.8042$};
               \draw (0.1pt,{5/12}) -- (-0.9pt,{5/12}) node[left,font=\scriptsize] at (-0.01,{5/12}) {$-2.8038$};
               \draw (0.1pt,{11/15}) -- (-0.9pt,{11/15}) node[left,font=\scriptsize] at (-0.01, {11/15}) {$-2.8034$};
               \draw (0.1pt,{1.05}) -- (-0.9pt,{1.05}) node[left,font=\scriptsize] at (-0.01, {1.05}) {$-2.803$};
               
               \draw[gray,line width=1.5pt] 
               	({-0.0062500000001364242052659392356873},{0.26681495543425626237876713275909}) 
               	-- ({1.9187500000000454747350886464119},{1.0632353984801738988608121871948});
                
               \draw[black,line width=.8pt,dash pattern=on 3pt off 1.5pt on 0.5pt off 1.5pt] 
               	({0.50249999999994088284438475966454},{-0.048232223130526108434423804283142})
               	-- ({1.155625},{1.1540491091363946907222270965576});
	
               \draw[blue,line width=1pt,dash pattern=on 4pt off 2pt] 
               	({0.72937499999989086063578724861145},{-0.0097316826595488237217068672180176})
               	-- ({1.3824999999999363353708758950233},{1.1925496496078267227858304977417});
               
               \draw[semithick,
               		only marks,
               		mark options={solid, draw={rgb,1:red,0;green,0;blue,0},
			fill={rgb,1:red,0.45;green,0.45;blue,0.45}}] 
			plot[mark=*,mark size=0.9pt] 
			coordinates {
				({0.87075304172651613043854013085365},{0.62965295600588433444499969482422}) 
			};               

               \draw[semithick,
               		only marks,
               		mark options={solid, draw={rgb,1:red,0;green,0;blue,0},
			fill={rgb,1:red,0;green,0;blue,1}}] 
			plot[mark=*,mark size=0.9pt] 
			coordinates {
				({1.1364226174586065098992548882961},{0.73956707611432648263871669769287})
			};           
                            
               \node[font=\footnotesize] at (1.0,-0.28) {$\pi_t$ (percentage points from s.s.)};
               \node[rotate=90,font=\footnotesize] at (-0.5,0.55) {$c_t$ (\% deviation from s.s.)};
                
               \draw[gray, line width=1.5pt] (1.2,0.34) -- (1.33,0.34) node[right,font=\scriptsize,black] at (1.32,0.335) {AD};
               \draw[black,line width=.8pt,dash pattern=on 3pt off 1.5pt on 0.5pt off 1.5pt] (1.2,0.22) -- (1.33,0.22) node[right,font=\scriptsize,black] at (1.325,0.22) {AS ($g_t = 0\%$)};
               \draw[blue,line width=1pt,dash pattern=on 4pt off 2pt] (1.205,0.1) -- (1.335,0.1) node[right,font=\scriptsize,black] at (1.325,0.1) {AS ($g_t = 5\%$)};
                              
            \end{tikzpicture}
        }\hspace*{0.25em}
        \subcaptionbox{Expected consumption gaps\hspace*{-1.8em}}{
         \centering
         \begin{tikzpicture}[scale=3.15]
         
               \draw[black,thick] (0,0)--(0,1.2);
               \draw[black,thick] (0,0)--(2,0);
            
               \draw ({0*0.36+0.05},0.1pt) -- ({0*0.36+0.05},-1.2pt) node[anchor=north,font=\scriptsize] {09Q1};
               \draw ({1*0.36+0.05},0.1pt) -- ({1*0.36+0.05},-1.2pt) node[anchor=north,font=\scriptsize] {09Q2};
               \draw ({2*0.36+0.05},0.1pt) -- ({2*0.36+0.05},-1.2pt) node[anchor=north,font=\scriptsize] {09Q3};
               \draw ({3*0.36+0.05},0.1pt) -- ({3*0.36+0.05},-1.2pt) node[anchor=north,font=\scriptsize] {09Q4};
               \draw ({4*0.36+0.05},0.1pt) -- ({4*0.36+0.05},-1.2pt) node[anchor=north,font=\scriptsize] {10Q1};
               \draw ({5*0.36+0.05},0.1pt) -- ({5*0.36+0.05},-1.2pt) node[anchor=north,font=\scriptsize] {10Q2};
            
               \draw (0.1pt,{-6*0.15+1.05}) -- (-0.9pt,{-6*0.15+1.05}) node[left,font=\scriptsize] at (-0.01,{-6*0.15+1.05}) {$-6$};
               \draw (0.1pt,{-4*0.15+1.05}) -- (-0.9pt,{-4*0.15+1.05}) node[left,font=\scriptsize] at (-0.01,{-4*0.15+1.05}) {$-4$};
               \draw (0.1pt,{-2*0.15+1.05}) -- (-0.9pt,{-2*0.15+1.05}) node[left,font=\scriptsize] at (-0.01, {-2*0.15+1.05}) {$-2$};
               \draw (0.1pt,{0*0.15+1.05}) -- (-0.9pt,{0*0.15+1.05}) node[left,font=\scriptsize] at (-0.01, {0*0.15+1.05}) {$0$};

               \fill[fill={rgb,1:red,0.9;green,0.9;blue,0.9}, opacity=0.5]
               		({0*0.36+0.025}, {-2.4658*0.15+1.05})
               		-- ({1*0.36+0.05}, {-2.8478*0.15+1.05})
               		-- ({2*0.36+0.05}, {-2.9626*0.15+1.05})
               		-- ({3*0.36+0.05}, {-2.7518*0.15+1.05})
               		-- ({4*0.36+0.05}, {-2.7823*0.15+1.05})
               		-- ({5*0.36+0.075}, {-2.1909*0.15+1.05})
               		-- ({5*0.36+0.075}, {-5.5717*0.15+1.05})
               		-- ({4*0.36+0.05}, {-5.6995*0.15+1.05})
               		-- ({3*0.36+0.05}, {-5.4140*0.15+1.05})
               		-- ({2*0.36+0.05}, {-5.0148*0.15+1.05})
               		-- ({1*0.36+0.05}, {-4.2342*0.15+1.05})
               		-- ({0*0.36+0.025}, {-3.1928*0.15+1.05})
               		-- cycle;
		
               \draw[line width=.3pt,draw={rgb,1:red,0.5;green,0.5;blue,0.5}] plot
			coordinates {
				({0*0.36+0.025},{-3.1928*0.15+1.05})
				({1*0.36+0.05},{-4.2342*0.15+1.05})
				({2*0.36+0.05},{-5.0148*0.15+1.05})
				({3*0.36+0.05},{-5.4140*0.15+1.05})
				({4*0.36+0.05},{-5.6995*0.15+1.05})
				({5*0.36+0.075},{-5.5717*0.15+1.05})
			};

               \draw[line width=.3pt, draw={rgb,1:red,0.5;green,0.5;blue,0.5}] plot
			coordinates {
				({0*0.36+0.025},{-2.4658*0.15+1.05})
				({1*0.36+0.05},{-2.8478*0.15+1.05})
				({2*0.36+0.05},{-2.9626*0.15+1.05})
				({3*0.36+0.05},{-2.7518*0.15+1.05})
				({4*0.36+0.05},{-2.7823*0.15+1.05})
				({5*0.36+0.075},{-2.1909*0.15+1.05})
			};
			
               \draw[line width = .5pt,
               		dash pattern=on 3pt off 1.5pt on 0.5pt off 1.5pt,
               		mark options={solid, line width = 0.5pt,draw={rgb,1:red,0;green,0;blue,0},
			fill={rgb,1:red,0.99;green,0.99;blue,0.99}}] 
			plot[mark=square*,mark size=.8pt] 
			coordinates {
				({0*0.36+0.05},{-2.9329*0.15+1.05})
				({1*0.36+0.05},{-3.6594*0.15+1.05})
				({2*0.36+0.05},{-3.9789*0.15+1.05})
				({3*0.36+0.05},{-4.1591*0.15+1.05})
				({4*0.36+0.05},{-4.1896*0.15+1.05})
				({5*0.36+0.05},{-3.8552*0.15+1.05})
			};      

               \draw[thick,
               		mark options={solid, draw={rgb,1:red,0;green,0;blue,0},
			fill={rgb,1:red,0.518;green,0.596;blue,0.686}}] 
			plot[mark=*,mark size=0.85pt] 
			coordinates {
				({0*0.36+0.05},{-2.8035*0.15+1.05})
				({1*0.36+0.05},{-3.9228*0.15+1.05})
				({2*0.36+0.05},{-4.1954*0.15+1.05})
				({3*0.36+0.05},{-4.1009*0.15+1.05})
				({4*0.36+0.05},{-3.9132*0.15+1.05})
				({5*0.36+0.05},{-3.7232*0.15+1.05})
			};
            
                \node[font=\footnotesize] at (0.98,-0.28) {Quarter};
        
         \end{tikzpicture}
         \vspace{0.4cm}
        }\quad
        \hspace*{0.42cm}\subcaptionbox{Expected inflation gaps\hspace*{-2.5em}}{
         \centering
         \begin{tikzpicture}[scale=3.15]
            
               \draw[black,thick] (0,0)--(0,1.2);
               \draw[black,thick] (0,0)--(2,0);
            
               \draw ({0*0.36+0.05},0.1pt) -- ({0*0.36+0.05},-1.2pt) node[anchor=north,font=\scriptsize] {09Q1};
               \draw ({1*0.36+0.05},0.1pt) -- ({1*0.36+0.05},-1.2pt) node[anchor=north,font=\scriptsize] {09Q2};
               \draw ({2*0.36+0.05},0.1pt) -- ({2*0.36+0.05},-1.2pt) node[anchor=north,font=\scriptsize] {09Q3};
               \draw ({3*0.36+0.05},0.1pt) -- ({3*0.36+0.05},-1.2pt) node[anchor=north,font=\scriptsize] {09Q4};
               \draw ({4*0.36+0.05},0.1pt) -- ({4*0.36+0.05},-1.2pt) node[anchor=north,font=\scriptsize] {10Q1};
               \draw ({5*0.36+0.05},0.1pt) -- ({5*0.36+0.05},-1.2pt) node[anchor=north,font=\scriptsize] {10Q2};
            
               \draw (0.1pt,{-2.25*0.4+1.05}) -- (-0.9pt,{-2.25*0.4+1.05}) node[left,font=\scriptsize] at (-0.01,{-2.25*0.4+1.05}) {$-2.25$};
               \draw (0.1pt,{-1.5*0.4+1.05}) -- (-0.9pt,{-1.5*0.4+1.05}) node[left,font=\scriptsize] at (-0.01,{-1.5*0.4+1.05}) {$-1.5$};
               \draw (0.1pt,{-0.75*0.4+1.05}) -- (-0.9pt,{-0.75*0.4+1.05}) node[left,font=\scriptsize] at (-0.01, {-0.75*0.4+1.05}) {$-0.75$};
               \draw (0.1pt,{0*0.4+1.05}) -- (-0.9pt,{0*0.4+1.05}) node[left,font=\scriptsize] at (-0.01, {0*0.4+1.05}) {$0$};
            
               \fill[fill={rgb,1:red,0.9;green,0.9;blue,0.9}, opacity=0.5]
               		({0*0.36+0.025}, {-0.4375*0.4+1.05})
               		-- ({1*0.36+0.05}, {-0.0464*0.4+1.05})
               		-- ({2*0.36+0.05}, {0.2255*0.4+1.05})
               		-- ({3*0.36+0.05}, {0.0255*0.4+1.05})
               		-- ({4*0.36+0.05}, {0.1250*0.4+1.05})
               		-- ({5*0.36+0.075}, {0.0825*0.4+1.05})
               		-- ({5*0.36+0.075}, {-0.4374*0.4+1.05})
               		-- ({4*0.36+0.05}, {-0.4945*0.4+1.05})
               		-- ({3*0.36+0.05}, {-0.4878*0.4+1.05})
               		-- ({2*0.36+0.05}, {-0.4840*0.4+1.05})
               		-- ({1*0.36+0.05}, {-0.8257*0.4+1.05})
               		-- ({0*0.36+0.025}, {-1.9078*0.4+1.05})
               		-- cycle;
		
               \draw[line width=.3pt,draw={rgb,1:red,0.5;green,0.5;blue,0.5}] plot
			coordinates {
				({0*0.36+0.025},{-1.9078*0.4+1.05})
				({1*0.36+0.05},{-0.8257*0.4+1.05})
				({2*0.36+0.05},{-0.4840*0.4+1.05})
				({3*0.36+0.05},{-0.4878*0.4+1.05})
				({4*0.36+0.05},{-0.4945*0.4+1.05})
				({5*0.36+0.075},{-0.4374*0.4+1.05})
			};
			
               \draw[line width=.3pt, draw={rgb,1:red,0.5;green,0.5;blue,0.5}] plot
			coordinates {
				({0*0.36+0.025},{-0.4375*0.4+1.05})
				({1*0.36+0.05},{-0.0464*0.4+1.05})
				({2*0.36+0.05},{0.2255*0.4+1.05})
				({3*0.36+0.05},{0.0255*0.4+1.05})
				({4*0.36+0.05},{0.1250*0.4+1.05})
				({5*0.36+0.075},{0.0825*0.4+1.05})
			};

               \draw[line width = .5pt,
               		dash pattern=on 3pt off 1.5pt on 0.5pt off 1.5pt,
               		mark options={solid, line width = 0.5pt,draw={rgb,1:red,0;green,0;blue,0},
			fill={rgb,1:red,0.99;green,0.99;blue,0.99}}] 
			plot[mark=square*,mark size=.8pt] 
			coordinates {
				({0*0.36+0.05},{-1.2799*0.4+1.05})
				({1*0.36+0.05},{-0.3956*0.4+1.05})
				({2*0.36+0.05},{-0.1859*0.4+1.05})
				({3*0.36+0.05},{-0.1500*0.4+1.05})
				({4*0.36+0.05},{-0.1125*0.4+1.05})
				({5*0.36+0.05},{-0.1250*0.4+1.05})
			};

               \draw[semithick,
               		mark options={solid, draw={rgb,1:red,0;green,0;blue,0},
			fill={rgb,1:red,0.518;green,0.596;blue,0.686}}] 
			plot[mark=*,mark size=0.85pt] 
			coordinates {
				({0*0.36+0.05},{-1.0906*0.4+1.05})
				({1*0.36+0.05},{-0.4978*0.4+1.05})
				({2*0.36+0.05},{-0.2316*0.4+1.05})
				({3*0.36+0.05},{-0.1329*0.4+1.05})
				({4*0.36+0.05},{-0.1092*0.4+1.05})
				({5*0.36+0.05},{-0.1063*0.4+1.05})
			};

                \node[font=\footnotesize] at (0.98,-0.28) {Quarter};

                \draw[semithick] (0.9,0.35) -- (1.03,0.35) node[right,font=\scriptsize,black] at (1.03,0.35) {Model};
                \draw[only marks,mark=*,mark size=0.85pt,
                		mark options={draw={rgb,1:red,0;green,0;blue,0},fill={rgb,1:red,0.518;green,0.596;blue,0.686}}] plot coordinates {(0.965,0.35)};
                \draw[line width = .5pt, dash pattern=on 3pt off 1.5pt on 0.5pt off 1.5pt] (0.9,0.225) -- (1.03,0.225) node[right,font=\scriptsize,black] at (1.03,0.225) {SPF median};
                \draw[only marks,mark=square*,mark size=0.8pt,
                		mark options={solid, line width = 0.5pt,draw={rgb,1:red,0;green,0;blue,0},fill={rgb,1:red,0.99;green,0.99;blue,0.99}}] plot coordinates {(0.968,0.225)};
                \draw[line width=.3pt, draw={rgb,1:red,0.5;green,0.5;blue,0.5}] (0.9,0.1) -- (1.03,0.1) node[right,font=\scriptsize,black] at (1.03,0.1) {SPF 10--90 percentiles};

         \end{tikzpicture}
        }\hspace*{0.25em}     
        \hspace*{0.12cm}\subcaptionbox{Expected nominal rates\hspace*{-2em}}{
         \centering
         \begin{tikzpicture}[scale=3.15]
        
               \draw[black,thick] (0,0)--(0,1.2);
               \draw[black,thick] (0,0)--(2,0);
            
               \draw ({0*0.36+0.05},0.1pt) -- ({0*0.36+0.05},-1.2pt) node[anchor=north,font=\scriptsize] {09Q1};
               \draw ({1*0.36+0.05},0.1pt) -- ({1*0.36+0.05},-1.2pt) node[anchor=north,font=\scriptsize] {09Q2};
               \draw ({2*0.36+0.05},0.1pt) -- ({2*0.36+0.05},-1.2pt) node[anchor=north,font=\scriptsize] {09Q3};
               \draw ({3*0.36+0.05},0.1pt) -- ({3*0.36+0.05},-1.2pt) node[anchor=north,font=\scriptsize] {09Q4};
               \draw ({4*0.36+0.05},0.1pt) -- ({4*0.36+0.05},-1.2pt) node[anchor=north,font=\scriptsize] {10Q1};
               \draw ({5*0.36+0.05},0.1pt) -- ({5*0.36+0.05},-1.2pt) node[anchor=north,font=\scriptsize] {10Q2};
            
               \draw (0.1pt,{0*1+0.15}) -- (-0.9pt,{0*1+0.15}) node[left,font=\scriptsize] at (-0.01,{0*1+0.15}) {$0$};
               \draw (0.1pt,{0.3*1+0.15}) -- (-0.9pt,{0.3*1+0.15}) node[left,font=\scriptsize] at (-0.01,{0.3*1+0.15}) {$0.3$};
               \draw (0.1pt,{0.6*1+0.15}) -- (-0.9pt,{0.6*1+0.15}) node[left,font=\scriptsize] at (-0.01, {0.6*1+0.15}) {$0.6$};
               \draw (0.1pt,{0.9*1+0.15}) -- (-0.9pt,{0.9*1+0.15}) node[left,font=\scriptsize] at (-0.01, {0.9*1+0.15}) {$0.9$};
            
               \fill[fill={rgb,1:red,0.9;green,0.9;blue,0.9}, opacity=0.5]
               		({0*0.36+0.025}, {0*1+0.15})
               		-- ({1*0.36+0.05}, {0.0212*1+0.15})
               		-- ({2*0.36+0.05}, {0.08752*1+0.15})
               		-- ({3*0.36+0.05}, {0.35622*1+0.15})
               		-- ({4*0.36+0.05}, {0.59133*1+0.15})
               		-- ({5*0.36+0.075}, {0.840665*1+0.15})
               		-- ({5*0.36+0.075}, {0*1+0.15})
               		-- ({4*0.36+0.05}, {0*1+0.15})
               		-- ({3*0.36+0.05}, {0*1+0.15})
               		-- ({2*0.36+0.05}, {0*1+0.15})
               		-- ({1*0.36+0.05}, {0*1+0.15})
               		-- ({0*0.36+0.025}, {0*1+0.15})
               		-- cycle;
		
               \draw[line width=.3pt,draw={rgb,1:red,0.5;green,0.5;blue,0.5}] plot
			coordinates {
				({0*0.36+0.025},{0*1+0.15})
				({1*0.36+0.05},{0*1+0.15})
				({2*0.36+0.05},{0*1+0.15})
				({3*0.36+0.05},{0*1+0.15})
				({4*0.36+0.05},{0*1+0.15})
				({5*0.36+0.075},{0*1+0.15})
			};
			
               \draw[line width=.3pt, draw={rgb,1:red,0.5;green,0.5;blue,0.5}] plot
			coordinates {
				({0*0.36+0.025},{0*1+0.15})
				({1*0.36+0.05},{0.0212*1+0.15})
				({2*0.36+0.05},{0.08752*1+0.15})
				({3*0.36+0.05},{0.35622*1+0.15})
				({4*0.36+0.05},{0.59133*1+0.15})
				({5*0.36+0.075},{0.840665*1+0.15})
			};

               \draw[line width = .5pt,
               		dash pattern=on 3pt off 1.5pt on 0.5pt off 1.5pt,
               		mark options={solid, line width = 0.5pt, draw={rgb,1:red,0;green,0;blue,0},
			fill={rgb,1:red,0.99;green,0.99;blue,0.99}}] 
			plot[mark=square*,mark size=.8pt] 
			coordinates {
				({0*0.36+0.05},{0*1+0.15})
				({1*0.36+0.05},{0*1+0.15})
				({2*0.36+0.05},{0*1+0.15})
				({3*0.36+0.05},{0*1+0.15})
				({4*0.36+0.05},{0.0750*1+0.15})
				({5*0.36+0.05},{0.2975*1+0.15})
			};

               \draw[semithick,
               		mark options={solid, draw={rgb,1:red,0;green,0;blue,0},
			fill={rgb,1:red,0.518;green,0.596;blue,0.686}}] 
			plot[mark=*,mark size=0.85pt] 
			coordinates {
				({0*0.36+0.05},{0*1+0.15})
				({1*0.36+0.05},{0*1+0.15})
				({2*0.36+0.05},{0*1+0.15})
				({3*0.36+0.05},{0*1+0.15})
				({4*0.36+0.05},{0.1033*1+0.15})
				({5*0.36+0.05},{0.2284*1+0.15})
			};

                \node[font=\footnotesize] at (0.98,-0.28) {Quarter};
            
         \end{tikzpicture}
        }
        \end{minipage}
        };
        \draw[semithick,dash pattern=on 1pt off 1pt] ({-3.3},{3.4}) -- ({1.40},{3.4});
        \end{tikzpicture}
\begin{minipage}{0.95\linewidth}
\setstretch{0.93}
{\footnotesize
\medskip
\emph{Notes:} Panel (a) presents the AD and AS lines implied by an NK model with habit formation in the ELB state. These lines are derived from the Markov restrictions in Appendix \ref{apsec:app_markovres_shortspell}, with all parameters set to the minimum-distance estimates. The initial Markov state of the risk premium shock ($s_{\xi,1}$) is re-calibrated so that equilibrium consumption without government spending (gray dot) matches the model-implied consumption gap in 2009:Q1\textemdash the first blue dot in Panel (b). Panels (b), (c), and (d) plot the estimated conditional expectations for consumption gaps, inflation gaps, and nominal rates, respectively, under the assumption that the risk premium shock hits in 2009:Q1. The model-implied estimates are overlaid on the median (white boxes), 10th-, and 90th-percentile conditional forecasts of professional forecasters as of 2009:Q1.
}
\end{minipage}
\end{figure}

One issue that arises when taking the model to the data on the Great Recession is that deviations from that potential trend in the data can be quite sizable. At the same time, our model is piecewise linear: linear at the ELB and linear outside the ELB. We want to make sure that non-linear Euler equation errors are sufficiently close to zero.\footnote{Here we mean Euler equation in the general sense of equations having conditional expectations in them, not just the consumption Euler equation.} In the words of \cite{Eggertsson2019log}, the piecewise linear model that we consider is a mis-specified version of the true, non-linear model. Our procedure is designed to ensure that our piece-wise linear model is a good approximation of the true non-linear model. In order to deal with that issue, we use a penalized minimum-distance estimation. More specifically, let us define $\theta^{\text{MD}}$ as the vector of parameters that we estimate. We then set out to minimize the following objective function:
\begin{align*}
\theta^{\text{MD}} \defeq \argmin_{\theta} G(\theta)\mathcal{W}  G(\theta)' + \tau_{\pazocal{E}}\cdot \pazocal{E}  + \tau_\ell\cdot \mathbb{1}(\ell-\ell^d),
\end{align*}
where $G(\theta)$ collects the difference between model- and data-based expectations. $\mathcal{W}$ is the weighting matrix, $\tau_{\pazocal{E}}\geq 0$ is a tuning parameter that governs the weight of squared non-linear Euler equation errors $\pazocal{E} $, while $\tau_\ell$ penalizes squared deviations of the duration $\ell$ from its data counterpart $\ell^d$. In practice, we set $\tau_{\pazocal{E}}=\tau_\ell=1000$, which ensures that our non-linear Euler equation errors are approximately of the order of $10^{-3}$ and that the ELB binds for the required number of periods in expectation. We provide more details on the estimation procedure, the parameter estimates as well as their confidence bands in the online appendix and focus on the visual fit here. The latter is reported in Figure \ref{fig:AS_AD_US} alongside the implied supply/demand diagram for ease of interpretation.  

One feature that stands out from Figure \ref{fig:AS_AD_US} is that the model is able to almost perfectly match the data. In particular, the presence of habits allows the model to match the fact that $\E_t \mathbf{C}_{t+1} < c_t$.\footnote{In the online appendix, we also provide more evidence along these lines. First, we show that this also holds true at the onset of the Great Recession at the individual forecaster level: on average, if a forecaster nowcasts a lower consumption respective to trend, he/she will forecast even lower consumption for the next quarter. We also show that this is not specific to the Great Recession. Using the main business cycle shock computed in \cite{Angeletos2020business}, we show that, conditional on a realization of this shock, expected consumption reacts more than actual/realized consumption.} Note that this cannot happen in the simple New Keynesian model typically used in the MC-CF literature because in these models $c_t<\E_t \mathbf{C}_{t+1} = p_s c_t < 0$. Because our model with habits is able to replicate this, it is a more reasonable laboratory to study the effects of government spending in a large recession.

\subsection{The fiscal multiplier in short- vs long-lived ELB spells}
\vspace*{-0.5em}

Armed with our estimation results, we can now answer the following question: are the early stages of the U.S Great Recession best represented by \cite{Eggertsson2011} or \cite{Mertens2014}-type dynamics? To answer this question, we provide an exact representation of the model under the assumption of $\ell=1$ for expositional purposes in the top left panel of Figure \ref{fig:AS_AD_US}.\footnote{A detailed explanation of how we compute these supply/demand lines is in the online appendix.} One can see that in that case the slope of the AD line is clearly positive and slopes less than the AS line: the U.S fits the dynamics reported in \cite{Eggertsson2011}. Regarding the implications for the government spending multiplier, we consider a rather large increase in government spending for ease of exposition. In that case, the familiar story arises: the AS line shifts to the right and slides along an upward sloping AD line: consumption is crowded in and the government spending multiplier on output is larger than 1. This increase in consumption is associated with higher inflation through higher marginal costs.

In the online appendix, we re-estimate the model without matching the hump-shaped path of expected consumption. In that case, the model-implied consumption gap is upward-sloping, which is completely at odds with the conditional forecasts of the professional forecasters. Given those parameter estimates, the AD line is \textit{steeper} than the AS line, resulting in a different policy implication than those shown here: consumption is crowded out.

\begin{figure}[ht]
	\centering
	\caption{Impact multipliers}\label{fig:M_ell_compar_2}
	\hspace*{-.4cm}\subcaptionbox{Output\hspace*{-2.5em}}{
            \centering
            \begin{tikzpicture}[scale=3.4]
            
               \draw[black,thick] (0,0)--(0,1.2);
               \draw[black,thick] (0,0)--(2,0);
            
               \draw ({1*0.55-0.35},0.1pt) -- ({1*0.55-0.35},-1.2pt) node[anchor=north,font=\scriptsize] {1};
               \draw ({2*0.55-0.35},0.1pt) -- ({2*0.55-0.35},-1.2pt) node[anchor=north,font=\scriptsize] {2};
               \draw ({3*0.55-0.35},0.1pt) -- ({3*0.55-0.35},-1.2pt) node[anchor=north,font=\scriptsize] {3};
               \draw ({4*0.55-0.35},0.1pt) -- ({4*0.55-0.35},-1.2pt) node[anchor=north,font=\scriptsize] {4};
            
               \draw (0.1pt,{0.4*1.5-0.45}) -- (-0.9pt,{0.4*1.5-0.45}) node[left,font=\scriptsize] at (-0.01,{0.4*1.5-0.45}) {$0.4$};
               \draw (0.1pt,{0.6*1.5-0.45}) -- (-0.9pt,{0.6*1.5-0.45}) node[left,font=\scriptsize] at (-0.01,{0.6*1.5-0.45}) {$0.6$};
               \draw (0.1pt,{0.8*1.5-0.45}) -- (-0.9pt,{0.8*1.5-0.45}) node[left,font=\scriptsize] at (-0.01, {0.8*1.5-0.45}) {$0.8$};
               \draw (0.1pt,{1*1.5-0.45}) -- (-0.9pt,{1*1.5-0.45}) node[left,font=\scriptsize] at (-0.01, {1*1.5-0.45}) {$1$};

               \draw[thick,
               		mark options={solid, draw={rgb,1:red,0;green,0;blue,0},
			fill={rgb,1:red,0.518;green,0.596;blue,0.686}}] 
			plot[mark=*,mark size=0.85pt] 
			coordinates {
				({1*0.55-0.35},{1.0002*1.5-0.45})
				({2*0.55-0.35},{1.0005*1.5-0.45})
				({3*0.55-0.35},{1.0009*1.5-0.45})
				({4*0.55-0.35},{1.0015*1.5-0.45})
			};
		
               \draw[line width = .5pt,
               		dash pattern=on 3pt off 1.5pt on 0.5pt off 1.5pt,
               		mark options={solid, line width = 0.5pt,draw={rgb,1:red,0;green,0;blue,0},
			fill={rgb,1:red,0.9;green,0.9;blue,0.9}}] 
			plot[mark=square*,mark size=.8pt] 
			coordinates {
				({1*0.55-0.35},{0.5209*1.5-0.45})
				({2*0.55-0.35},{0.6105*1.5-0.45})
				({3*0.55-0.35},{0.6447*1.5-0.45})
				({4*0.55-0.35},{0.6583*1.5-0.45})
			};      
            
                \node[font=\footnotesize] at (1,-0.28) {Expected ELB duration};
                
                \draw[semithick] (0.76,0.1) -- (0.9,0.1) node[right,font=\scriptsize,black] at (0.89,0.105) {$\mathcal{M}_y^f(\ell;\theta)$};
                \draw[only marks,mark=*,mark size=0.85pt,
                		mark options={draw={rgb,1:red,0;green,0;blue,0},fill={rgb,1:red,0.518;green,0.596;blue,0.686}}] plot coordinates {(0.83,0.1)};
                \draw[line width = .5pt, dash pattern=on 3pt off 1.5pt on 0.5pt off 1.5pt] (1.35,0.1) -- (1.5,0.1) node[right,font=\scriptsize,black] at (1.49,0.105) {$\mathcal{M}_y^\phi(\ell;\theta)$};
                \draw[only marks,mark=square*,mark size=0.8pt,
                		mark options={solid, line width = 0.5pt,draw={rgb,1:red,0;green,0;blue,0},fill={rgb,1:red,0.9;green,0.9;blue,0.9}}] plot coordinates {(1.425,0.1)};
             
            \end{tikzpicture}
        }\hspace*{0.2em}
        \subcaptionbox{Inflation\hspace*{-2.5em}}{
         \centering
         \begin{tikzpicture}[scale=3.4]
         
               \draw[black,thick] (0,0)--(0,1.2);
               \draw[black,thick] (0,0)--(2,0);
            
               \draw ({1*0.55-0.35},0.1pt) -- ({1*0.55-0.35},-1.2pt) node[anchor=north,font=\scriptsize] {1};
               \draw ({2*0.55-0.35},0.1pt) -- ({2*0.55-0.35},-1.2pt) node[anchor=north,font=\scriptsize] {2};
               \draw ({3*0.55-0.35},0.1pt) -- ({3*0.55-0.35},-1.2pt) node[anchor=north,font=\scriptsize] {3};
               \draw ({4*0.55-0.35},0.1pt) -- ({4*0.55-0.35},-1.2pt) node[anchor=north,font=\scriptsize] {4};
            
               \draw (0.1pt,{-0.03*30+1.05}) -- (-0.9pt,{-0.03*30+1.05}) node[left,font=\scriptsize] at (-0.01,{-0.03*30+1.05}) {$-0.03$};
               \draw (0.1pt,{-0.02*30+1.05}) -- (-0.9pt,{-0.02*30+1.05}) node[left,font=\scriptsize] at (-0.01,{-0.02*30+1.05}) {$-0.02$};
               \draw (0.1pt,{-0.01*30+1.05}) -- (-0.9pt,{-0.01*30+1.05}) node[left,font=\scriptsize] at (-0.01, {-0.01*30+1.05}) {$-0.01$};
               \draw (0.1pt,{0*30+1.05}) -- (-0.9pt,{0*30+1.05}) node[left,font=\scriptsize] at (-0.01, {0*30+1.05}) {$0$};

               \draw[thick,
               		mark options={solid, draw={rgb,1:red,0;green,0;blue,0},
			fill={rgb,1:red,0.518;green,0.596;blue,0.686}}] 
			plot[mark=*,mark size=0.85pt] 
			coordinates {
				({1*0.55-0.35},{3.8643e-05*30+1.05})
				({2*0.55-0.35},{5.2251e-05*30+1.05})
				({3*0.55-0.35},{7.7161e-05*30+1.05})
				({4*0.55-0.35},{1.2528e-04*30+1.05})
			};
		
               \draw[line width = .5pt,
               		dash pattern=on 3pt off 1.5pt on 0.5pt off 1.5pt,
               		mark options={solid, line width = 0.5pt,draw={rgb,1:red,0;green,0;blue,0},
			fill={rgb,1:red,0.9;green,0.9;blue,0.9}}] 
			plot[mark=square*,mark size=.8pt] 
			coordinates {
				({1*0.55-0.35},{-0.0262*30+1.05})
				({2*0.55-0.35},{-0.0258*30+1.05})
				({3*0.55-0.35},{-0.0255*30+1.05})
				({4*0.55-0.35},{-0.0253*30+1.05})
			};      
            
                \node[font=\footnotesize] at (1,-0.28) {Expected ELB duration};
                
                \draw[semithick] (0.76,0.1) -- (0.9,0.1) node[right,font=\scriptsize,black] at (0.89,0.108) {$\mathcal{M}_\pi^f(\ell;\theta)$};
                \draw[only marks,mark=*,mark size=0.85pt,
                		mark options={draw={rgb,1:red,0;green,0;blue,0},fill={rgb,1:red,0.518;green,0.596;blue,0.686}}] plot coordinates {(0.83,0.1)};
                \draw[line width = .5pt, dash pattern=on 3pt off 1.5pt on 0.5pt off 1.5pt] (1.35,0.1) -- (1.5,0.1) node[right,font=\scriptsize,black] at (1.49,0.108) {$\mathcal{M}_\pi^\phi(\ell;\theta)$};
                \draw[only marks,mark=square*,mark size=0.8pt,
                		mark options={solid, line width = 0.5pt,draw={rgb,1:red,0;green,0;blue,0},fill={rgb,1:red,0.9;green,0.9;blue,0.9}}] plot coordinates {(1.425,0.1)};
        
         \end{tikzpicture}
        }
\begin{minipage}{0.95\linewidth}
\setstretch{0.93}
{\footnotesize
\medskip
\emph{Notes:} Panel (a) shows the impact output multipliers\textemdash\ obtained under the proposed method (blue dots) and the AR-NA (gray squares)\textemdash for cases in which the ELB is expected to bind for $\ell=1,2,3,$ and $4$ quarters. The multipliers are computed as $\mathcal{M}_y(\ell;\theta)=1+(s_c/s_g)\mathcal{M}_c(\ell;\theta)$, where $\mathcal{M}_c(\ell;\theta)$ denotes the impact multiplier for consumption. Panel (b) plots the impact multipliers for inflation, which are computed via the formulas in Proposition \ref{prop:impact_multiplier} directly.
}
\end{minipage}
\end{figure}
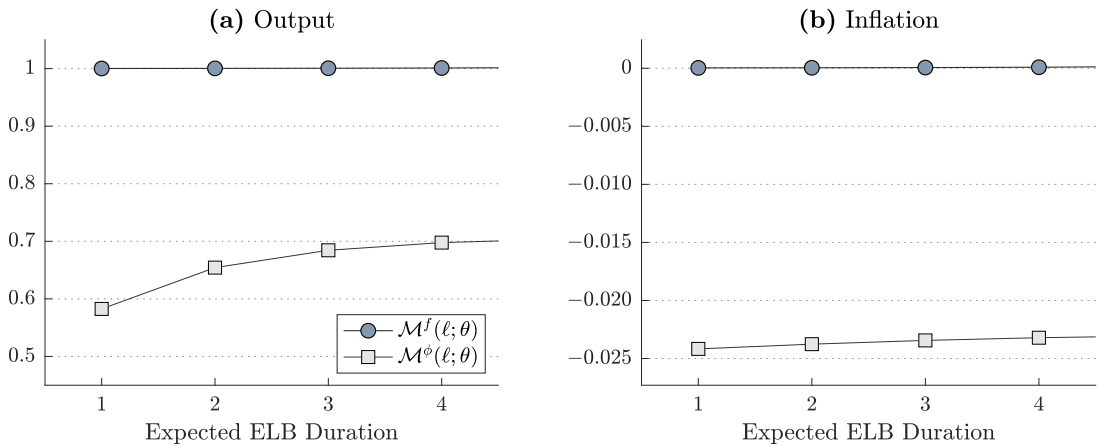

In Figure \ref{fig:AS_AD_US}, we have computed the AS and AD lines under the assumption of an expected ELB duration of $\ell=1$ for analytical tractability. As can be seen from the bottom right panel however, the expected duration in the data is actually $\ell=4$ quarters. Given that the objective of the current exercise is to gauge the effectiveness of fiscal policy at the ELB, we want to make sure that the conclusions drawn from the AS/AD graph are close to those that would arise in the case where the ELB is expected to bind for one year. To this effect, we report in Figure \ref{fig:M_ell_compar_2} the path of both the consumption and inflation multipliers for our method using the estimated parameters. For the sake of comparison, we also report the path of multipliers that AR-NA methods would produce for the same parameters.\footnote{We have re-estimated the model under the Taylor rule specification typically used in AR-NA methods and found qualitatively similar results.} Both of these impact multipliers are reported as a function of the expected duration of the ELB. 

There are many features worth flagging from Figure \ref{fig:M_ell_compar_2}. First, notice that the impact multipliers
computed using AR-NA and our methods have very different paths. For the duration of $\ell=4$ in the data, our impact multiplier is close to 1, while the one for AR-NA is closer to 0.6. What explains this discrepancy? Remember that the main difference across methods is the nature of the interest rate rule upon exit. If we use AR-NA, then the increase in government spending generates inflation in the short run. Given the presence of both exogenous and endogenous persistence, some of this inflation will be present upon exit and will force the Central Bank to increase interest rates. These higher future nominal rates will be anticipated by the representative, permanent income consumer: the increase in consumption will be dampened in the short run. In contrast, with our method this feedback from higher future interest rates will not be strong enough to overturn the expansionary effects of inflation during the ELB period and consumption will actually increase. This explains the discrepancy for an ELB duration of $\ell=1$. 

As the ELB lasts for longer, a bigger chunk of government spending happens at the ELB and a lower chunk outside. Given the configuration of the supply/demand graph in Figure \ref{fig:AS_AD_US}, one should expect the AR-NA multiplier to \textit{increase} with the duration of the trap. From Figure \ref{fig:M_ell_compar_2}, this is what is happening. After a number of periods however, the expected interest effect upon exit takes over and grows unboundedly large. In the limit where $\ell\to\infty$, the AR-NA multiplier diverges away to $-\infty$.

This is predicated on the fact that our parameter estimates imply $p_s>p^D$, which is the threshold above which the path of multipliers computed using AR-NA methods will diverge. 
In the standard New Keynesian model this feature is tightly linked with the magnitude of the slopes of AS and AD at the ELB. In the current framework where the expected duration is $\ell=1$, it turns out that the slopes reported in Figure \ref{fig:AS_AD_US} are \textit{not} informative. Can we still use AS and AD slopes to explain the instability of multipliers computed using existing piecewise linear methods? The answer is \textit{yes}, but only if we assume that $\ell\to\infty$.

For the sake of the argument, assume now that the ELB is expected to bind for an arbitrary long time. In that case, the value of $q$ (which governs the extent of endogenous persistence) will have to reflect that. More precisely, we now compute the Markov states and the last transition probability under the assumption that the ELB is binding forever. Let us denote the resulting value of the last transition probability as $q^\ast$. Except in some pathological cases, we will have $q\neq q^\ast$. We show in the online appendix that we can also cast this version of the model in a four state Markov chain framework. Given the value of $q^\ast$, we can compute the slopes of AS ($\pazocal{S}_{AS}(q^\ast)$) and AD ($\pazocal{S}_{AD}(q^\ast)$) at the ELB in the short run. In that case, we can prove that if $p_s$ is such that $\pazocal{S}_{AD}(q^\ast)>\pazocal{S}_{AS}(q^\ast)$, then the sequence of multipliers under AR-NA diverges. We establish this formally in the following proposition. 

\begin{prop}
\label{prop:p^Sinf_pD}
Let $\overline{p}_s(q)$ be the threshold probability such that $\pazocal{S}_{AD}(q)>\pazocal{S}_{AS}(q)$ if $p_s>\overline{p}_s(q)$. Likewise, let $\overline{p}_s(q^\ast)$ be the threshold probability such that $\pazocal{S}_{AD}(q^\ast)>\pazocal{S}_{AS}(q^\ast)$ if $p_s>\overline{p}_s(q^\ast)$. Then we have
\[
\overline{p}_s(q)\neq \overline{p}_s(q^\ast) = p^D.    
\]
In addition, if $h\to 0$ then we have $p^D=\overline{p}$ from the MC-CF literature. 
\end{prop}
\vspace*{-2em}
\begin{proof}
See Appendix \ref{apsec:proof_prop_2}.
\end{proof}

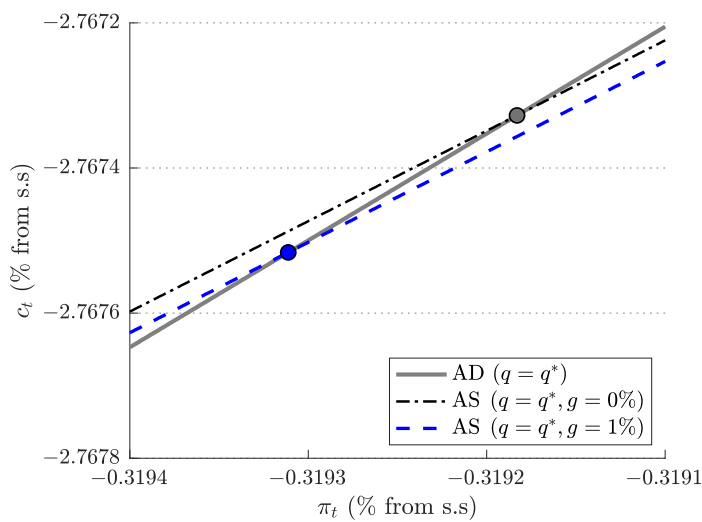
\begin{figure}[ht]
	\centering
	\caption{AD and AS lines when $\ell\to\infty$}\label{fig:AS_AD_US_qstar}
	\vspace*{-1em}
	\hspace*{-2.5em}\subcaptionbox*{}{
            \centering
            \begin{tikzpicture}[scale=3.8]
            
               \draw[black,thick] (0,0)--(0,1.2);
               \draw[black,thick] (0,0)--(2,0);
            
               \draw ({0.2},0.1pt) -- ({0.2},-1.2pt) node[anchor=north,font=\scriptsize] {$-0.2935$};
               \draw ({0.75},0.1pt) -- ({0.75},-1.2pt) node[anchor=north,font=\scriptsize] {$-0.293$};
               \draw ({1.3},0.1pt) -- ({1.3},-1.2pt) node[anchor=north,font=\scriptsize] {$-0.2925$};
               \draw ({1.85},0.1pt) -- ({1.85},-1.2pt) node[anchor=north,font=\scriptsize] {$-0.292$};
            
               \draw (0.1pt,{0.13}) -- (-0.9pt,{0.13}) node[left,font=\scriptsize] at (-0.01,{0.13}) {$-2.805$};
               \draw (0.1pt,{0.3675}) -- (-0.9pt,{0.3675}) node[left,font=\scriptsize] at (-0.01,{0.3675}) {$-2.8045$};
               \draw (0.1pt,{0.605}) -- (-0.9pt,{0.605}) node[left,font=\scriptsize] at (-0.01, {0.605}) {$-2.804$};
               \draw (0.1pt,{0.8425}) -- (-0.9pt,{0.8425}) node[left,font=\scriptsize] at (-0.01, {0.8425}) {$-2.8035$};
               \draw (0.1pt,{1.08}) -- (-0.9pt,{1.08}) node[left,font=\scriptsize] at (-0.01, {1.08}) {$-2.803$};

               \draw[gray,line width=1.5pt] 
               	({0.34299999999996089172782376408577},{0.00098410470718590659089386463165283}) 
               	-- ({1.96},{1.0697598618507981882430613040924});

               \draw[black,line width=.8pt,dash pattern=on 3pt off 1.5pt on 0.5pt off 1.5pt] 
               	({0.11200000000002319211489520967007},{-0.0035970879855540260905399918556213})
               	-- ({1.982},{1.0455162577402461465680971741676});

               \draw[blue,line width=1pt,dash pattern=on 4pt off 2pt] 
               	({0.28800000000001091393642127513885},{0.0027043062295888375956565141677856})
               	-- ({1.982},{0.95307757235764256620313972234726});

               \draw[semithick,
               		only marks,
               		mark options={solid, draw={rgb,1:red,0;green,0;blue,0},
			fill={rgb,1:red,0.45;green,0.45;blue,0.45}}] 
			plot[mark=*,mark size=0.9pt] 
			coordinates {
				({1.5939153502504836978914681822062},{0.82779177360362155013717710971832}) 
			};     
			
               \draw[semithick,
               		only marks,
               		mark options={solid, draw={rgb,1:red,0;green,0;blue,0},
			fill={rgb,1:red,0;green,0;blue,1}}] 
			plot[mark=*,mark size=0.9pt] 
			coordinates {
				({0.66896377747121960055665113031864},{0.21643381595504251890815794467926})
			};             

               \node[font=\footnotesize] at (1,-0.28) {$\pi_t$ (percentage points from s.s.)};
               \node[rotate=90,font=\footnotesize] at (-0.42,0.6) {$c_t$ (\% deviation from s.s.)};

               \draw[gray, line width=1.5pt] (1.2,0.34) -- (1.34,0.34) node[right,font=\scriptsize,black] at (1.33,0.335) {AD ($q^\ast$)};
               \draw[black,line width=.8pt,dash pattern=on 3pt off 1.5pt on 0.5pt off 1.5pt] (1.2,0.22) -- (1.34,0.22) node[right,font=\scriptsize,black] at (1.343,0.22) {AS ($q^\ast, g_t = 0\%$)};
               \draw[blue,line width=1pt,dash pattern=on 4pt off 2pt] (1.2,0.1) -- (1.34,0.1) node[right,font=\scriptsize,black] at (1.343,0.1) {AS ($q^\ast, g_t = 5\%$)};
 
            \end{tikzpicture}
        }
\begin{minipage}{0.95\linewidth}
\setstretch{0.93}
{\footnotesize
\medskip
\emph{Notes:} As in Figure \ref{fig:AS_AD_US}, the impact Markov state of the risk premium shock is re-calibrated such that equilibrium consumption with no government spending matches the model-implied consumption gap in 2009:Q1. All other parameters in the AD and AS equations are set to the minimum-distance estimates; explicit expressions for the AD and AS lines are provided in the online appendix. 
}
\end{minipage}
\end{figure}

The main take-away from Proposition \ref{prop:p^Sinf_pD} is that, just as in the simple model studied in the MC-CF literature, we can look at the slopes of AS and AD to gauge stability, but not just any slopes. In fact, the relevant slopes are the ones for which the ELB is expected to last for a very long time. Further, note that the last statement guarantees that our new threshold nests the one studied in the existing literature following \cite{Eggertsson2011} as a special case. We report these AS/AD lines under our estimated parameters for the U.S Great Recession in Figure \ref{fig:AS_AD_US_qstar}. Loosely speaking, these represent how the economy would react to a government spending shock in the short run if we increase the size of the shock $\xi_t$ to a very large value so that consumers and firms expect a much longer ELB period. 

From Figure \ref{fig:AS_AD_US_qstar}, note that the AD line slopes more than the AS line. This is due to the interaction of habits and passive monetary policy in the short run. We relegate a full description of the underlying intuition to the online appendix and focus on the consequences here. As in \cite{Bilbiie2022neo}, the bigger slope of AD tells us that expected income effects dominate. This is the reason why a small expected increase in the nominal interest rate upon exit percolates back and causes a large decrease in consumption on impact. In our method, this feedback is muted and the consumption multiplier converges to a finite value that is negative and which can be read off from Figure \ref{fig:AS_AD_US_qstar}: for a long ELB period, the U.S economy exhibits a response to government spending that follows the dynamics in \cite{Mertens2014}. It follows that the government spending multiplier flips qualitatively from $\ell=1$ to $\ell\to\infty$. In our case, this is due to a bifurcation at $\ell=7$ where the multiplier goes from 1.01 to 0.98. This has to be contrasted with the typical bifurcation/explosion that can be found in the MC-CF literature around $\overline{p}$ and where $p$ is a continuous variable. In our setup, $\ell$ is a discrete variable so that this explosion does not happen.\footnote{Given this logic, one could potentially recast the model in continuous time and derive a threshold value $\overline{\ell}$. In that case, it could be that the multiplier diverges as $\ell\to\overline{\ell}$. We leave this avenue for future research.}

Given how crucial the expected duration of the ELB period is, a natural next step is then to apply our methodology to the case of Japan, which has experienced the longest recorded ELB spell.

\section{The Japanese Example}\label{sec:Japan}
\vspace*{-0.8em}

\begin{figure}[ht]
	\centering
	\caption{Conditional expectations with endogenous persistence: Japan}\label{fig:TDSGE_h_GR_JPN}
	\begin{tikzpicture}
	\node (boxes)[inner sep=3.5pt] {
	\begin{minipage}{1\textwidth}
	\centering
	\hspace*{-.4cm}\subcaptionbox{AD and AS lines\hspace*{-5em}}{
            \centering
            \begin{tikzpicture}[scale=3.15]
            
                \draw[black,thick] (0,0)--(0,1.2);
                \draw[black,thick] (0,0)--(2,0);
                
               \draw ({0.10},0.1pt) -- ({0.10},-1.2pt) node[anchor=north,font=\scriptsize] {$-6.69835$};
               \draw ({0.55},0.1pt) -- ({0.55},-1.2pt) node[anchor=north,font=\scriptsize] {$-6.6983$};
               \draw ({1.00},0.1pt) -- ({1.00},-1.2pt) node[anchor=north,font=\scriptsize] {$-6.69825$};
               \draw ({1.45},0.1pt) -- ({1.45},-1.2pt) node[anchor=north,font=\scriptsize] {$-6.6982$};
               \draw ({1.90},0.1pt) -- ({1.90},-1.2pt) node[anchor=north,font=\scriptsize] {$-6.69815$};
            
               \draw (0.1pt,{0.14}) -- (-0.9pt,{0.14}) node[left,font=\scriptsize] at (-0.01,{0.13}) {$-7.54381$};
               \draw (0.1pt,{0.39}) -- (-0.9pt,{0.39}) node[left,font=\scriptsize] at (-0.01,{0.38}) {$-7.54371$};
               \draw (0.1pt,{0.64}) -- (-0.9pt,{0.64}) node[left,font=\scriptsize] at (-0.01,{0.63}) {$-7.54361$};
               \draw (0.1pt,{0.89}) -- (-0.9pt,{0.89}) node[left,font=\scriptsize] at (-0.01,{0.88}) {$-7.54351$};
               \draw (0.1pt,{1.14}) -- (-0.9pt,{1.14}) node[left,font=\scriptsize] at (-0.01,{1.13}) {$-7.54341$};
               
               \draw[gray,line width=1.5pt] 
               	({-0.0080000000016298145055770874023438},{0.22794150359914056025445461273193}) 
               	-- ({1.9899999999906867742538452148438},{1.0759898048127070069313049316406});
                
               \draw[black,line width=.8pt,dash pattern=on 3pt off 1.5pt on 0.5pt off 1.5pt] 
               	({-0.0080000000016298145055770874023438},{0.34483985461702104657888412475586})
               	-- ({1.9899999999906867742538452148438},{0.95532022211045841686427593231201});
	
               \draw[blue,line width=1pt,dash pattern=on 4pt off 2pt] 
               	({-0.0080000000016298145055770874023438},{0.29670325761981075629591941833496})
               	-- ({1.9899999999906867742538452148438},{0.90718362510961014777421951293945});
               
               \draw[semithick,
               		only marks,
               		mark options={solid, draw={rgb,1:red,0;green,0;blue,0},
			fill={rgb,1:red,0.45;green,0.45;blue,0.45}}] 
			plot[mark=*,mark size=0.9pt] 
			coordinates {
				({0.97514154548116493970155715942383},{0.64523455532980733551084995269775}) 
			};               

               \draw[semithick,
               		only marks,
               		mark options={solid, draw={rgb,1:red,0;green,0;blue,0},
			fill={rgb,1:red,0;green,0;blue,1}}] 
			plot[mark=*,mark size=0.9pt] 
			coordinates {
				({0.57030188782664481550455093383789},{0.47340092973172431811690330505371})
			};           
                            
               \node[font=\footnotesize] at (1.0,-0.28) {$\pi_t$ (percentage points from s.s.)};
               \node[rotate=90,font=\footnotesize] at (-0.53,0.55) {$c_t$ (\% deviation from s.s.)};
                
               \draw[gray, line width=1.5pt] (1.2,0.34) -- (1.33,0.34) node[right,font=\scriptsize,black] at (1.32,0.335) {AD};
               \draw[black,line width=.8pt,dash pattern=on 3pt off 1.5pt on 0.5pt off 1.5pt] (1.2,0.22) -- (1.33,0.22) node[right,font=\scriptsize,black] at (1.325,0.22) {AS ($g_t = 0\%$)};
               \draw[blue,line width=1pt,dash pattern=on 4pt off 2pt] (1.205,0.1) -- (1.335,0.1) node[right,font=\scriptsize,black] at (1.325,0.1) {AS ($g_t = 5\%$)};
                              
            \end{tikzpicture}
        }\hspace*{0.1em}
        \subcaptionbox{Expected output gaps\hspace*{-2em}}{
         \centering
         \begin{tikzpicture}[scale=3.15]
         
               \draw[black,thick] (0,0)--(0,1.2);
               \draw[black,thick] (0,0)--(2,0);
            
               \draw ({0*0.45+0.1},0.1pt) -- ({0*0.45+0.1},-1.2pt) node[anchor=north,font=\scriptsize] {09Q1};
               \draw ({1*0.45+0.1},0.1pt) -- ({1*0.45+0.1},-1.2pt) node[anchor=north,font=\scriptsize] {09Q3};
               \draw ({2*0.45+0.1},0.1pt) -- ({2*0.45+0.1},-1.2pt) node[anchor=north,font=\scriptsize] {10Q1};
               \draw ({3*0.45+0.1},0.1pt) -- ({3*0.45+0.1},-1.2pt) node[anchor=north,font=\scriptsize] {10Q3};
               \draw ({4*0.45+0.1},0.1pt) -- ({4*0.45+0.1},-1.2pt) node[anchor=north,font=\scriptsize] {11Q1};
            
               \draw (0.1pt,{-12*0.125+1.6}) -- (-0.9pt,{-12*0.125+1.6}) node[left,font=\scriptsize] at (-0.01,{-12*0.125+1.6}) {$-12$};
               \draw (0.1pt,{-10*0.125+1.6}) -- (-0.9pt,{-10*0.125+1.6}) node[left,font=\scriptsize] at (-0.01,{-10*0.125+1.6}) {$-10$};
               \draw (0.1pt,{-8*0.125+1.6}) -- (-0.9pt,{-8*0.125+1.6}) node[left,font=\scriptsize] at (-0.01, {-8*0.125+1.6}) {$-8$};
               \draw (0.1pt,{-6*0.125+1.6}) -- (-0.9pt,{-6*0.125+1.6}) node[left,font=\scriptsize] at (-0.01, {-6*0.125+1.6}) {$-6$};
               \draw (0.1pt,{-4*0.125+1.6}) -- (-0.9pt,{-4*0.125+1.6}) node[left,font=\scriptsize] at (-0.01, {-4*0.125+1.6}) {$-4$};

               \fill[fill={rgb,1:red,0.9;green,0.9;blue,0.9}, opacity=0.5]
               		({0*0.45+0.075},    {-5.87108*0.125+1.6})
               		-- ({0.5*0.45+0.1},  {-6.12974*0.125+1.6})
               		-- ({1*0.45+0.1},     {-5.99580*0.125+1.6})
               		-- ({1.5*0.45+0.1},  {-5.69135*0.125+1.6})
               		-- ({2*0.45+0.1},     {-5.30772*0.125+1.6})
               		-- ({2.5*0.45+0.1},  {-5.11686*0.125+1.6})
               		-- ({3*0.45+0.1},     {-4.95120*0.125+1.6})
               		-- ({3.5*0.45+0.1},  {-4.70376*0.125+1.6})
               		-- ({4*0.45+0.125}, {-4.46735*0.125+1.6})
               		-- ({4*0.45+0.125}, {-11.3619*0.125+1.6})
               		-- ({3.5*0.45+0.1},  {-11.2743*0.125+1.6})
               		-- ({3*0.45+0.1},     {-11.1778*0.125+1.6})
               		-- ({2.5*0.45+0.1},  {-10.9774*0.125+1.6})
               		-- ({2*0.45+0.1},     {-10.6652*0.125+1.6})
               		-- ({1.5*0.45+0.1},  {-10.2935*0.125+1.6})
               		-- ({1*0.45+0.1},     {-9.80252*0.125+1.6})
               		-- ({0.5*0.45+0.1},  {-8.98835*0.125+1.6})
               		-- ({0*0.45+0.075}, {-7.79461*0.125+1.6})
               		-- cycle;
		
               \draw[line width=.3pt,draw={rgb,1:red,0.5;green,0.5;blue,0.5}] plot
			coordinates {
               		({0*0.45+0.075}, {-7.79461*0.125+1.6})
               		({0.5*0.45+0.1},  {-8.98835*0.125+1.6})
               		({1*0.45+0.1}, 	  {-9.80252*0.125+1.6})
               		({1.5*0.45+0.1},  {-10.2935*0.125+1.6})
               		({2*0.45+0.1}, 	  {-10.6652*0.125+1.6})
               		({2.5*0.45+0.1},  {-10.9774*0.125+1.6})
               		({3*0.45+0.1}, 	  {-11.1778*0.125+1.6})
               		({3.5*0.45+0.1},  {-11.2743*0.125+1.6})
               		({4*0.45+0.125}, {-11.3619*0.125+1.6})
			};

               \draw[line width=.3pt, draw={rgb,1:red,0.5;green,0.5;blue,0.5}] plot
			coordinates {
               		({0*0.45+0.075}, {-5.87108*0.125+1.6})
               		({0.5*0.45+0.1},  {-6.12974*0.125+1.6})
               		({1*0.45+0.1},     {-5.99580*0.125+1.6})
               		({1.5*0.45+0.1},  {-5.69135*0.125+1.6})
               		({2*0.45+0.1},     {-5.30772*0.125+1.6})
               		({2.5*0.45+0.1},  {-5.11686*0.125+1.6})
               		({3*0.45+0.1},     {-4.95120*0.125+1.6})
               		({3.5*0.45+0.1},  {-4.70376*0.125+1.6})
               		({4*0.45+0.125}, {-4.46735*0.125+1.6})
			};
			
               \draw[line width = .5pt,
               		dash pattern=on 3pt off 1.5pt on 0.5pt off 1.5pt,
               		mark options={solid, line width = 0.5pt,draw={rgb,1:red,0;green,0;blue,0},
			fill={rgb,1:red,0.99;green,0.99;blue,0.99}}] 
			plot[mark=square*,mark size=.8pt] 
			coordinates {
				({0*0.45+0.1},	{-6.53203*0.125+1.6})
				({0.5*0.45+0.1},{-7.27834*0.125+1.6})
				({1*0.45+0.1},	{-7.69260*0.125+1.6})
				({1.5*0.45+0.1},{-7.88665*0.125+1.6})
				({2*0.45+0.1},	{-8.01182*0.125+1.6})
				({2.5*0.45+0.1},{-8.12542*0.125+1.6})
				({3*0.45+0.1},	{-8.14808*0.125+1.6})
				({3.5*0.45+0.1},{-8.11413*0.125+1.6})
				({4*0.45+0.1},	{-8.03495*0.125+1.6})
			};      
			
               \draw[thick,
               		mark options={solid, draw={rgb,1:red,0;green,0;blue,0},
			fill={rgb,1:red,0.518;green,0.596;blue,0.686}}] 
			plot[mark=*,mark size=0.85pt] 
			coordinates {
				({0*0.45+0.1},	{-6.74501*0.125+1.6})
				({0.5*0.45+0.1},{-7.19021*0.125+1.6})
				({1*0.45+0.1},	{-7.53384*0.125+1.6})
				({1.5*0.45+0.1},{-7.78972*0.125+1.6})
				({2*0.45+0.1},	{-7.97002*0.125+1.6})
				({2.5*0.45+0.1},{-8.08552*0.125+1.6})
				({3*0.45+0.1},	{-8.14571*0.125+1.6})
				({3.5*0.45+0.1},{-8.15899*0.125+1.6})
				({4*0.45+0.1},	{-8.13278*0.125+1.6})
			};
            
               \draw[line width = .5pt,
               		dash pattern=on 3pt off 1.5pt on 0.5pt off 1.5pt,
               		mark options={solid, line width = 1.25pt,draw={rgb,1:red,0;green,0;blue,0},
			fill={rgb,1:red,0.99;green,0.99;blue,0.99}, rotate=180}] 
			plot[mark=triangle*,mark size=.8pt] 
			coordinates {
				({0*0.45+0.1},{-7.54361*0.125+1.6})
			};
            
                \node[font=\footnotesize] at (0.98,-0.28) {Quarter};
        
         \end{tikzpicture}
         \vspace{0.4cm}
        }\quad
        \hspace*{0.7cm}\subcaptionbox{Expected inflation gaps\hspace*{-2.5em}}{
         \centering
         \begin{tikzpicture}[scale=3.15]
            
               \draw[black,thick] (0,0)--(0,1.2);
               \draw[black,thick] (0,0)--(2,0);
            
               \draw ({0*0.45+0.1},0.1pt) -- ({0*0.45+0.1},-1.2pt) node[anchor=north,font=\scriptsize] {09Q1};
               \draw ({1*0.45+0.1},0.1pt) -- ({1*0.45+0.1},-1.2pt) node[anchor=north,font=\scriptsize] {09Q3};
               \draw ({2*0.45+0.1},0.1pt) -- ({2*0.45+0.1},-1.2pt) node[anchor=north,font=\scriptsize] {10Q1};
               \draw ({3*0.45+0.1},0.1pt) -- ({3*0.45+0.1},-1.2pt) node[anchor=north,font=\scriptsize] {10Q3};
               \draw ({4*0.45+0.1},0.1pt) -- ({4*0.45+0.1},-1.2pt) node[anchor=north,font=\scriptsize] {11Q1};
            
               \draw (0.1pt,{-1.5*0.5+0.85}) -- (-0.9pt,{-1.5*0.5+0.85}) node[left,font=\scriptsize] at (-0.01,{-1.5*0.5+0.85}) {$-1.5$};
               \draw (0.1pt,{-1.0*0.5+0.85}) -- (-0.9pt,{-1.0*0.5+0.85}) node[left,font=\scriptsize] at (-0.01,{-1.0*0.5+0.85}) {$-1.0$};
               \draw (0.1pt,{-0.5*0.5+0.85}) -- (-0.9pt,{-0.5*0.5+0.85}) node[left,font=\scriptsize] at (-0.01, {-0.5*0.5+0.85}) {$-0.5$};
               \draw (0.1pt,{-0.0*0.5+0.85}) -- (-0.9pt,{-0.0*0.5+0.85}) node[left,font=\scriptsize] at (-0.01, {-0.0*0.5+0.85}) {$0$};
               \draw (0.1pt,{ 0.5*0.5+0.85}) -- (-0.9pt,{ 0.5*0.5+0.85}) node[left,font=\scriptsize] at (-0.01, { 0.5*0.5+0.85}) {$0.5$};
            
               \fill[fill={rgb,1:red,0.9;green,0.9;blue,0.9}, opacity=0.5]
               		({0*0.45+0.075},    {-1.13863*0.5+0.85})
               		-- ({0.5*0.45+0.1},  {-0.77707*0.5+0.85})
               		-- ({1*0.45+0.1},     {-0.64906*0.5+0.85})
               		-- ({1.5*0.45+0.1},  {-0.44471*0.5+0.85})
               		-- ({2*0.45+0.1},     {-0.27162*0.5+0.85})
               		-- ({2.5*0.45+0.1},  {-0.24160*0.5+0.85})
               		-- ({3*0.45+0.1},     {-0.16919*0.5+0.85})
               		-- ({3.5*0.45+0.1},  {-0.11685*0.5+0.85})
               		-- ({4*0.45+0.125}, {-0.09195*0.5+0.85})
               		-- ({4*0.45+0.125}, {-0.44471*0.5+0.85})
               		-- ({3.5*0.45+0.1},  {-0.44722*0.5+0.85})
               		-- ({3*0.45+0.1},     {-0.44974*0.5+0.85})
               		-- ({2.5*0.45+0.1},  {-0.49757*0.5+0.85})
               		-- ({2*0.45+0.1},     {-0.57070*0.5+0.85})
               		-- ({1.5*0.45+0.1},  {-0.69717*0.5+0.85})
               		-- ({1*0.45+0.1},     {-0.84704*0.5+0.85})
               		-- ({0.5*0.45+0.1},  {-1.01009*0.5+0.85})
               		-- ({0*0.45+0.075}, {-1.33559*0.5+0.85})
               		-- cycle;
		
               \draw[line width=.3pt,draw={rgb,1:red,0.5;green,0.5;blue,0.5}] plot
			coordinates {
               		({0*0.45+0.075}, {-1.33559*0.5+0.85})
               		({0.5*0.45+0.1},  {-1.01009*0.5+0.85})
               		({1*0.45+0.1}, 	  {-0.84704*0.5+0.85})
               		({1.5*0.45+0.1},  {-0.69717*0.5+0.85})
               		({2*0.45+0.1}, 	  {-0.57070*0.5+0.85})
               		({2.5*0.45+0.1},  {-0.49757*0.5+0.85})
               		({3*0.45+0.1}, 	  {-0.44974*0.5+0.85})
               		({3.5*0.45+0.1},  {-0.44722*0.5+0.85})
               		({4*0.45+0.125}, {-0.44471*0.5+0.85})
			};

               \draw[line width=.3pt, draw={rgb,1:red,0.5;green,0.5;blue,0.5}] plot
			coordinates {
               		({0*0.45+0.075}, {-1.13863*0.5+0.85})
               		({0.5*0.45+0.1},  {-0.77707*0.5+0.85})
               		({1*0.45+0.1},     {-0.64906*0.5+0.85})
               		({1.5*0.45+0.1},  {-0.44471*0.5+0.85})
               		({2*0.45+0.1},     {-0.27162*0.5+0.85})
               		({2.5*0.45+0.1},  {-0.24160*0.5+0.85})
               		({3*0.45+0.1},     {-0.16919*0.5+0.85})
               		({3.5*0.45+0.1},  {-0.11685*0.5+0.85})
               		({4*0.45+0.125}, {-0.09195*0.5+0.85})
			};
					
               \draw[line width = .5pt,
               		dash pattern=on 3pt off 1.5pt on 0.5pt off 1.5pt,
               		mark options={solid, line width = 0.5pt,draw={rgb,1:red,0;green,0;blue,0},
			fill={rgb,1:red,0.99;green,0.99;blue,0.99}}] 
			plot[mark=square*,mark size=.8pt] 
			coordinates {
				({0*0.45+0.1},	{-1.23454*0.5+0.85})
				({0.5*0.45+0.1},{-0.90864*0.5+0.85})
				({1*0.45+0.1},	{-0.74790*0.5+0.85})
				({1.5*0.45+0.1},{-0.54546*0.5+0.85})
				({2*0.45+0.1},	{-0.43213*0.5+0.85})
				({2.5*0.45+0.1},{-0.36934*0.5+0.85})
				({3*0.45+0.1},	{-0.31919*0.5+0.85})
				({3.5*0.45+0.1},{-0.29414*0.5+0.85})
				({4*0.45+0.1},	{-0.26911*0.5+0.85})
			};      
				
               \draw[thick,
               		mark options={solid, draw={rgb,1:red,0;green,0;blue,0},
			fill={rgb,1:red,0.518;green,0.596;blue,0.686}}] 
			plot[mark=*,mark size=0.85pt] 
			coordinates {
				({0*0.45+0.1},	{-1.33865*0.5+0.85})
				({0.5*0.45+0.1},{-0.67626*0.5+0.85})
				({1*0.45+0.1},	{-0.60190*0.5+0.85})
				({1.5*0.45+0.1},{-0.53548*0.5+0.85})
				({2*0.45+0.1},	{-0.47612*0.5+0.85})
				({2.5*0.45+0.1},{-0.42300*0.5+0.85})
				({3*0.45+0.1},	{-0.37543*0.5+0.85})
				({3.5*0.45+0.1},{-0.33275*0.5+0.85})
				({4*0.45+0.1},	{-0.29441*0.5+0.85})
			};
			
                \node[font=\footnotesize] at (0.98,-0.28) {Quarter};

                \draw[semithick] (0.9,0.35) -- (1.03,0.35) node[right,font=\scriptsize,black] at (1.03,0.35) {Model};
                \draw[only marks,mark=*,mark size=0.85pt,
                		mark options={draw={rgb,1:red,0;green,0;blue,0},fill={rgb,1:red,0.518;green,0.596;blue,0.686}}] plot coordinates {(0.965,0.35)};
                \draw[line width = .5pt, dash pattern=on 3pt off 1.5pt on 0.5pt off 1.5pt] (0.9,0.225) -- (1.03,0.225) node[right,font=\scriptsize,black] at (1.03,0.225) {ESP median};
                \draw[only marks,mark=square*,mark size=0.8pt,
                		mark options={solid, line width = 0.5pt,draw={rgb,1:red,0;green,0;blue,0},fill={rgb,1:red,0.99;green,0.99;blue,0.99}}] plot coordinates {(0.968,0.225)};
                \draw[line width=.3pt, draw={rgb,1:red,0.5;green,0.5;blue,0.5}] (0.9,0.1) -- (1.03,0.1) node[right,font=\scriptsize,black] at (1.03,0.1) {ESP 10--90 percentiles};

         \end{tikzpicture}
        }\hspace*{0.1em}     
        \hspace*{0.12cm}\subcaptionbox{Expected nominal rates\hspace*{-2em}}{
         \centering
         \begin{tikzpicture}[scale=3.15]
        
               \draw[black,thick] (0,0)--(0,1.2);
               \draw[black,thick] (0,0)--(2,0);
            
               \draw ({0*0.45+0.1},0.1pt) -- ({0*0.45+0.1},-1.2pt) node[anchor=north,font=\scriptsize] {09Q1};
               \draw ({1*0.45+0.1},0.1pt) -- ({1*0.45+0.1},-1.2pt) node[anchor=north,font=\scriptsize] {09Q3};
               \draw ({2*0.45+0.1},0.1pt) -- ({2*0.45+0.1},-1.2pt) node[anchor=north,font=\scriptsize] {10Q1};
               \draw ({3*0.45+0.1},0.1pt) -- ({3*0.45+0.1},-1.2pt) node[anchor=north,font=\scriptsize] {10Q3};
               \draw ({4*0.45+0.1},0.1pt) -- ({4*0.45+0.1},-1.2pt) node[anchor=north,font=\scriptsize] {11Q1};
            
               \draw (0.1pt,{-1.5*0.5+0.85}) -- (-0.9pt,{-1.5*0.5+0.85}) node[left,font=\scriptsize] at (-0.01,{-1.5*0.5+0.85}) {$-1.5$};
               \draw (0.1pt,{-1.0*0.5+0.85}) -- (-0.9pt,{-1.0*0.5+0.85}) node[left,font=\scriptsize] at (-0.01,{-1.0*0.5+0.85}) {$-1.0$};
               \draw (0.1pt,{-0.5*0.5+0.85}) -- (-0.9pt,{-0.5*0.5+0.85}) node[left,font=\scriptsize] at (-0.01, {-0.5*0.5+0.85}) {$-0.5$};
               \draw (0.1pt,{-0.0*0.5+0.85}) -- (-0.9pt,{-0.0*0.5+0.85}) node[left,font=\scriptsize] at (-0.01, {-0.0*0.5+0.85}) {$0$};
               \draw (0.1pt,{ 0.5*0.5+0.85}) -- (-0.9pt,{ 0.5*0.5+0.85}) node[left,font=\scriptsize] at (-0.01, { 0.5*0.5+0.85}) {$0.5$};
            
               \draw[thick,
               		mark options={solid, draw={rgb,1:red,0;green,0;blue,0},
			fill={rgb,1:red,0.518;green,0.596;blue,0.686}}] 
			plot[mark=*,mark size=0.85pt] 
			coordinates {
				({0*0.45+0.1},	{0*0.5+0.85})
				({0.5*0.45+0.1},{0*0.5+0.85})
				({1*0.45+0.1},	{0*0.5+0.85})
				({1.5*0.45+0.1},{0*0.5+0.85})
				({2*0.45+0.1},	{0*0.5+0.85})
				({2.5*0.45+0.1},{0*0.5+0.85})
				({3*0.45+0.1},	{0*0.5+0.85})
				({3.5*0.45+0.1},{0*0.5+0.85})
				({4*0.45+0.1},	{0*0.5+0.85})
			};
			
                \node[font=\footnotesize] at (0.98,-0.28) {Quarter};
            
         \end{tikzpicture}
        }
        \end{minipage}
        };
        \draw[semithick,dash pattern=on 1pt off 1pt] ({-3.05},{3.4}) -- ({1.8},{3.4});
        \end{tikzpicture}
\begin{minipage}{0.95\linewidth}
\setstretch{0.93}
{\footnotesize
\medskip
\emph{Notes:} As in Figure \ref{fig:AS_AD_US}, Panel (a) plots the AD and AS lines derived from the Markov restrictions in Appendix \ref{apsec:app_markovres_shortspell}. The parameters are set to the minimum-distance estimates, and the Markov state $s_{\xi,1}$ is calibrated so that equilibrium consumption without government spending matches the model-implied consumption gap in 2009-Q1\textemdash i.e., $\triangledown$ in Panel (b). See notes to Figure \ref{fig:AS_AD_US} for Panels (b)--(d).
}
\end{minipage}
\end{figure}
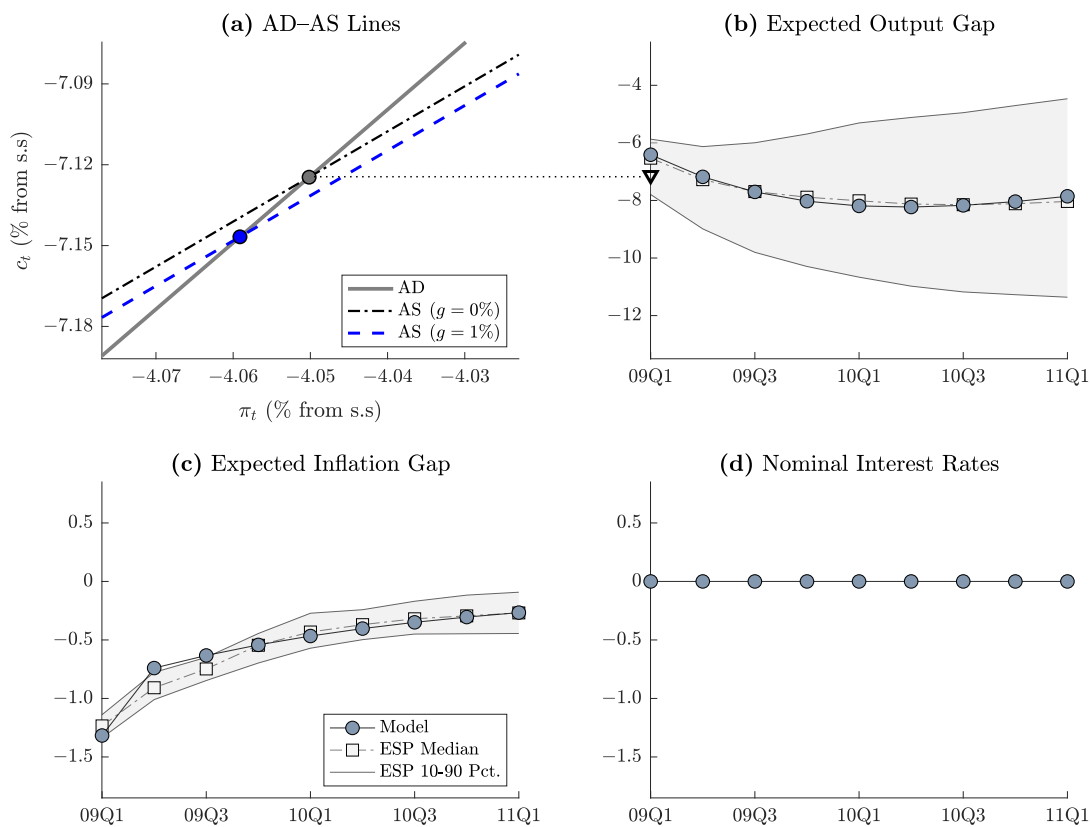

To the best of our knowledge, there does not exist SPF data for Japan so we follow \cite{Miyamoto2018} and use data from the Japan Center for Economic Research (JCER). The data is detailed in the online appendix. Using the same methodology as the one underlying Figure \ref{fig:AS_AD_US}, we fit the model to our Japanese expectations data. We do not have data for consumption but only for output, so we match output from the model instead now. In addition, while we do have data for a longer horizon (9 quarters) compared to the U.S case, we unfortunately do not have data for the expected nominal interest rate. Note however that this nominal rate had been stuck at essentially zero for a decade prior to the Great Recession. As a result, we assume that forecasters in our sample expect a zero interest rate for the whole 9 quarters going forward.\footnote{We have explored different values of $\ell^d$ ranging from 4 to 20 quarters but found that a duration of 9 quarters provides the best empirical fit.} Our approach is complementary with the one based on the non-linear Phillips curve in \cite{Cao2023nonlinear} in that we can generate a fairly long ELB episode without explosive dynamics. With this in mind, we report the results of this experiment in Figure \ref{fig:TDSGE_h_GR_JPN} and relegate the estimation results to the online appendix. 

\begin{figure}[ht]
	\centering
	\caption{Impact multipliers: Japan}\label{fig:M_ell_compar_2_JPN}
	\hspace*{-.4cm}\subcaptionbox{Output\hspace*{-2.5em}}{
            \centering
            \begin{tikzpicture}[scale=3.4]
            
               \draw[black,thick] (0,0)--(0,1.2);
               \draw[black,thick] (0,0)--(2,0);
            
               \draw ({1*0.225-0.125},0.1pt) -- ({1*0.225-0.125},-1.2pt) node[anchor=north,font=\scriptsize] {1};
               \draw ({3*0.225-0.125},0.1pt) -- ({3*0.225-0.125},-1.2pt) node[anchor=north,font=\scriptsize] {3};
               \draw ({5*0.225-0.125},0.1pt) -- ({5*0.225-0.125},-1.2pt) node[anchor=north,font=\scriptsize] {5};
               \draw ({7*0.225-0.125},0.1pt) -- ({7*0.225-0.125},-1.2pt) node[anchor=north,font=\scriptsize] {7};
               \draw ({9*0.225-0.125},0.1pt) -- ({9*0.225-0.125},-1.2pt) node[anchor=north,font=\scriptsize] {9};
            
               \draw (0.1pt,{0.97*29-27.95}) -- (-0.9pt,{0.97*29-27.95}) node[left,font=\scriptsize] at (-0.01,{0.97*29-27.95}) {$0.97$};
               \draw (0.1pt,{0.98*29-27.95}) -- (-0.9pt,{0.98*29-27.95}) node[left,font=\scriptsize] at (-0.01,{0.98*29-27.95}) {$0.98$};
               \draw (0.1pt,{0.99*29-27.95}) -- (-0.9pt,{0.99*29-27.95}) node[left,font=\scriptsize] at (-0.01,{0.99*29-27.95}) {$0.99$};
               \draw (0.1pt,{1.00*29-27.95}) -- (-0.9pt,{1.00*29-27.95}) node[left,font=\scriptsize] at (-0.01,{1.00*29-27.95}) {$1$};

               \draw[thick,
               		mark options={solid, draw={rgb,1:red,0;green,0;blue,0},
			fill={rgb,1:red,0.518;green,0.596;blue,0.686}}] 
			plot[mark=*,mark size=0.85pt] 
			coordinates {
				({1*0.225-0.125},{0.999883895111317*29-27.95})
				({2*0.225-0.125},{0.999874262565838*29-27.95})
				({3*0.225-0.125},{0.999862830618241*29-27.95})
				({4*0.225-0.125},{0.999849156202125*29-27.95})
				({5*0.225-0.125},{0.999832623280505*29-27.95})
				({6*0.225-0.125},{0.999812353029942*29-27.95})
				({7*0.225-0.125},{0.999787050743430*29-27.95})
				({8*0.225-0.125},{0.999754730695030*29-27.95})
				({9*0.225-0.125},{0.999712188288747*29-27.95})
			};
		
               \draw[line width = .5pt,
               		dash pattern=on 3pt off 1.5pt on 0.5pt off 1.5pt,
               		mark options={solid, line width = 0.5pt,draw={rgb,1:red,0;green,0;blue,0},
			fill={rgb,1:red,0.9;green,0.9;blue,0.9}}] 
			plot[mark=square*,mark size=.8pt] 
			coordinates {
				({1*0.225-0.125},{0.969716537001115*29-27.95})
				({2*0.225-0.125},{0.968846280081826*29-27.95})
				({3*0.225-0.125},{0.968112909284676*29-27.95})
				({4*0.225-0.125},{0.967496004718632*29-27.95})
				({5*0.225-0.125},{0.966978453423903*29-27.95})
				({6*0.225-0.125},{0.966546012029417*29-27.95})
				({7*0.225-0.125},{0.966186883846472*29-27.95})
				({8*0.225-0.125},{0.965891328336680*29-27.95})
				({9*0.225-0.125},{0.965651312114936*29-27.95})
			};      
            
                \node[font=\footnotesize] at (1,-0.28) {Expected ELB duration};
                
                \draw[semithick] (0.76,0.28) -- (0.9,0.28) node[right,font=\scriptsize,black] at (0.89,0.285) {$\mathcal{M}_y^f(\ell;\theta)$};
                \draw[only marks,mark=*,mark size=0.85pt,
                		mark options={draw={rgb,1:red,0;green,0;blue,0},fill={rgb,1:red,0.518;green,0.596;blue,0.686}}] plot coordinates {(0.83,0.285)};
                \draw[line width = .5pt, dash pattern=on 3pt off 1.5pt on 0.5pt off 1.5pt] (1.35,0.28) -- (1.5,0.28) node[right,font=\scriptsize,black] at (1.49,0.285) {$\mathcal{M}_y^\phi(\ell;\theta)$};
                \draw[only marks,mark=square*,mark size=0.8pt,
                		mark options={solid, line width = 0.5pt,draw={rgb,1:red,0;green,0;blue,0},fill={rgb,1:red,0.9;green,0.9;blue,0.9}}] plot coordinates {(1.425,0.28)};
             
            \end{tikzpicture}
        }\hspace*{0.2em}
        \subcaptionbox{Inflation\hspace*{-2.5em}}{
         \centering
         \begin{tikzpicture}[scale=3.4]
         
               \draw[black,thick] (0,0)--(0,1.2);
               \draw[black,thick] (0,0)--(2,0);
            
               \draw ({1*0.225-0.125},0.1pt) -- ({1*0.225-0.125},-1.2pt) node[anchor=north,font=\scriptsize] {1};
               \draw ({3*0.225-0.125},0.1pt) -- ({3*0.225-0.125},-1.2pt) node[anchor=north,font=\scriptsize] {3};
               \draw ({5*0.225-0.125},0.1pt) -- ({5*0.225-0.125},-1.2pt) node[anchor=north,font=\scriptsize] {5};
               \draw ({7*0.225-0.125},0.1pt) -- ({7*0.225-0.125},-1.2pt) node[anchor=north,font=\scriptsize] {7};
               \draw ({9*0.225-0.125},0.1pt) -- ({9*0.225-0.125},-1.2pt) node[anchor=north,font=\scriptsize] {9};
            
               \draw (0.1pt,{-0.003*290+1.05}) -- (-0.9pt,{-0.003*290+1.05}) node[left,font=\scriptsize] at (-0.01,{-0.003*290+1.05}) {$-0.003$};
               \draw (0.1pt,{-0.002*290+1.05}) -- (-0.9pt,{-0.002*290+1.05}) node[left,font=\scriptsize] at (-0.01,{-0.002*290+1.05}) {$-0.002$};
               \draw (0.1pt,{-0.001*290+1.05}) -- (-0.9pt,{-0.001*290+1.05}) node[left,font=\scriptsize] at (-0.01,{-0.001*290+1.05}) {$-0.001$};
               \draw (0.1pt,{-0.000*290+1.05}) -- (-0.9pt,{-0.000*290+1.05}) node[left,font=\scriptsize] at (-0.01,{-0.000*290+1.05}) {$0$};

               \draw[thick,
               		mark options={solid, draw={rgb,1:red,0;green,0;blue,0},
			fill={rgb,1:red,0.518;green,0.596;blue,0.686}}] 
			plot[mark=*,mark size=0.85pt] 
			coordinates {
				({1*0.225-0.125},{-0.0000089964*290+1.05})
				({2*0.225-0.125},{-0.0000099999*290+1.05})
				({3*0.225-0.125},{-0.0000111614*290+1.05})
				({4*0.225-0.125},{-0.0000125247*290+1.05})
				({5*0.225-0.125},{-0.0000141497*290+1.05})
				({6*0.225-0.125},{-0.0000161208*290+1.05})
				({7*0.225-0.125},{-0.0000185618*290+1.05})
				({8*0.225-0.125},{-0.0000216617*290+1.05})
				({9*0.225-0.125},{-0.0000257250*290+1.05})
			};

               \draw[line width = .5pt,
               		dash pattern=on 3pt off 1.5pt on 0.5pt off 1.5pt,
               		mark options={solid, line width = 0.5pt,draw={rgb,1:red,0;green,0;blue,0},
			fill={rgb,1:red,0.9;green,0.9;blue,0.9}}] 
			plot[mark=square*,mark size=.8pt] 
			coordinates {
				({1*0.225-0.125},{-0.003151584576625*290+1.05})
				({2*0.225-0.125},{-0.003162566297616*290+1.05})
				({3*0.225-0.125},{-0.003176578605883*290+1.05})
				({4*0.225-0.125},{-0.003192367305541*290+1.05})
				({5*0.225-0.125},{-0.003208921837895*290+1.05})
				({6*0.225-0.125},{-0.003225439743394*290+1.05})
				({7*0.225-0.125},{-0.003241293934848*290+1.05})
				({8*0.225-0.125},{-0.003256003486014*290+1.05})
				({9*0.225-0.125},{-0.003269208146018*290+1.05})
			};          
             
                \node[font=\footnotesize] at (1,-0.28) {Expected ELB duration};
                
                \draw[semithick] (0.76,0.28) -- (0.9,0.28) node[right,font=\scriptsize,black] at (0.89,0.288) {$\mathcal{M}_\pi^f(\ell;\theta)$};
                \draw[only marks,mark=*,mark size=0.85pt,
                		mark options={draw={rgb,1:red,0;green,0;blue,0},fill={rgb,1:red,0.518;green,0.596;blue,0.686}}] plot coordinates {(0.83,0.28)};
                \draw[line width = .5pt, dash pattern=on 3pt off 1.5pt on 0.5pt off 1.5pt] (1.35,0.28) -- (1.5,0.28) node[right,font=\scriptsize,black] at (1.49,0.288) {$\mathcal{M}_\pi^\phi(\ell;\theta)$};
                \draw[only marks,mark=square*,mark size=0.8pt,
                		mark options={solid, line width = 0.5pt,draw={rgb,1:red,0;green,0;blue,0},fill={rgb,1:red,0.9;green,0.9;blue,0.9}}] plot coordinates {(1.425,0.28)};
        
         \end{tikzpicture}
        }
\begin{minipage}{0.95\linewidth}
\setstretch{0.95}
{\footnotesize
\medskip
\emph{Notes:} See notes to Figure \ref{fig:M_ell_compar_2}.
}
\end{minipage}
\end{figure}
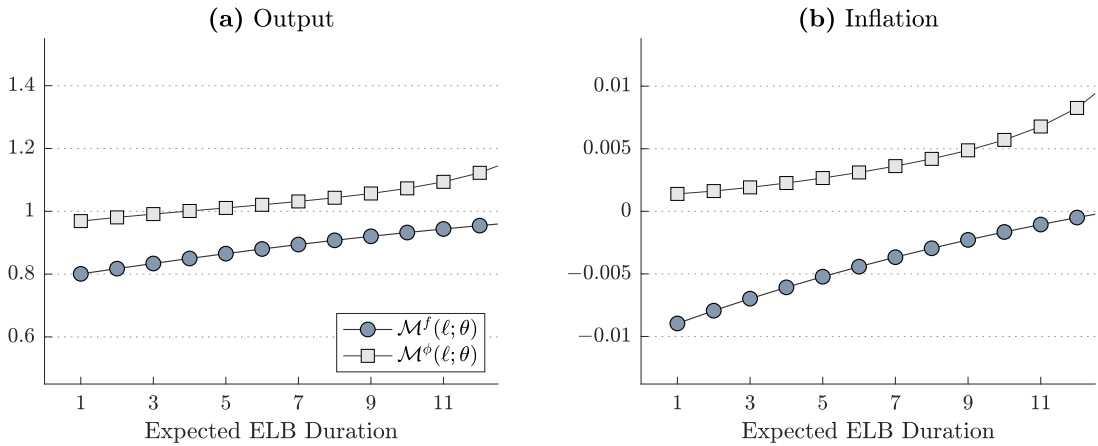

In line with our earlier findings, one clearly sees that expected output displays a hump-shape. In addition, notice that while inflation still looks like an $AR(1)$, it now reacts much less relative to real activity. Another feature of expected inflation is that it is quite persistent. Overall, our simple model with consumption habits still does a very good job in matching the expectations data closely. What kind of supply and demand lines at the ELB do rationalize these impulse response functions? The answer lies in the top-left corner of Figure \ref{fig:TDSGE_h_GR_JPN}: for the expected ELB duration of $\ell=9$ quarters, the AD line slopes more than the AS line at the ELB. In that situation, an increase in government spending shifts the AS line to the right and generates lower inflation and consumption according to the effects described in \cite{Mertens2014} and \cite{Bilbiie2022neo}. This is further evidenced in the path of output/inflation multipliers as a function of $\ell$ that we report in Figure \ref{fig:M_ell_compar_2_JPN}.

Notice that the multiplier under our method is once again higher than the one computed using AR-NA methods, but only slightly so. Following the intuition developed in the U.S case, this is because some of the increase in inflation due to government spending in the short run will result in higher nominal interest rates upon exit with a Taylor rule. These higher expected nominal rates in the future have a negative impact on consumption today. Using our method, the endogenous peg upon exit will mute these effects and results in a higher impact multiplier. Notice however that this effect is quantitatively small: both multipliers actually hover and quickly stabilize around 1. 

Our method provides a clear intuition for why this happens. The parameter configuration that best matches the Japanese data is such that $p_s<p^D$: the AD line slopes \textit{less} than the AS line in the hypothetical case where $\ell\to\infty$. Given our previous discussion, in that case the income/wealth effects are not strong enough to make the impact multiplier arbitrarily large as a function of $\ell$. In fact, given that $p_s<p^D$ we can guarantee that the multipliers computed using both methods will agree in the limit as $\ell$ is growing: they will both give a consumption multiplier above zero. In that case, the fact that the slope of AD is less steep than that for AS in the hypothetical case where $\ell\to\infty$ means that, for a long expected duration, the economy reacts to a government spending shock as in \cite{Eggertsson2011}: the AS line shifts along an AD line that is less steep, which result in a \textit{crowding in} of private consumption.

Our result that the consumption spending multiplier is positive in Japan in the case of an arbitrarily long liquidity trap is in sharp contrast with the results in \cite{Mertens2014}, \cite{Aruoba2018} and \cite{Bilbiie2022neo}. All three use a standard New Keynesian model without habit formation and in which the sunspot regime is caused by a two state Markov chain. In that framework, one can show that what matters is the persistence of the underlying shock and not the expected duration of the ELB epsisode. In both \cite{Mertens2014} and \cite{Bilbiie2022neo}, the expected duration is $\ell=1$, the persistence is very close to 1 and the realized duration is immaterial. \cite{Aruoba2018} also do estimate a probability to stay in the sunspot regime that is close to 1. In our case, the persistence of the underlying demand shock is almost zero and the persistence of the government spending shock is given by $p_s\simeq 0.94$. In contrast, the realized number of ELB periods can be large in our case because of both endogenous inertia through habit formation and the size of the demand shock in spite of its low persistence. This is the reason why we can have a positive consumption multiplier in the context of a long ELB. 

To sum it up, we have found dynamics that are somewhat different between the U.S and Japan. In the U.S case, we have found that the recession is caused by a relatively persistent demand shock and that one should expect a positive consumption multiplier for a short ELB duration. For the case of Japan, we have found that the recession is essentially given by a comparatively larger but almost one-off demand shock. In that context, the long duration of the ELB period is mostly due to the presence of endogenous inertia through external habit formation and our method gives a consumption multiplier that is slightly negative. In both cases however, we find a short run output multiplier that is very close to 1, which aligns well with the available empirical evidence \textemdash see \cite{Barro2011}, \cite{Ramey2011a,Ramey2011b} and more recently \cite{Ramey2018}. We do not find any evidence of policy puzzles except for the U.S case when we use the AR-NA method for a long ELB period: in that case the multiplier diverges away to $-\infty$.

\section{Conclusion}\label{sec:ccl}
\vspace*{-0.8em}

We have shown that, while extremely useful in clarifying the mechanisms at the ELB, standard three-equations New Keynesian models rely crucially on expectations dynamics which, by construction, cannot match the expectations data from the Great Recession. Against this backdrop, we have developed a method that is both $(i)$ able to replicate the salient features of these expectations and $(ii)$ is guaranteed to produce reasonable policy multipliers. Using our method, we have provided a set of tools to analyze all the properties of these models in detail. Finally, our results speak to the literature about the puzzles in the New Keynesian model. We have considered a model that is very standard in that it does not feature tractable heterogeneity, imperfect information, an OLG structure or even behavioral expectations. Even then, by taking the model to the data we have found impact output multipliers that are largely in line with what can be found in the empirical literature. Indeed, we have found no evidence of puzzling features in our simple model with external habit formation, except if we solve it using available piece-wise linear perfect foresight methods.

{
\small
\setstretch{1}
\bibliographystyle{apalike2}
\bibliography{EPELB}
}


\newpage
\appendix
\setlength{\abovedisplayskip}{1ex}
\setlength{\belowdisplayskip}{1ex}

\renewcommand\thesection{\Alph{section}}
\renewcommand{\theequation}{\thesection.\arabic{equation}}
\setcounter{equation}{0}

\section{Proof of Proposition \ref{prop:impact_multiplier}}\label{apsec:proof_prop_1}
\vspace*{-1.1em}

Suppose the ELB binds for $\ell$ periods in expectation, after which nominal interest rates follow the endogenous peg rule.\footnote{The proof for the AR-NA case is in the online appendix; the approach is analogous to that in this section.
} Then, the first $\ell$ states of the Markov chains $\mathbf{X}_{t+n}$, $\mathbf{Y}_{1,t+n},\dots,\mathbf{Y}_{N,t+n}$ are uniquely determined by $\ell\times(N+1)$ linear restrictions, given by:
\begin{align}
s_{x,1} & = Ds_{Y,1}, \label{apeq_prop1:x_restriction_s1}\\
s_{x,m+1} & = \varrho s_{x,m}+Ds_{Y,m+1}, \label{apeq_prop1:x_restriction_sm}\\
s_{Y,m} & = \mathbf{A}^\ast s_{Y,m+1} + B^\ast s_{x,m} + p_b^{m-1}C_b^\ast s_{w_b,1} + p_s^{m-1}C_s^\ast s_{w_s,1} + E^\ast, \label{apeq_prop1:Y_restriction_sm}\\
0_{N\times1} & = \boldsymbol{\Omega}^\ast_Ys_{Y,\ell}+\Omega^\ast_xs_{x,\ell} + p_b^{\ell-1}\Omega^\ast_{w_b}s_{w_b,1} + p_s^{\ell-1}C_s^\ast s_{w_s,1}+E^\ast, \label{apeq_prop1:Y_restriction_sL}
\end{align}
for $m = 1,2,\dots,\ell-1$.\footnote{$\mathbf{X}_t$, $\mathbf{Y}_{1,t}$, \dots, $\mathbf{Y}_{N,t}$ represent the Markov chains of the $(N+1)$ endogenous variables described in equation \eqref{eq:fwd_1} of the paper. $s_{Y,m} \equiv [s_{Y_1,m},\dots,s_{Y_N,m}]^\top$ contains the $m$th states for $\mathbf{Y}_{1,t},\dots,\mathbf{Y}_{N,t}$, and $0_{N\times1}$ is the $N\times1$ zero vector. The matrices $\boldsymbol{\Omega}_Y^\ast$ ($N\times N$), $\Omega_x^\ast$ ($N\times1$), and $\Omega_{w_b}^\ast$ ($N\times1$) contain the model parameters, with $\mathbf{\Omega}_Y^\ast = -(\mathbf{I}_N-p_s\mathbf{A}^\ast) + p_sq\mathbf{A}^\ast(\mathbf{I}_N-q\mathbf{A})^{-1}BD/\varrho$, $\Omega_x^\ast=B^\ast-(p_s-\varrho)q\mathbf{A}^\ast(\mathbf{I}_N-q\mathbf{A})^{-1}B/\varrho$, while $\Omega_{w_b}^\ast$ is given in the online appendix. The derivations of \eqref{apeq_prop1:x_restriction_s1}--\eqref{apeq_prop1:Y_restriction_sL} are also relegated to the online appendix. Throughout this section, we assume that $\mathbf{A}^\ast$ and $\mathbf{I}_N-q\mathbf{A}$ are nonsingular matrices.
}
Definition \ref{def:impact_multiplier} of the paper implies that we can compute $\mathcal{M}(\ell;\theta)$ via a two-step process: (i) use \eqref{apeq_prop1:x_restriction_s1}--\eqref{apeq_prop1:Y_restriction_sL} to express $s_{Y,1}$ in terms of $s_{w_s,1}$, $s_{w_b,1}$, and $E^\ast$; (ii) calculate $\mathcal{M}(\ell;\theta)$ as $\partial s_{Y,1}/\partial s_{w_s,1}$. We apply these steps to prove, by induction, that $\mathcal{M}(\ell;\theta)$ satisfies the following recurrence for all $\ell>1$:
\begin{align}
\pazocal{M}(\ell;\theta) & = (\mathbf{A}^\ast)^{-1}\pazocal{X}_{\ell-1}\left[C_s^\ast + p_s\mathbf{A}^\ast\pazocal{M}(\ell-1;\theta)\right],\label{apeq_prop1:caliM_equation} 
\end{align}
given $\mathcal{M}(1;\theta)$ and a sequence of nonsingular matrices $\mathcal{X}_1,\dots,\mathcal{X}_{\ell-1}$ such that:
\begin{align}
\pazocal{X}_{\ell-1} & = \mathbf{A}^\ast(\mathbf{I}_{N} - B^\ast D+\varrho\mathbf{A}^\ast-\varrho\pazocal{X}_{\ell-2})^{-1} \quad\text{ for all \ } \ell>2. \label{apeq_prop1:caliX_equation}
\end{align}
To verify the induction base case, we derive $\mathcal{M}(1;\theta)$, $\mathcal{M}(2;\theta)$, and $\mathcal{M}(3;\theta)$ as follows:
\begin{align}
\mathcal{M}(1;\theta) & = \mathcal{M}^f(1;\theta) \equiv (-\boldsymbol{\Omega}_Y^\ast-\Omega_x^\ast D)^{-1}C_s^\ast \equiv \boldsymbol{\Omega}^fC_s^\ast, \label{apeq_prop1:caliM_1}\\
\mathcal{M}(2;\theta) & = (\mathbf{I}_N - B^\ast D - \varrho\mathbf{A}^\ast\boldsymbol{\Omega}^f\Omega_x^\ast D)^{-1}[C_s^\ast + p_s\mathbf{A}^\ast\mathcal{M}(1;\theta)] \notag \\
& \equiv (\mathbf{A}^\ast)^{-1}\mathcal{X}_1^f[C_s^\ast + p_s\mathbf{A}^\ast\mathcal{M}^f(1;\theta)], \label{apeq_prop1:caliM_2}
\end{align}
where we set $\mathcal{X}_1 = \mathcal{X}^f_1 \equiv \mathbf{A}^\ast(\mathbf{I}_N - B^\ast D - \varrho\mathbf{A}^\ast\boldsymbol{\Omega}^f\Omega_x^\ast D)^{-1}$ and use \eqref{apeq_prop1:caliM_1} to yield \eqref{apeq_prop1:caliM_2}. (Superscript $f$ on $\mathcal{M}(1;\theta)$ and $\mathcal{X}_1$ indicate initial condition expressions obtained under the proposed solution method.) Note that \eqref{apeq_prop1:caliM_2} verifies the recurrence \eqref{apeq_prop1:caliM_equation} for $\ell=2$. To derive $\mathcal{M}(3;\theta)$, we first use \eqref{apeq_prop1:x_restriction_sm} with $m=2$ to remove $s_{x,3}$ in \eqref{apeq_prop1:Y_restriction_sL}. We then obtain $s_{Y,3} = p_s^2\mathcal{M}^f(1;\theta)s_{w_s,1} + \varrho\boldsymbol{\Omega}^f\Omega_x^\ast s_{x,2} + \text{t.i.p.}$, where t.i.p.\ contains terms involving $s_{w_b,1}$ and $E^\ast$. Using this, alongside \eqref{apeq_prop1:x_restriction_s1}, \eqref{apeq_prop1:x_restriction_sm} for $m=1$, \eqref{apeq_prop1:Y_restriction_sm} for $m=2$, and $\mathcal{X}_1^f$, we yield:
\begin{align*}
s_{Y,2} 
& = p_s(\mathbf{A}^\ast)^{-1}\mathcal{X}_1^f[C_s^\ast + p_s\mathbf{A}^\ast\mathcal{M}^f(1;\theta)]s_{w_s,1} + \varrho(\mathbf{A}^\ast)^{-1}\mathcal{X}_1^f(\varrho\mathbf{A}^\ast\boldsymbol{\Omega}^f\Omega_x^\ast + B^\ast)Ds_{Y,1} + \text{t.i.p.} \\
& = p_s\mathcal{M}(2;\theta)s_{w_s,1} + \varrho(\mathbf{A}^\ast)^{-1}(\mathcal{X}_1^f-\mathbf{A}^\ast)s_{Y,1} + \text{t.i.p.}
\end{align*}
The second equality follows from \eqref{apeq_prop1:caliM_2}, and the fact that:
\begin{align}
\label{apeq_prop1:simplify_caliX1}
\mathcal{X}_1^f(\varrho\mathbf{A}^\ast\boldsymbol{\Omega}^f\Omega_x^\ast D + B^\ast D) = \mathcal{X}_1^f\big[\mathbf{I}_N - (\mathcal{X}^f_1)^{-1}\mathbf{A}^\ast\big] = \mathcal{X}_1^f - \mathbf{A}^\ast.
\end{align}
Now, using \eqref{apeq_prop1:Y_restriction_sm} with $m=1$ and defining $\mathcal{X}_2$ according to \eqref{apeq_prop1:caliX_equation}, we obtain:
\begin{align}
\mathcal{M}(3;\theta) = (\mathbf{A}^\ast)^{-1}\mathcal{X}_2[C_s^\ast+p_s\mathbf{A}^\ast\mathcal{M}(2;\theta)]. \label{apeq_prop1:caliM_3}
\end{align}
This completes the base case of the induction, where \eqref{apeq_prop1:caliM_equation} holds given \eqref{apeq_prop1:caliX_equation}.

Suppose now that \eqref{apeq_prop1:caliM_equation} holds for all $1 < \ell \leq n$ and that \eqref{apeq_prop1:caliX_equation} holds for all $2 < \ell\leq n+1$, with $n\geq3$. To complete the proof, we need to show that the recurrence in \eqref{apeq_prop1:caliM_equation} also holds for $\ell=n+1$. Setting $\ell=n+1$, we write $s_{Y,m}$ (for $m=n,n-1,\dots,1$) as:\footnote{We adopt the notation $\prod_{j=1}^{n}\mathcal{X}_j = \mathcal{X}_n\mathcal{X}_{n-1}\dots\mathcal{X}_1$. Also, $\prod_{j=a+1}^{a}\mathcal{X}_j=\mathbf{I}_N$ and $\sum_{k=1}^0\prod_{j=k+1}^{n-m}(\varrho\mathcal{X}_j)=\mathbf{0}_{N}$.} 
\begin{align}
\hspace*{-2em} s_{Y,m} & = p_s^{m-1}[C_s^\ast + p_s\mathbf{A}^\ast\mathcal{M}(n-m+1;\theta)]s_{w_s,1} \notag \\
& \qquad\quad + \Big[\varrho^{n-m}\prod_{j=1}^{n-m}\mathcal{X}_j(\varrho\mathbf{A}^\ast\boldsymbol{\Omega}^f\Omega_x^\ast+B^\ast) + \sum_{k=1}^{n-m}\prod_{j=k+1}^{n-m}(\varrho\mathcal{X}_j)B^\ast\Big]s_{x,m} + \text{t.i.p.} \label{apeq_prop1:Ym_expression}
\end{align}
\vspace*{-.8em}
To see this, we first express \eqref{apeq_prop1:Ym_expression} as follows:
\begin{align}
s_{Y,m} & = p_s^{m-1}[C_s^\ast + p_s\mathbf{A}^\ast\mathcal{M}(n-m+1;\theta)]s_{w_s,1}  \label{apeq_prop1:Ym_expression_long}\\
& \qquad\quad + (\varrho\mathcal{X}_{n-m}(\dots(\varrho\mathcal{X}_2(\varrho\mathcal{X}_1^f(\varrho\mathbf{A}^\ast\boldsymbol{\Omega}^f\Omega_x^\ast+\underbracket{ B^\ast)+B^\ast)+B^\ast)\dots)+B^\ast}_{(n-m+1) \text{ nested terms}})s_{x,m} + \text{t.i.p.}
\notag
\end{align}
Let $m=n$. Then, \eqref{apeq_prop1:Ym_expression_long} holds true by substituting \eqref{apeq_prop1:x_restriction_sm} and \eqref{apeq_prop1:Y_restriction_sm} for $m=n$ into \eqref{apeq_prop1:Y_restriction_sL}. Suppose \eqref{apeq_prop1:Ym_expression_long} holds for some $m=\bar{m}$ with $1<\bar{m}< n$. We now show that \eqref{apeq_prop1:Ym_expression_long} also holds for $m=\bar{m}-1$. Replacing $s_{x,\bar{m}}$ in \eqref{apeq_prop1:Ym_expression_long} using \eqref{apeq_prop1:x_restriction_sm} allows us to derive:
\begin{align*}
s_{Y,\bar{m}} & = p_s^{\bar{m}-1}[C_s^\ast + p_s\mathbf{A}^\ast\mathcal{M}(n-\bar{m}+1;\theta)]s_{w_s,1} \\
& \quad + \varrho(\varrho\mathcal{X}_{n-\bar{m}}(\dots(\varrho\mathcal{X}_2(\varrho\mathcal{X}^f_1(\varrho\mathbf{A}^\ast\boldsymbol{\Omega}^f\Omega_x^\ast + B^\ast)+B^\ast)+B^\ast)\dots)+B^\ast)s_{x,\bar{m}-1} \\
& \quad + (\underbracket{\varrho\mathcal{X}_{n-\bar{m}}(\dots(\underbracket{\varrho\mathcal{X}_2(\underbracket{\varrho\mathcal{X}^f_1(\varrho\mathbf{A}^\ast\boldsymbol{\Omega}^f\Omega_x^\ast D+ B^\ast D)+B^\ast D}_{ = \varrho\mathcal{X}^f_1-\varrho\mathbf{A}^\ast+B^\ast D \text{ \ by \eqref{apeq_prop1:simplify_caliX1}.}})+B^\ast D}_{= \varrho\mathcal{X}_2-\varrho\mathbf{A}^\ast+B^\ast D \text{ \ by \eqref{apeq_prop1:caliX_equation} for $\ell=3$.}})\dots)+B^\ast D}_{= \varrho\mathcal{X}_{n-\bar{m}}-\varrho\mathbf{A}^\ast+B^\ast D \text{ \ by \eqref{apeq_prop1:caliX_equation} for $\ell=n-\bar{m}+1$.}})s_{Y,\bar{m}} + \text{t.i.p.}\\
& = p_s^{\bar{m}-1}\mathcal{M}(n-\bar{m}+2;\theta)s_{w_s,1} + \varrho(\mathbf{A}^\ast)^{-1}\mathcal{X}_{n-\bar{m}+1}(\dots(\varrho\mathbf{A}^\ast\boldsymbol{\Omega}^f\Omega_x^\ast + B^\ast)\dots)s_{x,\bar{m}-1} + \text{t.i.p.}
\end{align*}
The second equality follows from the fact that \eqref{apeq_prop1:caliX_equation} holds for all $\ell=3,\dots,n-\bar{m}+2$ and that \eqref{apeq_prop1:caliM_equation} holds for $\ell=n-\bar{m}+2$ (induction hypothesis). Substituting the above expression for $s_{Y,\bar{m}}$ into \eqref{apeq_prop1:Y_restriction_sm} with $m=\bar{m}-1$ gives:
\begin{align}
s_{Y,\bar{m}-1} & = p_s^{\bar{m}-2}[C_s^\ast + p_s\mathbf{A}^\ast\mathcal{M}(n-\bar{m}+2;\theta)]s_{w_s,1} \notag\\
& \qquad\quad + (\varrho\mathcal{X}_{n-\bar{m}+1}(\dots(\varrho\mathbf{A}^\ast\boldsymbol{\Omega}^f\Omega_x^\ast + \underbracket{B^\ast)\dots)+B^\ast}_{(n-\bar{m}+2) \text{ nested terms}})s_{x,\bar{m}-1} + \text{t.i.p.} \label{apeq_prop1:intermediatestep_Ym}
\end{align}
This implies that \eqref{apeq_prop1:Ym_expression_long}, and hence \eqref{apeq_prop1:Ym_expression}, hold for $m=\bar{m}-1$. Since \eqref{apeq_prop1:x_restriction_sm} and \eqref{apeq_prop1:Y_restriction_sm} hold for all $m=1,\dots,n$, we deduce that \eqref{apeq_prop1:Ym_expression} holds for $m=n,\dots,1$. Now, set $m=1$ in \eqref{apeq_prop1:Ym_expression}, and use \eqref{apeq_prop1:x_restriction_s1} to express $s_{x,1}$ in terms of $s_{Y,1}$, we obtain:
\begin{align}
s_{Y,1} & =  [C_s^\ast + p_s\mathbf{A}^\ast\mathcal{M}(n;\theta)]s_{w_s,1}  \notag\\
& \qquad + (\varrho\mathcal{X}_{n-1}(\dots(\varrho\mathcal{X}^f_1(\varrho\mathbf{A}^\ast\boldsymbol{\Omega}^f\Omega_x^\ast D+B^\ast D)+B^\ast D)\dots)+B^\ast D)s_{Y,1} + \text{t.i.p.} \notag\\
& = [C_s^\ast + p_s\mathbf{A}^\ast\mathcal{M}(n;\theta)]s_{w_s,1}  + (\varrho\mathcal{X}_{n-1}- \varrho\mathbf{A}^\ast+B^\ast D)s_{Y,1} + \text{t.i.p.} \notag\\
& = (\mathbf{A}^\ast)^{-1}\mathcal{X}_n[C_s^\ast + p_s\mathbf{A}^\ast\mathcal{M}(n;\theta)]s_{w_s,1} + \text{t.i.p.}  \label{apeq_prop1:caliM_n1}
\end{align}
The second equality follows by applying a strategy similar to the derivation for \eqref{apeq_prop1:intermediatestep_Ym}. The third equality follows from the induction hypothesis that \eqref{apeq_prop1:caliX_equation} holds for $\ell=n+1$. Since \eqref{apeq_prop1:caliM_n1} implies that \eqref{apeq_prop1:caliM_equation} holds for $\ell=n+1$, we conclude by induction that the recurrence \eqref{apeq_prop1:caliM_equation} holds for all $\ell>1$, given that the sequence $\{\mathcal{X}_\ell\}$ is updated according to \eqref{apeq_prop1:caliX_equation} from $\ell\geq2$ onward. Hence, the proof is complete.\qed 

\vspace*{-1.5em}

\setcounter{equation}{0}
\section{Limiting Behavior of \texorpdfstring{$\{\mathcal{X}_j\}$}{Xj}}\label{apsec:dynamics_X}
\vspace*{-1.1em}

In Proposition \ref{prop:impact_multiplier}, we show that the sequence $\{\mathcal{X}_j\}$ satisfies:
\begin{align*}
\mathcal{X}_{j+1} & = \mathbf{A}^\ast\left(\mathbf{I}_N - B^\ast D + \varrho\mathbf{A}^\ast - \varrho\mathcal{X}_{j}\right)^{-1}
\end{align*}
for $j\in\mathbb{Z}^+$, where $\mathbb{Z}^+$ denotes the set of positive integers. Rearranging this gives:
\begin{align}
\label{apeq_dynx:recurrence_x}
\mathcal{X}_{j+1}\mathbf{\Psi}_1 - \mathcal{X}_{j+1}\mathcal{X}_{j}\mathbf{\Psi}_2 & = \mathbf{I}_N,
\end{align}
with coefficients $\mathbf{\Psi}_1=\left(\mathbf{I}_N - B^\ast D + \varrho\mathbf{A}^\ast \right)(\mathbf{A}^\ast)^{-1}$ and $\mathbf{\Psi}_2 = \varrho(\mathbf{A}^\ast)^{-1}$. Consider a nonsingular matrix $\mathcal{K}_j$ such that $\mathcal{X}_j=(\mathcal{K}_j)^{-1}\mathcal{K}_{j-1}$ for $j\in\mathbb{Z}^+$.\footnote{If $\mathcal{X}_j^{-1}$ exists, we can set $\mathcal{K}_{j}=\mathcal{K}_{j-1}\mathcal{X}_{j}^{-1}$ with $\mathcal{K}_0=\mathbf{I}_N$. Then, $\mathcal{K}_1^{-1}$ exists, and inductively, $\mathcal{K}_j^{-1}$ exists.} With this, \eqref{apeq_dynx:recurrence_x} simplifies to the following second-order linear matrix recurrence:
\begin{align}
\label{apeq_dynx:recurrence_k}
\mathcal{K}_{j+1}^\top - \mathbf{\Psi}_1^\top\mathcal{K}_j^\top + \mathbf{\Psi}_2^\top\mathcal{K}_{j-1}^\top & = \mathbf{0}_N,
\end{align}
with initial conditions $\mathcal{K}^\top_0=\mathbf{I}_N$ and $\mathcal{K}^\top_1=(\mathcal{X}_1^\top)^{-1}$. ($\mathcal{K}_j^\top$ is the transpose of $\mathcal{K}_j$.) Now, the next result provides a general solution to the recurrence in \eqref{apeq_dynx:recurrence_k}.

\begin{lemma}\label{aplemma_dynx:solution_k}
Let $\mathcal{S}_1$ and $\mathcal{S}_2$ be the dominant and minimal solutions to the matrix polynomial:
\begin{align}
\label{apeq_dynx:qme_K}
\mathcal{Q}(\mathcal{S}) = \mathcal{S}^2 - \mathbf{\Psi}_1^\top\mathcal{S} + \mathbf{\Psi}_2^\top = \mathbf{0}_N.
\end{align}
Then, the following results hold: (i) The block Vandermonde matrix $\mathcal{V}(\mathcal{S}_1,\mathcal{S}_2)$ is nonsingular,
\begin{align*}
\det\Big(\mathcal{V}(\mathcal{S}_1,\mathcal{S}_2)\Big) \neq0 \quad\text{ for }\quad \mathcal{V}(\mathcal{S}_1,\mathcal{S}_2) =
\begin{bmatrix}
\mathbf{I}_N & \mathbf{I}_N \\ \mathcal{S}_1 & \mathcal{S}_2
\end{bmatrix}.
\end{align*}
(ii) The general solution to \eqref{apeq_dynx:recurrence_k} is $\mathcal{K}_j^\top = \mathcal{S}_1^j\mathcal{C}_1+\mathcal{S}_2^j\mathcal{C}_2$, where $\mathcal{C}_2=(\mathcal{S}_2-\mathcal{S}_1)^{-1}(\mathcal{K}_1^\top-\mathcal{S}_1)$ and $\mathcal{C}_1=\mathbf{I}_N-\mathcal{C}_2$.
\end{lemma}
\vspace*{-2em}
\begin{proof}
See \citeauthor{Higham2000numerical} (\citeyear{Higham2000numerical}, Theorems 8 and 7) for (i) and (ii), respectively. Since $\det(\mathcal{V}(\mathcal{S}_1,\mathcal{S}_2))\neq0$, we have $\det(\mathcal{S}_2-\mathcal{S}_1)\neq0$ and thus, $\mathcal{C}_2$ is well-defined. 
\end{proof}
\vspace*{-0.25em}
Given the relevance of the dominant and minimal solutions in Lemma \ref{aplemma_dynx:solution_k}, we now define these solution concepts following \citeauthor{Higham2000numerical} (\citeyear{Higham2000numerical}, Definition 5).
\begin{definition}\label{apdef_dynx:dominant_minimal_solutions}
Since $\mathcal{Q}(\mathcal{S})$ is a monic polynomial, it has exactly $2N$ finite eigenvalues, which we order by absolute value:
\begin{align}
\label{apeq_dynx:ordered_eigenvalues}
|\lambda_1| \geq |\lambda_2| \geq \dots \geq |\lambda_{2N}|.
\end{align}
Let $\mathcal{S}_1$ and $\mathcal{S}_2$ be two solutions of $\mathcal{Q}(\mathcal{S})$ where $\lambda(\mathcal{S}_1) = \{\lambda_i\}_{i=1}^N$ and $\lambda(\mathcal{S}_2) = \{\lambda_i\}_{i=N+1}^{2N}$. Then, $\mathcal{S}_1$ ($\mathcal{S}_2$) is the dominant (minimal) solution of $\mathcal{Q}(\mathcal{S})$ if $|\lambda_N|>|\lambda_{N+1}|$.
\end{definition}
\vspace*{-0.25em}
Definition \ref{apdef_dynx:dominant_minimal_solutions} implies that if both $\mathcal{S}_1$ and $\mathcal{S}_2$ exist, then $\lambda(\mathcal{S}_1)\cap\lambda(\mathcal{S}_2)=\emptyset$, since:
\begin{align}
\label{apeq_dynx:dominant_minimal_solutions_implication}
\min\{|\lambda_i|:\lambda_i\in\lambda(\mathcal{S}_1)\} > \max\{|\lambda_i|:\lambda_i\in\lambda(\mathcal{S}_2)\}.\text{\footnotemark}
\end{align}
\footnotetext{Moreover, the dominant and minimal solutions, if exist, are unique (\citeauthor{Gohberg2009matrix}, \citeyear{Gohberg2009matrix}, Theorem 4.1).}
The next result provides the sufficient conditions for the existence of these solutions.

\begin{lemma}\label{aplemma_dynx:existence_dominant_minimal_solution}
The quadratic eigenvalue problem associated with $\mathcal{Q}(\mathcal{S})$ is given by:
\begin{align}
\label{apeq_dynx:qep_K}
\mathcal{Q}(\lambda)v = (\lambda^2\mathbf{I}_N-\lambda\mathbf{\Psi}_1^\top+\mathbf{\Psi}_2^\top)v = \mathbf{0}_N.\text{\footnotemark}
\end{align}
\footnotetext{To see the connection between $\mathcal{Q}(\mathcal{S})$ and $\mathcal{Q}(\lambda)$, note that if $\mathcal{K}$ is a solution of $\mathcal{Q}(\mathcal{S})$, then any eigenpair $(\lambda_i,v_i)$ of $\mathcal{K}$ is a solution to $\mathcal{Q}(\lambda_i)v_i=\mathbf{0}_N$. To see this, note that: $\mathcal{Q}(\lambda_i)=(\mathbf{\Psi}_1^\top-\mathcal{K}-\lambda_i\mathbf{I}_N)(\mathcal{K}-\lambda_i\mathbf{I}_N)$. Since $\mathcal{K}v_i = \lambda_iv_i$, we have $(\mathcal{K}-\lambda_i\mathbf{I}_N)v_i=\mathbf{0}_N$ and $\mathcal{Q}(\lambda_i)v_i=\mathbf{0}_N$. Hence, the claim is shown.}
Suppose that the eigenvalues of $\mathcal{Q}(\lambda)$ is ordered as in \eqref{apeq_dynx:ordered_eigenvalues}, with $|\lambda_N| > |\lambda_{N+1}|$. In addition, suppose that there are two sets of linearly independent eigenvectors, $\{v_i\}_{i=1}^N$ and $\{v_i\}_{i=N+1}^{2N}$, corresponding to $\{\lambda_i\}_{i=1}^N$ and $\{\lambda_i\}_{i=N+1}^{2N}$.\footnote{If $\mathcal{Q}(\lambda)$ has $M$ distinct eigenvalues, where $N\leq M\leq 2N$, and the corresponding set of $M$ eigenvectors satisfies the Haar condition (i.e., each subset of $N$ eigenvectors is linearly independent), then the second condition in Lemma \ref{aplemma_dynx:existence_dominant_minimal_solution} is automatically satisfied.} Then, the dominant and minimal solutions exist.
\end{lemma}
\vspace*{-2em}
\begin{proof}
See \citeauthor{Higham2000numerical} (\citeyear{Higham2000numerical}, Theorem 6).
\end{proof}
\vspace*{-0.8em}

Before we present the limiting behavior of $\{\mathcal{X}_j\}$, we need one more result below: 

\begin{lemma}\label{aplemma_dynx:dominant_minimal_solution_relationship}
Let $\mathcal{S}_1$ and $\mathcal{S}_2$ denote the dominant and minimal solutions of $\mathcal{Q}(\mathcal{S})$, respectively. Then, $\mathcal{S}_1$ is nonsingular and, for any matrix norm, we have $\lim_{j\to\infty}\big|\big|\mathcal{S}_2^j\big|\big| \cdot \big|\big|\mathcal{S}_1^{-j}\big|\big|=0$.
\end{lemma}
\vspace*{-2em}
\begin{proof}
This result follows directly from \citeauthor{Gohberg2009matrix} (\citeyear{Gohberg2009matrix}, Lemma 4.9).
\end{proof}

\renewcommand{\theprop}{B.1} 
\begin{prop}
\label{approp_dynx:limit_x}
Let $\mathcal{S}_1$ and $\mathcal{S}_2$ be the dominant and minimal solutions of $\mathcal{Q}(\mathcal{S})$, respectively. If $\mathcal{C}_1$ is nonsingular, and $\mathcal{X}_j=\mathcal{K}_j^{-1}\mathcal{K}_{j-1}$ is defined for all $j\in\mathbb{Z}^+$, then 
\begin{align}
\label{apeq_dynx:limit_x}
\lim_{j\to\infty}\mathcal{X}_j = (\mathcal{S}_1^\top)^{-1} \equiv \underline{\mathcal{X}}.
\end{align}
\end{prop}
\vspace*{-2em}
\begin{proof}
From Lemma \ref{aplemma_dynx:solution_k} (ii), we have:
\begin{align*}
\mathcal{X}_j 
& =  \Big[\mathcal{C}_1^\top(\mathcal{S}_1^\top)^j + \mathcal{C}_2^\top(\mathcal{S}_2^\top)^j\Big]^{-1}\Big[\mathcal{C}_1^\top(\mathcal{S}_1^\top)^{j-1} + \mathcal{C}_2^\top(\mathcal{S}_2^\top)^{j-1}\Big] \\
& =  \Big(\Big[\mathcal{C}_1^\top(\mathcal{S}_1^\top)^j\Big]\Big[\mathbf{I}_N + (\mathcal{S}_1^\top)^{-j}(\mathcal{C}_1^\top)^{-1}\mathcal{C}_2^\top(\mathcal{S}_2^\top)^j\Big]\Big)^{-1}  \\ &\qquad\qquad\times\Big(\Big[\mathcal{C}_1^\top(\mathcal{S}_1^\top)^{j-1}\Big]\Big[\mathbf{I}_N + (\mathcal{S}_1^\top)^{-j+1}(\mathcal{C}_1^\top)^{-1}\mathcal{C}_2^\top(\mathcal{S}_2^\top)^{j-1}\Big]\Big)\\
& = \Big[\mathbf{I}_N + (\mathcal{S}_1^\top)^{-j}(\mathcal{C}_1^\top)^{-1}\mathcal{C}_2^\top(\mathcal{S}_2^\top)^j\Big]^{-1}(\mathcal{S}_1^\top)^{-1}\Big[\mathbf{I}_N + (\mathcal{S}_1^\top)^{-j+1}(\mathcal{C}_1^\top)^{-1}\mathcal{C}_2^\top(\mathcal{S}_2^\top)^{j-1}\Big].
\end{align*}
Using Lemma \ref{aplemma_dynx:dominant_minimal_solution_relationship} and the submultiplicative property of a matrix norm, it follows that:
\begin{align*}
 \big|\big| (\mathcal{S}_1^\top)^{-j}(\mathcal{C}_1^\top)^{-1}\mathcal{C}_2^\top(\mathcal{S}_2^\top)^j\big|\big| & \leq \big|\big|(\mathcal{S}_2^\top)^j\big|\big| \cdot \big|\big| (\mathcal{S}_1^\top)^{-j}\big|\big|\cdot\big|\big|(\mathcal{C}_1^\top)^{-1}\big|\big|\cdot\big|\big|\mathcal{C}_2^\top\big|\big| \to 0,
\end{align*}
which implies $\lim_{j\to\infty}(\mathcal{S}_1^\top)^{-j}(\mathcal{C}_1^\top)^{-1}\mathcal{C}_2^\top(\mathcal{S}_2^\top)^j = \mathbf{0}_{N}$. By similar argument, we have:
\begin{align*}
\lim_{j\to\infty}(\mathcal{S}_1^\top)^{-j+1}(\mathcal{C}_1^\top)^{-1}\mathcal{C}_2^\top(\mathcal{S}_2^\top)^{j-1} = \mathbf{0}_N.
\end{align*}
Hence, we have $\lim_{j\to\infty}\mathcal{X}_j = (\mathcal{S}_1^\top)^{-1}$, and the proof is complete.
\end{proof}

\vspace*{-0.8em}

Moreover, if $\mathcal{S}_2$ is nonsingular and $\overline{\mathcal{X}}\equiv(\mathcal{S}_2^\top)^{-1}$, then we have:
\begin{align*}
\min\{|\lambda_i|:\lambda_i\in\lambda(\overline{\mathcal{X}})\} > \max\{|\lambda_i|:\lambda_i\in\lambda(\underline{\mathcal{X}})\}.
\end{align*}
Thus, the sequence $\{\mathcal{X}_j\}$ converges to the minimal solution of a quadratic matrix equation with dominant and minimal solutions $\overline{\mathcal{X}}$ and $\underline{\mathcal{X}}$, respectively.
As will be clear in Appendix \ref{apsec:proof_thm_1}, a necessary condition for $\{\mathcal{M}(j;\theta)\}$ to have an \textit{economically-relevant} limit, i.e., $\lim_{j\to\infty}\mathcal{M}(j;\theta)\in\mathbb{R}^{N\times1}$, is $\underline{\mathcal{X}}\in\mathbb{R}^{N\times N}$. Thus, the following assumption is useful:
\begin{Ass}\label{apass_dynx:fixed_points_X}
The dominant solution of $\mathcal{Q}(\mathcal{S})$ consists of real entries, i.e., $\mathcal{S}_1\in\mathbb{R}^{N\times N}$.
\end{Ass}

\vspace*{-1.5em}

\setcounter{equation}{0}
\section{Proof of Theorem \ref{thm:saddle}}\label{apsec:proof_thm_1}
\vspace*{-1.1em}

In Proposition \ref{prop:impact_multiplier}, we show that $\mathcal{M}(\ell;\theta)\equiv\mathcal{M}_\ell$ satisfies the following recurrence:
\begin{align}
\label{apeq_thm1:intermediatestep_1}
\mathbf{A}^\ast\pazocal{M}_\ell & = \pazocal{X}_{\ell-1}\big(C_s^\ast + p_s\mathbf{A}^\ast\pazocal{M}_{\ell-1}\big) \text{ \ \ for all \ } \ell>1,
\end{align}
given an initial condition $\mathcal{M}_1$ and a sequence of nonsingular matrices $\mathcal{X}_1,\mathcal{X}_2, \dots, \mathcal{X}_{\ell-1}$. Pre-multiplying both sides of \eqref{apeq_thm1:intermediatestep_1} by $\left[\prod_{j=1}^{\ell-1}(p_s\mathcal{X}_j)\right]^{-1}$, and defining $\mathcal{V}_\ell$ by:\footnote{We adopt the notations $\prod_{j=1}^{\ell-1}(p_s\mathcal{X}_j) = (p_s\mathcal{X}_{\ell-1})(p_s\mathcal{X}_{\ell-2})\dots(p_s\mathcal{X}_2)(p_s\mathcal{X}_1)$ and $\prod_{j=1}^{0}(p_s\mathcal{X}_j) = \mathbf{I}_N$. Given $\ell\in\mathbb{Z}^+$, the matrix product $\prod_{j=1}^{\ell-1}(p_s\mathcal{X}_j)$ is nonsingular since each $\mathcal{X}_j$ is nonsingular.
}
\begin{align}
\label{apeq_thm1:intermediatestep_2}
\mathcal{V}_\ell = \Big[\prod_{j=1}^{\ell-1}(p_s\mathcal{X}_j)\Big]^{-1}\mathbf{A}^\ast\pazocal{M}_\ell,
\end{align}
we can rewrite \eqref{apeq_thm1:intermediatestep_1} as:
\begin{align}
\label{apeq_thm1:intermediatestep_3}
\mathcal{V}_\ell - \mathcal{V}_{\ell-1} = \Big[\prod_{j=1}^{\ell-2}(p_s\mathcal{X}_j)\Big]^{-1}\frac{1}{p_s}C_s^\ast  \text{ \ \ for \ } \ell\geq2,
\end{align}
Now, note that $\mathcal{V}_\ell$ can be expressed as a telescoping sum: $\mathcal{V}_\ell = (\mathcal{V}_\ell-\mathcal{V}_{\ell-1})+(\mathcal{V}_{\ell-1}-\mathcal{V}_{\ell-2})+\dots+(\mathcal{V}_3-\mathcal{V}_2)+\mathcal{V}_2$. Substituting \eqref{apeq_thm1:intermediatestep_3} into each term on the right-hand side of the telescoping sum, and using \eqref{apeq_thm1:intermediatestep_2} to replace $\mathcal{V}_\ell$ on the left-hand side, we obtain:
\begin{align}
\mathbf{A}^\ast\pazocal{M}_\ell & = \left\{\sum_{i=3}^{\ell}\Big[\prod_{j=1}^{\ell-1}(p_s\mathcal{X}_j)\Big]\Big[\prod_{j=1}^{i-2}(p_s\mathcal{X}_j)\Big]^{-1}\right\}\frac{1}{p_s}C_s^\ast + \Big[\prod_{j=2}^{\ell-1}(p_s\mathcal{X}_j)\Big]\mathbf{A}^\ast\pazocal{M}_2 \notag\\ 
& = \left\{\sum_{i=3}^{\ell}\Big[\prod_{j=i-1}^{\ell-1}(p_s\mathcal{X}_j)\Big]\right\}\frac{1}{p_s}C_s^\ast + \Big[\prod_{j=2}^{\ell-1}(p_s\mathcal{X}_j)\Big](p_s\mathcal{X}_1)\Big[\frac{1}{p_s}C_s^\ast + \mathbf{A}^\ast\pazocal{M}_1\Big] \notag \\
& = \left\{\sum_{i=2}^{\ell}\Big[\prod_{j=i-1}^{\ell-1}(p_s\mathcal{X}_j)\Big]\right\}\frac{1}{p_s}C_s^\ast + \Big[\prod_{j=1}^{\ell-1}(p_s\mathcal{X}_j)\Big]\mathbf{A}^\ast\pazocal{M}_1, \label{apeq_thm1:intermediatestep_4}
\end{align}
where the second equality follows from \eqref{apeq_thm1:intermediatestep_1}. In Appendix \ref{apsec:dynamics_X}, we show that $\mathcal{X}_j$ can be written as $\mathcal{X}_j= \mathcal{K}_j^{-1}\mathcal{K}_{j-1}$ for all $j\in\mathbb{Z}^+$, with $\mathcal{K}_0=\mathbf{I}_N$ and $\mathcal{K}_1=\mathcal{X}_1^{-1}$. Using this, we get $\prod_{j=1}^{\ell-1}(p_s\mathcal{X}_j) = p_s^{\ell-1}\mathcal{K}_{\ell-1}^{-1}$ and $\prod_{j=i-1}^{\ell-1}(p_s\mathcal{X}_j) = p_s^{\ell-i+1}\mathcal{K}_{\ell-1}^{-1}\mathcal{K}_{i-2}$ for $2\leq i\leq\ell$. Thus:
\begin{align}
\mathbf{A}^\ast\mathcal{M}_\ell & = p_s^{\ell-2}\mathcal{K}_{\ell-1}^{-1}\Big(\sum_{i=1}^{\ell-1}p_s^{-(i-1)}\mathcal{K}_{i-1}C_s^\ast + p_s\mathbf{A}^\ast\mathcal{M}_1\Big) \notag\\
& = p_s^{\ell-2}\mathcal{K}_{\ell-1}^{-1}\Big(\sum_{i=1}^{\ell-1}p_s^{-(i-1)}\Big[\mathcal{C}_1^\top(\mathcal{S}_1^\top)^{i-1}+\mathcal{C}_2^\top(\mathcal{S}_2^\top)^{i-1}\Big]C_s^\ast + p_s\mathbf{A}^\ast\mathcal{M}_1\Big) \notag\\
& = p_s^{\ell-2}\mathcal{K}_{\ell-1}^{-1}\Big(\mathcal{C}_1^\top\Big[\mathbf{I}_N-(\mathcal{S}_1^\top/p_s)^{\ell-1}\Big]\Big[\mathbf{I}_N-(\mathcal{S}_1^\top/p_s)\Big]^{-1}C_s^\ast \notag\\
&\qquad\qquad\qquad\quad + \mathcal{C}_2^\top\Big[\mathbf{I}_N-(\mathcal{S}_2^\top/p_s)^{\ell-1}\Big]\Big[\mathbf{I}_N-(\mathcal{S}_2^\top/p_s)\Big]^{-1}C_s^\ast + p_s\mathbf{A}^\ast\mathcal{M}_1\Big) \notag\\
& = \underbracket{p_s^{\ell-1}\mathcal{K}_{\ell-1}^{-1}\big\{\mathcal{C}_1^\top(p_s\mathbf{I}_N-\mathcal{S}_1^\top)^{-1}C_s^\ast + \mathcal{C}_2^\top(p_s\mathbf{I}_N-\mathcal{S}_2^\top)^{-1}C_s^\ast + \mathbf{A}^\ast\mathcal{M}_1\big\}}_{\text{first term}} \notag \\
-& \underbracket{\mathcal{K}_{\ell-1}^{-1}\mathcal{C}_1^\top(\mathcal{S}_1^\top)^{\ell-1}(p_s\mathbf{I}_N-\mathcal{S}_1^\top)^{-1}C_s^\ast}_{\text{second term}} - \underbracket{\mathcal{K}_{\ell-1}^{-1}\mathcal{C}_2^\top(\mathcal{S}_2^\top)^{\ell-1}(p_s\mathbf{I}_N-\mathcal{S}_2^\top)^{-1}C_s^\ast}_{\text{third term}}
\label{apeq_thm1:intermediatestep_5}
\end{align}
The second equality follows from the general solution for $\mathcal{K}_{i-1}$ (see Lemma \ref{aplemma_dynx:solution_k}), while the third equality holds because $\mathbf{I}_N-(\mathcal{S}_1^\top/p_s)$ and $\mathbf{I}_N-(\mathcal{S}_2^\top/p_s)$ are nonsingular, as the exogenous parameter $p_s$ is not an eigenvalue of either $\mathcal{S}_1$ or $\mathcal{S}_2$. Suppose now that $\mathcal{C}_1$ is nonsingular, then $\mathcal{K}_{\ell-1}^{-1}\mathcal{C}_1^\top(\mathcal{S}_1)^{\ell-1} \to \mathbf{I}_N$ in \eqref{apeq_thm1:intermediatestep_5}. Moreover, since $\mathcal{K}_{\ell-1}^{-1}\mathcal{C}_1^\top(\mathcal{S}^\top_1)^{\ell-1}$ $= \big(\mathbf{I}_N + (\mathcal{S}^\top_1)^{-(\ell-1)}(\mathcal{C}_1^\top)^{-1}\mathcal{C}_2^\top(\mathcal{S}^\top_2)^{\ell-1}\big)^{-1}$
and $\lim_{\ell\to\infty}(\mathcal{S}^\top_1)^{-(\ell-1)}(\mathcal{C}_1^\top)^{-1}\mathcal{C}_2^\top(\mathcal{S}^\top_2)^{\ell-1}=\mathbf{0}_N$ (as shown in the proof of Proposition \ref{approp_dynx:limit_x}), we deduce $\mathcal{K}_{\ell-1}^{-1}\mathcal{C}_1^\top(\mathcal{S}_1)^{\ell-1}\to\mathbf{I}_N$, and thus, the claim is verified. Moreover, since $\mathcal{K}_{\ell-1}^{-1}\mathcal{K}_{\ell-1} = \mathbf{I}_N$, it follows directly that $\mathcal{K}_{\ell-1}^{-1}\mathcal{C}_2^\top(\mathcal{S}^\top_2)^{\ell-1} = \mathbf{I}_N - \mathcal{K}_{\ell-1}^{-1}\mathcal{C}_1^\top(\mathcal{S}^\top_1)^{\ell-1}$. This implies $\mathcal{K}_{\ell-1}^{-1}\mathcal{C}_2^\top(\mathcal{S}^\top_2)^{\ell-1} \to \mathbf{0}_N$. Thus, the second and third terms in \eqref{apeq_thm1:intermediatestep_5} converge to:
\begin{align}
-(p_s\mathbf{I}_N-\mathcal{S}_1^\top)^{-1}C_s^\ast = -(p_s\mathbf{I}_N-\underline{\mathcal{X}}^{-1})^{-1}C_s^\ast = (\mathbf{I}_N-p_s\underline{\mathcal{X}})^{-1}\underline{\mathcal{X}}C_s^\ast,
\label{apeq_thm1:intermediatestep_6}
\end{align}
where we use the fact that $\underline{\mathcal{X}}=(\mathcal{S}_1^\top)^{-1}$ in the derivation above (see Proposition \ref{approp_dynx:limit_x}). To inspect the limiting behavior of the first term in \eqref{apeq_thm1:intermediatestep_5}, we consider two cases: (i) the largest eigenvalue of $p_s\underline{\mathcal{X}}$ (in absolute terms) is strictly below one, and (ii) it is strictly above one. In the first case, the dynamics of $\{\mathcal{M}_\ell\}$ behave like a \textit{sink}; in the second case, they resemble a \textit{saddle}. We treat these cases sequentially below.
\vspace*{-0.2em}

\noindent\textbf{Sink Dynamics.} Since we can write:
\begin{align*}
p_s^{\ell-1}\mathcal{K}_{\ell-1}^{-1} & = p_s^{\ell-1}\Big[\mathbf{I}_N+(\mathcal{S}_1^\top)^{-(\ell-1)}(\mathcal{C}_1^\top)^{-1}\mathcal{C}_2^\top(\mathcal{S}^\top_2)^{\ell-1}\Big]^{-1}(\mathcal{S}^\top_1)^{-(\ell-1)}(\mathcal{C}_1^\top)^{-1},
\end{align*}
where $(\mathcal{S}^\top_1)^{-(\ell-1)}(\mathcal{C}_1^\top)^{-1}\mathcal{C}_2^\top(\mathcal{S}^\top_2)^{\ell-1}\to\mathbf{0}_N$ and $(\mathcal{S}_1^\top/p_s)^{-(\ell-1)} = (p_s\underline{\mathcal{X}})^{\ell-1}\to\mathbf{0}_N$ (as all eigenvalues of $p_s\underline{\mathcal{X}}$ fall within the unit interval), it follows that $p_s^{\ell-1}\mathcal{K}_{\ell-1}^{-1}\to\mathbf{0}_N$ and hence, the first term in \eqref{apeq_thm1:intermediatestep_5} converges to $\mathbf{0}_N$. Regardless of $\mathcal{M}_1$, we then have:
\begin{align}
\mathcal{M}(\ell;\theta) \to (\mathbf{A}^\ast)^{-1} (\mathbf{I}_N-p_s\underline{\mathcal{X}})^{-1}\underline{\mathcal{X}}C_s^\ast = [\mathbf{I}_N-p_s(\mathbf{A}^\ast)^{-1}\underline{\mathcal{X}}\mathbf{A}^\ast]^{-1}(\mathbf{A}^\ast)^{-1}\underline{\mathcal{X}}C_s^\ast.
\label{apeq_thm1:intermediatestep_7}
\end{align}
\vspace*{-2.5em}

\noindent\textbf{Saddle Dynamics.} If at least one eigenvalue of $p_s\underline{\mathcal{X}}$ has an absolute value larger than one, then the term $p_s^{\ell-1}\mathcal{K}_{\ell-1}^{-1}$ in \eqref{apeq_thm1:intermediatestep_5} diverges. In this case, $\{\mathcal{M}_\ell\}$ converges to \eqref{apeq_thm1:intermediatestep_7} if and only if the expression within the curly brackets of the first term in \eqref{apeq_thm1:intermediatestep_5} equals the zero vector. That is:
\begin{align}
\mathbf{A}^\ast\pazocal{M}_1 & = -\big[\mathcal{C}_1^\top (p_s\mathbf{I}_N - \mathcal{S}_1^\top)^{-1} + \mathcal{C}_2^\top (p_s\mathbf{I}_N - \mathcal{S}_2^\top)^{-1}\big]C_s^\ast.
\label{apeq_thm1:intermediatestep_8}
\end{align}
In what follows, we show that $\mathcal{M}_1=\mathcal{M}^f_1$ and $\mathcal{X}_1=\mathcal{X}^f_1$, as given in \eqref{apeq_prop1:caliM_1}--\eqref{apeq_prop1:caliM_2}, satisfy \eqref{apeq_thm1:intermediatestep_8}, while the AR-NA initial conditions do not.
We begin by simplifying the terms:
\begin{align}
&\hspace{-.8em} \mathcal{C}_1^\top(p_s\mathbf{I}_N-\mathcal{S}_1^\top)^{-1}+\mathcal{C}_2^\top(p_s\mathbf{I}_N-\mathcal{S}_2^\top)^{-1} \notag\\
& = -\mathcal{C}_2^\top\big[(p_s\mathbf{I}_N-\mathcal{S}_1^\top)^{-1}-(p_s\mathbf{I}_N-\mathcal{S}_2^\top)^{-1}\big] + (p_s\mathbf{I}_N-\mathcal{S}_1^\top)^{-1} \label{apeq_thm1:intermediatestep_9}\\
& = -\mathcal{C}_2^\top(p_s\mathbf{I}_N-\mathcal{S}_2^\top)^{-1}(\mathcal{S}_1^\top-\mathcal{S}_2^\top)(p_s\mathbf{I}_N-\mathcal{S}_1^\top)^{-1} + (p_s\mathbf{I}_N-\mathcal{S}_1^\top)^{-1} \label{apeq_thm1:intermediatestep_10} \\
& = \big[\mathbf{I}_N+(\mathcal{K}_1-\mathcal{S}_1^\top)(\mathcal{S}_2^\top-\mathcal{S}_1^\top)^{-1}(p_s\mathbf{I}_N-\mathcal{S}_2^\top)^{-1}(\mathcal{S}_2^\top-\mathcal{S}_1^\top)\big](p_s\mathbf{I}_N-\mathcal{S}_1^\top)^{-1} \label{apeq_thm1:intermediatestep_11}\\ 
& = \big[\mathbf{I}_N+(\mathcal{K}_1-\mathcal{S}_1^\top)(p_s\mathbf{I}_N+\mathcal{S}_1^\top-\mathbf{\Psi}_1)^{-1} \big](p_s\mathbf{I}_N-\mathcal{S}_1^\top)^{-1}\label{apeq_thm1:intermediatestep_12}\\
& = \big(p_s\mathbf{I}_N-\mathbf{\Psi}_1+\mathcal{K}_1\big)\big[(p_s\mathbf{I}_N-\mathcal{S}_1^\top)(p_s\mathbf{I}_N+\mathcal{S}_1^\top-\mathbf{\Psi}_1)\big]^{-1} \notag\\
& = \big[p_s\mathbf{I}_N-\mathbf{\Psi}_1+\mathcal{K}_1\big]\big[p_s(p_s\mathbf{I}_N-\mathbf{\Psi}_1)+\mathbf{\Psi}_2\big]^{-1}
 \label{apeq_thm1:intermediatestep_13} \\
& = \big[p_s\mathbf{I}_N-\mathbf{\Psi}_1+\mathcal{K}_1\big]\mathbf{A}^\ast\big[\underbracket{p_s(p_s-\varrho)\mathbf{A}^\ast-(p_s-\varrho)\mathbf{I}_N+p_sB^\ast D}_{\mathcal{G}}\big]^{-1}.\label{apeq_thm1:intermediatestep_14}
\end{align}
To derive \eqref{apeq_thm1:intermediatestep_9}, we use the fact that $\mathcal{C}_1^\top=\mathbf{I}_N-\mathcal{C}_2^\top$ (see Lemma \ref{aplemma_dynx:solution_k}). To obtain \eqref{apeq_thm1:intermediatestep_10}, first observe that $\mathcal{S}_1^\top-\mathcal{S}_2^\top = (p_s\mathbf{I}_N-\mathcal{S}_2^\top)-(p_s\mathbf{I}_N-\mathcal{S}_1^\top)$. Then, pre-multiply both sides by $(p_s\mathbf{I}_N-\mathcal{S}_2^\top)^{-1}$ and post-multiply them by $(p_s\mathbf{I}_N-\mathcal{S}_1^\top)^{-1}$ to yield the result. Next, \eqref{apeq_thm1:intermediatestep_11} follows from the definition of $\mathcal{C}_2^\top$ (see Lemma \ref{aplemma_dynx:solution_k}). To establish \eqref{apeq_thm1:intermediatestep_12}, we proceed in three steps: (i) since $(\mathcal{S}_j^\top)^2-\mathcal{S}_j^\top\mathbf{\Psi}_1+\mathbf{\Psi}_2=\mathbf{0}_N$ for $j=1,2$ (Lemma \ref{aplemma_dynx:solution_k}), we take difference of these two quadratic matrix equations to yield $(\mathcal{S}_2^\top-\mathcal{S}_1^\top)^{-1}[(\mathcal{S}_2^\top)^2-(\mathcal{S}_1^\top)^2]=\mathbf{\Psi}_1$; (ii) use the result from step (i) to get $(\mathcal{S}_2^\top-\mathcal{S}_1^\top)(\mathcal{S}_1^\top-\mathbf{\Psi}_1)=-\mathcal{S}_2^\top(\mathcal{S}_2^\top-\mathcal{S}_1^\top)$, which implies $(\mathcal{S}_1^\top-\mathbf{\Psi}_1)=(\mathcal{S}_2^\top-\mathcal{S}_1^\top)^{-1}\mathcal{S}_2^\top(\mathcal{S}_2^\top-\mathcal{S}_1^\top)$; (iii) finally, use the fact that $(\mathcal{S}_2^\top-\mathcal{S}_1^\top)^{-1}(p_s\mathbf{I}_N-\mathcal{S}_2^\top)^{-1}(\mathcal{S}_2^\top-\mathcal{S}_1^\top)=\big[p_s\mathbf{I}_N - (\mathcal{S}_2^\top-\mathcal{S}_1^\top)^{-1}\mathcal{S}_2^\top(\mathcal{S}_2^\top-\mathcal{S}_1^\top)\big]^{-1}$ and the result from step (ii) to derive \eqref{apeq_thm1:intermediatestep_12}. Now, since $(\mathcal{S}_2^\top-\mathcal{S}_1^\top)$ and $(p_s\mathbf{I}_N-\mathcal{S}_2^\top)$ are both nonsingular, $p_s\mathbf{I}_N+\mathcal{S}_1^\top-\mathbf{\Psi}_1$ in \eqref{apeq_thm1:intermediatestep_12} is also nonsingular. The penultimate line in the above follows from noting that $(\mathcal{S}_1^\top)^2-\mathcal{S}_1^\top\mathbf{\Psi}_1=-\mathbf{\Psi}_2$, and the last equality follows from the definitions of $\mathbf{\Psi}_1$ and $\mathbf{\Psi}_2$ (see Appendix \ref{apsec:dynamics_X}). By construction, we have $\mathcal{K}_1=(\mathcal{X}_1)^{-1}$. Setting $\mathcal{X}_1=\mathcal{X}_1^f$ using the expression from \eqref{apeq_prop1:caliM_2}, we derive that:
\begin{align*}
p_s\mathbf{I}_N-\mathbf{\Psi}_1+(\mathcal{X}_1^f)^{-1} = \big\{p_s\mathbf{A}^\ast-\varrho\mathbf{A}^\ast\mathbf{\Omega}^f + p_s\mathbf{A}^\ast\mathbf{\Omega}^f\big[\varrho\mathbf{A}^\ast+q\mathbf{A}^\ast(\mathbf{I}_N-q\mathbf{A})^{-1}BD\big]\big\}(\mathbf{A}^\ast)^{-1},
\end{align*}
\vspace*{-5em}
where we use the expression for $\Omega^\ast_x$ (as given in footnote 24) to simplify the equation. Therefore, \eqref{apeq_thm1:intermediatestep_14} simplifies to:
\begin{align*}
& \big\{p_s\mathbf{A}^\ast-\varrho\mathbf{A}^\ast\mathbf{\Omega}^f + p_s\mathbf{A}^\ast\mathbf{\Omega}^f\big[\varrho\mathbf{A}^\ast+q\mathbf{A}^\ast(\mathbf{I}_N-q\mathbf{A})^{-1}BD\big]\big\}\mathcal{G}^{-1} \\
&\quad  = \big\{p_s\mathbf{A}^\ast-\varrho\mathbf{A}^\ast\mathbf{\Omega}^f + \varrho p_s\mathbf{A}^\ast\mathbf{\Omega}^f\mathbf{A}^\ast-p_s\mathbf{A}^\ast\mathbf{\Omega}^f\big[(\mathbf{\Omega}^f)^{-1}-\mathbf{I}_N+p_s\mathbf{A}^\ast+B^\ast D\big]\big\}\mathcal{G}^{-1} \\
&\quad = \mathbf{A}^\ast\mathbf{\Omega}^f\big\{-\varrho\mathbf{I}_N + \varrho p_s\mathbf{A}^\ast+p_s\big[\mathbf{I}_N-p_s\mathbf{A}^\ast-B^\ast D\big]\big\}\mathcal{G}^{-1} =  -\mathbf{A}^\ast\mathbf{\Omega}^f\mathcal{G}\mathcal{G}^{-1},
\end{align*}
which implies that the right-hand side of \eqref{apeq_thm1:intermediatestep_8} simplifies to $\mathbf{A}^\ast\mathbf{\Omega}^fC_s^\ast$. For \eqref{apeq_thm1:intermediatestep_8} to hold, $\mathcal{M}_1$ must equal $\mathbf{\Omega}^fC_s^\ast (=\mathcal{M}_1^f)$. This completes the proof that $(\mathcal{M}^f_1,\mathcal{X}^f_1)$ satisfies \eqref{apeq_thm1:intermediatestep_8}. Therefore, the initial conditions under our method guarantee that $\{\mathcal{M}_\ell\}$ converges as $\ell\to\infty$.\ We now turn to the AR-NA initial conditions.\ In the online appendix, we show that $\mathcal{M}^\phi_1=\mathcal{F}^{-1}(C_s^\ast+p_s\mathbf{A}^\ast\mathbf{\Omega}^\phi C_s)$, where $\mathcal{F} = \mathbf{I}_N-B^\ast D-\varrho\mathbf{A}^\ast\mathbf{\Omega}^\phi\Omega_xD$, $\mathbf{\Omega}^\phi=(-\mathbf{\Omega}_Y$ $-\Omega_xD)^{-1}$, $\mathbf{\Omega}_Y=-(\mathbf{I}_N-p_s\mathbf{A})+p_sq\mathbf{A}(\mathbf{I}_N-q\mathbf{A})^{-1}BD/\varrho$, and $\Omega_x= (\varrho-p_s)q\mathbf{A}(\mathbf{I}_N-$ $q\mathbf{A})^{-1}B/\varrho+B$; also, $\mathcal{X}_1^\phi = \mathbf{A}^\ast(\mathbf{I}_N-B^\ast D+\varrho\mathbf{A}^\ast-\varrho\mathbf{A}^\ast[\mathbf{I}_N-B^\ast D-\varrho\mathbf{A}^\ast\mathbf{\Omega}^\phi\Omega_xD]^{-1})^{-1}$. The right-hand side of \eqref{apeq_thm1:intermediatestep_8} becomes $\mathbf{A}^\ast\mathcal{F}^{-1}C_s^\ast + p_s\mathbf{A}^\ast\mathcal{F}^{-1}\mathbf{A}^\ast\Omega^\phi[p_s(p_s-\varrho)\mathbf{A}+p_sBD$ $-(p_s-\varrho)\mathbf{I}_N]\mathcal{G}^{-1}C_s^\ast$. Since $\mathbf{A}\neq\mathbf{A}^\ast$, $B\neq B^\ast$, and $C_s^\ast\neq C_s$, the right-hand side of \eqref{apeq_thm1:intermediatestep_8} does not equal the left-hand side.\ Hence, $\{\mathcal{M}_\ell\}$ diverges under these initial conditions, which completes the proof.
\vspace*{-0.2em}

\noindent\textbf{Nesting the MC-CF literature}. In the online appendix, we clarify that our framework nests a version without endogenous persistence in three ways: (i) $D=0_{1\times N}$, (ii) $\varrho=0$, or (iii) $\varrho\neq0$ and $B^\ast=B=0_{N\times1}$. In addition, we show that under MC-CF, $\mathcal{M}=(\mathbf{I}_N-$ $p_s\mathbf{A}^\ast)^{-1}C_s^\ast$ in cases (i) and (iii), whereas $\mathcal{M}=(\mathbf{I}_N-p_s\mathbf{A}^\ast-B^\ast D)^{-1}C_s^\ast$ in case (ii). In what follows, we show that the expression for $\mathcal{M}$ in \eqref{apeq_thm1:intermediatestep_7} reduces to the ones obtained in the MC-CF literature. We consider cases (i) and (iii) first. In both cases, we have $\mathcal{X}^f_1$ $=\mathcal{X}^\phi_1=\mathbf{A}^\ast$. The recursion in \eqref{apeq_prop1:caliX_equation} then implies $\underline{\mathcal{X}}=\mathbf{A}^\ast$. Substituting this into \eqref{apeq_thm1:intermediatestep_7} yields $\mathcal{M}=(\mathbf{I}_N-p_s\mathbf{A}^\ast)^{-1}C_s^\ast$. Next, for case (ii), we have $\underline{\mathcal{X}}=\mathbf{A}^\ast(\mathbf{I}_N-B^\ast D)^{-1}$. The expression in \eqref{apeq_thm1:intermediatestep_7} then reduces to $\mathcal{M}=(\mathbf{I}_N-p_s\mathbf{A}^\ast-B^\ast D)^{-1}C_s^\ast$, as desired.\qed

\vspace*{-1em}
\setcounter{equation}{0}
\section{Markov Restrictions in a Short-Lived ELB Spell}\label{apsec:app_markovres_shortspell}
\vspace*{-1.1em}

In Section \ref{subsec:application} of the paper, we consider the following model with consumption habits:
\begin{align}
\label{apeq:habitsmodel_backward}
c_{t+n} 		& = hc_{t+n-1} + \frac{1-h}{\sigma}\lambda_{t+n}, \\
\lambda_{t+n} 	& = \mathbb{E}_{t}\lambda_{t+n+1} - (r_{t+n} - \mathbb{E}_{t}\pi_{t+n+1} - \xi_{t+n}), \label{apeq:habitsmodel_euler}\\
\pi_{t+n} 		& = \beta\mathbb{E}_{t}\pi_{t+n+1} + \kappa\eta s_cc_{t+n} + \kappa\eta s_gg_{t+n} + \kappa\lambda_{t+n}, \label{apeq:habitsmodel_inflation}
\end{align}
where interest rates follow $r_{t+n}=\underline{r}$ for $n=0,1,\dots,\ell-1$, and $r_{t+n} = f(n;\theta)$ for $n\geq\ell$. Now, we let $\ell=1$ and denote the Markov chains by $\mathbf{C}_{t+n}$, $\mathbf{\Lambda}_{t+n}$, $\mathbf{\Pi}_{t+n}$, $\mathbf{R}_{t+n}$, $\mathbf{\Xi}_{t+n}$, and $\mathbf{G}_{t+n}$. For each $\mathbf{Z}_{t+n} \in \{\mathbf{C}_{t+n},\mathbf{\Lambda}_{t+n},\mathbf{\Pi}_{t+n},\mathbf{R}_{t+n},\mathbf{\Xi}_{t+n},\mathbf{G}_{t+n}\}$, it is characterized by:
\begin{align}
\label{apeq:markovstructure_ell1}
u^\top = \begin{bmatrix}
1 \\
0 \\
0 \\
0
\end{bmatrix},\qquad
\mathcal{P}_1=
\begin{bmatrix}
p_s & 1-p_s & 0 & 0\\
0 & p_b & 1-p_b  & 0\\
0 &  0 & q & 1-q\\
0 & 0 & 0 & 1
\end{bmatrix},\qquad
S_z = 
\begin{bmatrix}
s_{z,1}\\
s_{z,2}\\
s_{z,3}\\
0
\end{bmatrix}.
\end{align}
This allows us to compute $\mathbb{E}_{t}\mathbf{Z}_{t+n} = u\pazocal{P}_1^{n}S_z$ for any $n\geq0$. Thus, $\mathbb{E}_t\mathbf{Z}_t=u\mathbf{I}_NS_z=s_{z,1}$ and $\mathbb{E}_t\mathbf{Z}_{t+1}=p_ss_{z,1}+(1-p_s)s_{z,2}$. Moreover, $\mathbb{E}_{t,2}\mathbf{Z}_{t+1} = u_2\pazocal{P}_1S_z = p_bs_{z,2}+(1-p_b)s_{z,3}$ and $\mathbb{E}_{t,3}\mathbf{Z}_{t+1} = qs_{z,3}$, where $u_2 = [0,1,0,0]$ and $u_3=[0,0,1,0]$. (Here, $\mathbb{E}_{t,i}$ denotes the expectation of $\mathbf{Z}_t$ conditional on being in state $i$ at time $t$.) We need to determine $3\times6$ unknown Markov states, along with the parameter $q$.\ To begin, note that $\{s_{\xi,1}, s_{g,1}\}$ are exogenously determined, whereas $\{s_{\xi,2},s_{\xi,3},s_{g,2},s_{g,3}\}$ are determined so that $\mathbb{E}_t\mathbf{\Xi}_{t+n} =$ $p_b^ns_{\xi,1}$ and $\mathbb{E}_t\mathbf{G}_{t+n} = p_s^ns_{g,1}$ for $n\geq0$.\footnote{Specifically, we have $s_{\xi,2}=\Gamma s_{\xi,1}$ with $\Gamma=(p_b-p_s)/(1-p_s)$; $s_{\xi,3}=s_{g,2}=s_{g,3}=0$.} Next, we note that $q$ satisfies $q = h + qD(\mathbf{I}_N-$ $q\mathbf{A})^{-1}B$, where coefficient matrices $\mathbf{A}$, $B$, and $D$ are given in equation \eqref{eq:fwd_2} of the paper.\footnote{This polynomial typically admits multiple solutions for $q$. We select the one consistent with the minimum state variable principle of \cite{M83}. See the online appendix for details of the selection procedure. Expressions for $\mathbf{A}$, $B$, and $D$ in the habits model are also provided in the online appendix.
} 
The Markov states associated with $\mathbf{R}_{t+n}$ are given by $s_{r,1}= \underline{r}$, whereas $s_{r,2}$ and $s_{r,3}$ are specified to match the Taylor rule $p(n;\theta)=\phi_\pi\pi_{t+n} + \phi_\xi\xi_{t+n} + \phi_y(s_cc_{t+n}+s_gg_{t+n})$, i.e.
\begin{align}
\label{apeq:markovstates_interests}
s_{r,2} & = \phi_\pi s_{\pi,2} + \phi_\xi \Gamma s_{\xi,1} + \phi_ys_cs_{c,2} \quad\text{and}\quad s_{r,3} = \phi_\pi s_{\pi,3} + \phi_ys_cs_{c,3}.\footnotemark
\end{align}
\footnotetext{See the online appendix for a discussion on the choices of $\{s_{r,1}, s_{r,2}, s_{r,3}\}$.}
We are left with 9 unknown states for $\mathbf{C}_{t+n}$, $\mathbf{\Lambda}_{t+n}$, and $\mathbf{\Pi}_{t+n}$. They are uniquely determined by the following linear restrictions:
\begin{align}
s_{c,1} & = \frac{1-h}{\sigma}s_{\lambda,1}, \label{apeq:ell1_mres1_backward}\\
p_ss_{c,1} + (1-p_s)s_{c,2} & = hs_{c,1} + \frac{1-h}{\sigma}[p_ss_{\lambda,1} + (1-p_s)s_{\lambda,2}], \label{apeq:ell1_mres2_backward}\\
p_bs_{c,2} + (1-p_b)s_{c,3} & = hs_{c,2} + \frac{1-h}{\sigma}[p_bs_{\lambda,2} + (1-p_b)s_{\lambda,3}], \label{apeq:ell1_mres3_backward}\\
s_{\lambda,1} & = p_ss_{\lambda,1}+(1-p_s)s_{\lambda,2} + p_ss_{\pi,1} + (1-p_s)s_{\pi,2} + s_{\xi,1} - \underline{r}, \label{apeq:ell1_mres1_euler}\\
(1-p_b)s_{\lambda,2} & = (1-p_b)s_{\lambda,3} + p_bs_{\pi,2} + (1-p_b)s_{\pi,3} + \Gamma s_{\xi,1} - s_{r,2}, \label{apeq:ell1_mres2_euler}\\
(1-q)s_{\lambda,3} & = -(\phi_\pi-q)s_{\pi,3} - \phi_ys_cs_{c,3}, \label{apeq:ell1_mres3_euler}\\
s_{\pi,1} & = \beta[p_ss_{\pi,1} + (1-p_s)s_{\pi,2}] + \kappa\eta s_cs_{c,1} + \kappa\eta s_gs_{g,1} + \kappa s_{\lambda,1}, \label{apeq:ell1_mres1_phillips}\\
s_{\pi,2} & = \beta[p_bs_{\pi,2} + (1-p_b)s_{\pi,3}] + \kappa\eta s_cs_{c,2} + \kappa s_{\lambda,2}, \label{apeq:ell1_mres2_phillips}\\
(1-\beta q)s_{\pi,3} & = \kappa s_{\lambda,3} + \kappa \eta s_cs_{c,3}. \label{apeq:ell1_mres3_phillips}
\end{align}
These restrictions are derived from equations (10)--(15) in the online appendix. Specifically, \eqref{apeq:ell1_mres1_backward}--\eqref{apeq:ell1_mres3_backward} are obtained by solving $\mathbb{E}_t\mathbf{C}_{t+n} = h\mathbb{E}_t\mathbf{C}_{t+n-1} + \frac{1-h}{\sigma}\mathbb{E}_t\mathbf{\Lambda}_{t+n}$ for $n=0,$ $1,2$, while \eqref{apeq:ell1_mres1_euler}--\eqref{apeq:ell1_mres3_euler} and \eqref{apeq:ell1_mres1_phillips}--\eqref{apeq:ell1_mres3_phillips} are derived by solving $\mathbb{E}_{t,i}\mathbf{\Lambda}_t = \mathbb{E}_{t,i}\mathbf{\Lambda}_{t+1} -$ $(\mathbb{E}_{t,i}\mathbf{R}_t - \mathbb{E}_{t,i}\mathbf{\Pi}_{t+1}-\mathbb{E}_{t,i}\mathbf{\Xi}_t)$ and $\mathbb{E}_{t,i}\mathbf{\Pi}_t = \beta\mathbb{E}_{t,i}\mathbf{\Pi}_{t+1} + \kappa\eta s_c\mathbb{E}_{t,i}\mathbf{C}_{t} + \kappa\eta s_g\mathbb{E}_{t,i}\mathbf{G}_t + \kappa\mathbb{E}_{t,i}\mathbf{\Lambda}_t$
for $i=1,2,3$, where $\mathbb{E}_{t,1}\mathbf{Z}_{t+n}\equiv\mathbb{E}_t\mathbf{Z}_{t+n}$.\ In the online appendix, we show that restrictions \eqref{apeq:ell1_mres1_backward}--\eqref{apeq:ell1_mres3_phillips}, along with the states for $\mathbf{R}_{t+n}$, $\mathbf{\Xi}_{t+n}$, $\mathbf{G}_{t+n}$, ensure that the expected paths of all Markov chains satisfy \eqref{apeq:habitsmodel_backward}--\eqref{apeq:habitsmodel_inflation} for all $n\geq0$.  If $h=0$, then $q=0$. Moreover, if $p_b=p_s$, then only $s_{z,1}\neq0$ for $\mathbf{Z}_{t+n} \in \{\mathbf{C}_{t+n},\mathbf{\Lambda}_{t+n},\mathbf{\Pi}_{t+n},\mathbf{R}_{t+n},\mathbf{\Xi}_{t+n},\mathbf{G}_{t+n}\}$, while $s_{z,2}$ $=s_{z,3}=0$. In this case, we are left with \eqref{apeq:ell1_mres1_backward}, \eqref{apeq:ell1_mres1_euler}, and \eqref{apeq:ell1_mres1_phillips}. Simplifying them, we obtain $s_{c,1} = p s_{c,1}- \frac{1}{\sigma}(\underline{r} - p_ss_{\pi,1} - s_{\xi,1})$ and $s_{\pi,1} = \beta p_ss_{\pi,1} + \kappa(\sigma + \eta s_c)s_{c,1} + \kappa\eta s_gs_{g,1}$. This system of two equations in turn matches the AD--AS equations that arise from the standard NK model studied in \cite{Eggertsson2011}. As in that paper, we use \eqref{apeq:ell1_mres1_backward}--\eqref{apeq:ell1_mres3_phillips} to compute the slopes of the AD--AS equations at the ELB, i.e., on impact of the risk premium shock. Explicit expressions of the AD--AS equations under \eqref{apeq:ell1_mres1_backward}--\eqref{apeq:ell1_mres3_phillips} are provided in the online appendix. We also derive a system of restrictions that applies when $\ell\to\infty$. In this case, we show in the online appendix that the parameter $q$ is replaced with $q^\ast$ and $s_{r,i}$ $=0$ for $i=1,2,3$.

\vspace*{-1em}

\setcounter{equation}{0}
\section{Proof of Proposition \ref{prop:p^Sinf_pD}}\label{apsec:proof_prop_2}
\vspace*{-1.1em}

When $\ell\to\infty$, the AD and AS equations under AR-NA can be derived as:
\begin{align*}
s_{c,1} & = \pazocal{S}_{AD}(q^\ast)s_{\pi,1} + \text{t.i.p.} \quad\text{and}\quad s_{c,1} = \pazocal{S}_{AS}(q^\ast)s_{\pi,1} + \pazocal{S}_G(q^\ast)s_{g,1} + \pazocal{S}_\Xi(q^\ast)s_{\xi,1},\footnotemark
\end{align*}
\footnotetext{Explicit expressions of the AD--AS equations for the habits model are provided in the online appendix.}
where $\text{t.i.p.}$ is independent of policy. Expressing these in matrices, we obtain:
\begin{align*}
\begin{bmatrix}
1 & -\pazocal{S}_{AD}(q^\ast) \\
1 & -\pazocal{S}_{AS}(q^\ast)
\end{bmatrix}
\begin{bmatrix}
s_{c,1} \\
s_{\pi,1}
\end{bmatrix}
& = 
\begin{bmatrix}
0 \\
\pazocal{S}_G(q^\ast)
\end{bmatrix}
s_{g,1} + 
\begin{bmatrix}
0 \\
\pazocal{S}_\Xi(q^\ast)
\end{bmatrix}
s_{\xi,1} + 
\begin{bmatrix}
\text{t.i.p.} \\
0
\end{bmatrix}.
\end{align*}
The multiplier $\mathcal{M}$ can then be derived as follows:
\begin{align}
\label{prop8:eq1}
\pazocal{M} & = 
\begin{bmatrix}
1 & -\pazocal{S}_{AD}(q^\ast) \\
1 & -\pazocal{S}_{AS}(q^\ast)
\end{bmatrix}^{-1}
\begin{bmatrix}
0 \\
\pazocal{S}_G(q^\ast)
\end{bmatrix}\equiv
\mathcal{U}_1^{-1}
\begin{bmatrix}
0 \\
\pazocal{S}_G(q^\ast)
\end{bmatrix}.
\end{align}
From Theorem \ref{thm:saddle}, we know that $\pazocal{M}$ can equivalently be expressed as:
\begin{align}
\label{prop8:eq2}
\pazocal{M} & = \big[\mathbf{I}_N - p_s(\mathbf{A}^\ast)^{-1}\underline{\mathcal{X}}\mathbf{A}^\ast\big]^{-1}(\mathbf{A}^\ast)^{-1}\underline{\mathcal{X}}C_s^\ast \equiv\mathcal{U}_2^{-1}(\mathbf{A}^\ast)^{-1}\underline{\mathcal{X}}C_s^\ast.
\end{align}
It suffices to prove that $p_s=p^D$ implies $p_s=\overline{p}_s(q^\ast)$. When $p_s=p^D$, i.e., $\rho(p_s\underline{\mathcal{X}})=1$, $\mathcal{U}_2$ is singular since $\rho\left((\mathbf{A}^\ast)^{-1}p_s\underline{\mathcal{X}}\mathbf{A}^\ast\right)= 1$.\ Thus, $\mathcal{M}$ is not defined under AR-NA.\ This implies that $\mathcal{U}_1$ must be singular. This occurs only if $p_s$ takes a value such that $\pazocal{S}_{AD}(q^\ast)$ $=\pazocal{S}_{AS}(q^\ast)$. Hence, we must have $p_s=\overline{p}_s(q^\ast)$, and the proof is complete.
\vspace*{-0.2em}

We note that $\overline{p}_s(q)\neq\overline{p}_s(q^\ast)$ follows directly from $q\neq q^\ast$, $\mathcal{S}_{AD}(q)\neq\mathcal{S}_{AD}(q^\ast)$, and $\mathcal{S}_{AS}(q)\neq\mathcal{S}_{AS}(q^\ast)$. When $h\to0$ in the habits model, the AD and AS equations reduce to the one studied in \cite{Eggertsson2011}. Thus, the threshold value $p^D$ is simply the one from the MC-CF literature.\qed

\end{document}